\documentclass[apj]{emulateapj}
\usepackage{multirow}
\usepackage{booktabs}
\usepackage{graphicx}
\usepackage{amsmath}
\usepackage{longtable}
\usepackage{rotating}
\usepackage{color}
\usepackage{enumerate}
\newcommand{\kep}{\emph{Kepler\ }}

\newcommand{\msun}{M_\odot}
\newcommand{\rmh}{R_{\rm H}}
\newcommand{\rgap}{r_{\rm gap}}
\newcommand{\rinner}{r_{\rm inner\ disk}}
\newcommand{\mj}{M_{\rm J}}
\newcommand{\mplanet}{M_{\rm p}}
\newcommand{\sigmagas}{\Sigma_{\rm gas}}
\newcommand{\deltagap}{\delta_{\rm gap}}
\newcommand{\deltagapmin}{\delta_{\rm gap,min}}
\newcommand{\deltagapi}{\delta_{\rm gap,i}}

\newcommand{\widthgapi}{\Delta_{\rm gap,i}}
\newcommand{\qcrit}{q_{\rm crit}}
\newcommand{\qgap}{q_{\rm gap}(\deltagapmin|\alpha,h/r)}
\newcommand{\ftdobs}{f_{\rm TD,observed}}
\newcommand{\ftdin}{f_{\rm TD,intrinsic}}
\newcommand{\fplanet}{f_{\rm p}}
\newcommand{\tu}{\tau_{\rm u}}
\newcommand{\tuq}{\tau_{\rm u,q}}
\newcommand{\icarus}{{\it Icarus}}

\begin{document}
\title{Stability and Occurrence Rate Constraints on the Planetary Sculpting Hypothesis for ``Transitional'' Disks}
\shorttitle{Planets in transitional disks}

\shortauthors{Dong \& Dawson}

\author{Ruobing Dong}
\affil{Nuclear Science Division, Lawrence Berkeley National Lab, Berkeley, CA 94720; rdong2013@berkeley.edu; Department of Astronomy, University of California, Berkeley; NASA Hubble Fellow}
\author{Rebekah Dawson}
\affil{Department of Astronomy, University of California, Berkeley; Miller Fellow; Department of Astronomy \& Astrophysics, The Pennsylvania State University; Center for Exoplanets and Habitable Worlds, The Pennsylvania State University}

\clearpage

\begin{abstract}
Transitional disks, protoplanetary disks with deep and wide central gaps, may be the result of planetary sculpting. By comparing numerical planet-opening-gap models with observed gaps, we find systems of 3--6 giant planets are needed in order to open gaps with the observed depths and widths. We explore the dynamical stability of such multi-planet systems using $N$-body simulations that incorporate prescriptions for gas effects. We find they can be stable over a typical disk lifetime, with the help of eccentricity damping from the residual gap gas that facilitates planets locking into mean motion resonances. However, in order to account for the occurrence rate of transitional disks, the planet sculpting scenario demands gap-opening-friendly disk conditions, in particular, a disk viscosity $\alpha\lesssim0.001$. In addition, the demography of giant planets at $\sim 3-30$ AU separations, poorly constrained by current data, has to largely follow occurrence rates extrapolated outward from radial velocity surveys, not the lower occurrence rates extrapolated inward from direct imaging surveys. Even with the most optimistic occurrence rates, transitional disks cannot be a common phase that most gas disks experience at the end of their life, as popularly assumed, simply because there are not enough planets to open these gaps. Finally, as consequences of demanding almost all giant planets at large separations participate in transitional disk sculpting, the majority of such planets must form early and end up in a chain of mean motion resonances at the end of disk lifetime.

\end{abstract}



\section{Introduction}\label{sec:intro}

Planets are born in protoplanetary disks. Planets can produce radial and azimuthal features in the disks observable in resolved images at various wavelengths \citep[e.g., simulations by][]{fouchet10, pinilla12-diffcavsize, gonzalez12, dejuanovelar13, dong15-gaps, juhasz15,dong15-spiralarms,pohl15}. Sculpting by planets is a leading hypothesis for the origin of transitional disks, protoplanetary disks with depleted inner cavities or gaps (see \citealt{espaillat14} for a recent review; hereafter we will refer to the depleted inner regions as gaps although some may be complete cavities). The existence of transitional disks was first suggested by the distinctive dip at $\sim1-10\micron$ in the spectral energy distributions \citep[SED;][]{storm89, skrutskie90, calvet02, espaillat07} of some young stars and subsequently confirmed in resolved observations in near-infrared (NIR) scattered light \citep[e.g.,][]{mayama12, muto12, hashimoto12, canovas13, garufi13, grady13, avenhaus14, tsukagoshi14} and mm dust continuum and molecular line emission \citep[e.g.,][]{andrews11, casassus13, fukagawa13, vandermarel13, perez14, zhang14}. 

The gaps in transitional disks are both deep and wide. In gas they typically extend several tens of AU \citep[e.g.,][]{zhang14,vandermarel15,vandermarel16}. Measuring their depths in gas has been more challenging, but recent ALMA observations have suggested that the gaps are depleted by at least a factor of 10--1000 \citep[e.g.,][]{vandermarel15, vandishoeck15, vandermarel16}. Non-dynamical mechanisms to open these gaps have been proposed, including photoevaporation \citep[e.g.,][]{clarke01,alexander06, owen12, rosotti13}, grain growth \citep[e.g.,][]{dullemond05, birnstiel12}, and the  magnetorotational instability \citep[e.g.,][]{chiang07}. However, all have major drawbacks and cannot explain at least a subset of protoplanetary disks \citep{espaillat14}. 

In this study, we will focus on the planetary sculpting scenario, in which the inner hole is a common gap opened by multiple planets \citep{dodsonrobinson11,zhu11}. Numerical simulations of the planetary sculpting scenario have reproduced all major observed aspects of transitional disks. \citet{dong15-gaps} showed that the width and depth of planet-opened common gaps can match observations at both NIR and mm wavelengths. In particular, observations have systematically found bigger gap sizes in the mm continuum than in the gas or scattered light \citep[e.g.,][]{dong12cavity, hashimoto15, zhang14, vandermarel15}. The planet-opening-gap scenario has successfully reproduced this feature by trapping the large dust at the pressure peak beyond the gas gap edge \citep{pinilla12-diffcavsize, zhu12, dejuanovelar13}. Millimeter observations have revealed large scale azimuthal asymmetries in a few systems, for which the mm dust ring is lopsided and most emission comes from one side \citep{vandermarel13, casassus13, isella13, perez14}. Vortex formation at the planet-induced gap edge has been proposed to explain these observations \citep[e.g.,][but also see \citealt{mittal15}]{zhu14stone, lyra13}. Finally, accretion rate analysis also suggests the presence of giant planets inside the gaps to alter the accretion flow onto the stars \citep{najita15}.

Despite the success of the planet sculpting hypothesis in explaining observations of transitional disks, only a handful of companion candidates have been discovered in direct imaging observations of transitional disks (e.g., \citealt{huelamo11, biller12, kraus12, quanz13-planet, brittain14, close14,  reggiani14}). Direct imaging of planets is difficult due to the low contrast ratio between the brightness of planets and their host stars, and the proximity of the planets to the stars\footnote{A planet at 30~AU is only $\sim$0.2$\arcsec$ away from its star at 140 pc, a typical distance for nearby resolved protoplanetary disks.}. In addition, most discovered companions are likely too far away from the outer gap edges to be solely responsible for the gap. So far the strongest direct evidence for the multi-planet sculpting scenario is the detection of three companions inside the wide gap in LkCa 15 \citep{sallum15}.
If the gaps in the younger protostellar disk HL Tau \citep{brogan15} are opened by sub-Jupiter mass planets \citep{dong15-gaps,dipierro15-hltau}, it may be a transitional disk ``embryo'' that will develop a wide common gap once the planets grow larger. The HR~8799 system, home to at least 4 giant planets at tens of AU orbiting a 30~Myr old young star \citep{marois08,marois10}, may be a later stage in the post-gas-disk era.  It remains an open question whether systems of massive giant planets are sufficiently common to account for the occurrence rate of transitional disks. 

Furthermore, it is unclear whether the large masses and close spacings of planets required to open common gaps are stable over transitional disks' $\sim$Myr lifetimes. \citet{duffell15dong} showed that a common gap is shallower than the gap open by a single planet of the same mass and the material between the two planets' gaps cannot be eroded away. Under typical disk conditions, a wide common gap with a depletion factor larger than 10 throughout the gap requires several massive planets with $q=\mplanet/M_\star\gtrsim2\times10^{-3}$ (or $\sim$2$\mj$ planets around a 1~$\msun$ star) spanning several to several tens of AU. The tension between close packing and stability has been explored for the HR 8799 system \citep{fabrycky10} and the HL Tau protoplanetary disk \citep{brogan15, tamayo15}, which features a series of gaps that may be sculpted by $\sim$~Saturn mass planets \citep{dong15-gaps, dipierro15-hltau}. \citet{sallum15} also suggested that the planets in LkCa 15's gap have to be either less massive than $5\mj$ or in 2:1 mean motion resonance in order for the system to be stable.

Here we assess the plausibility of the planetary sculpting hypothesis using constraints from stability and occurrence rates and explore the implications for the planetary systems' subsequent evolution. We start by defining the properties of observed gaps in transitional disks in Section~\ref{sec:gapproperties}. In Section~\ref{sec:planetrequirements} we synthesize the requirements for configurations of multi-planet systems that are able to open gaps consistent with transitional disk observations. We then carry out $N$-body simulations in Section~\ref{sec:stability} to explore whether such configurations can be stable. In Section~\ref{sec:statistics}, we compare the properties of the multi-planet systems that can create transitional disks with current statistics on the occurrence rate of giant planets at large separations. A summary and discussion are presented in Section~\ref{sec:summary}.

\section{Observed properties of the gaps}\label{sec:gapproperties}

Here we lay out the observed properties of gaps that planetary sculpting must account for. We focus primarily on the gaps opened in the gas (``gas gaps'') because the dynamics of dust is subject to dust-gas coupling in addition to disk-planet interactions. 

\subsection{Gap extent}

High resolution and high sensitivity resolved molecular gas observations are needed to determine the gap sizes. \citet{vandermarel15,vandermarel16} performed such observations for 8 systems and found a typical gas gap extent of $\sim$15-45 AU. We will use 30 AU as the fiducial value in the models throughout the paper. 

An important related quantity is the continuity of the gap, which constrains the spacing of the sculpting planets. The current image resolution does not rule out the possibility of undepleted narrow rings of material within the gap (e.g., between planets). However, because the total mass of gas in the gap is constrained (except the inner $\sim$~1AU or so, see below), undepleted rings cannot compromise a significant portion of the gap extent.

\subsection{Depth (depletion)}\label{sec:gapdepletion}

Here we parametrize the gas gap depletion as $\deltagap=\Sigma_0/\Sigma_{\rm gap}$, where $\Sigma_{\rm gap}$ is the gas surface density in the gap and $\Sigma_0$ is the undepleted value extrapolated from the outer disk. Observations use relatively rare species, such as $^{13}$CO and C$^{18}$O, to infer the total gas abundance at a given location\footnote{An additional layer of complication is that CO is condensable at the low temperatures in the outer parts of a disk, and this needs to be taken into account when using CO observations to constrain the density structure of the gas in total.}. To date, most studies have constrained only lower limits on $\deltagap$ (hereafter $\deltagapmin$). Getting robust constraints on $\Sigma_{\rm gap}$ has been difficult. The abundance of these rare species depends not only on the gas surface density but on the temperature conditions of the disk and chemical reactions with other species. Model degeneracies in the radial profile of the gas are also a challenge  (e.g., \citealt{bruderer14}).

Recently  $\deltagap$ (or $\deltagapmin$) has been measured in a few individual systems by combining the latest ALMA data with state of the art disk physical-chemical modeling tools. \citet{bruderer14} found that the IRS 48 system's mm data is consistent with $\deltagap \sim 10$ inside 60 AU and $\deltagap \sim 100$ inside 20 AU. \citet{perez15} found that in the HD 142527 system, the gas inside the $\sim100$~AU gap is depleted by $\deltagap\sim50$. Using ALMA CO observations, \citet{vandermarel15} determined $\deltagap \gtrsim 10$ for LkCa 15, $\deltagap \gtrsim 10^4$ for RXJ1615-3255, and $\deltagap \sim 10^5$ for J1604-2130; using $^{13}$CO and C$^{18}$O observations, \citet{vandermarel16} have recently pushed $\deltagap$ to $\sim5000$ for SAO 206462; $\sim100$ for SR 21; $\gtrsim10^4$ for DoAr 44; and $\gtrsim1000$ for IRS 48.

Alternatively, $\deltagap$ can be inferred from the gap depletion factor of the sub-$\micron$-sized grains -- which are traced by scattered light imaging -- because these fine grains are generally well-mixed with gas such that $\deltagap \approx \delta_{\rm fine\ grains}$. \citet{dong12pds70} found that the scattered light from the $\sim70$~AU gap in PDS~70 is consistent with $\delta_{\rm fine\ grains} \sim1000$. 

All the estimations of $\deltagap$ described above generally assume that the gap is uniformly (i.e., azimuthally symmetrically) depleted from a fiducial profile (usually taking the form of $\Sigma\propto1/r$ and sometimes including an exponential cut off at large radius) that describes the undepleted disk and that the gap edge is sharp. Azimuthal variations inside the gap, for example accretion streamers connecting the gap edge to the planets or the star \citep{casassus13}, may account for some remaining gas, though current data generally lack the needed angular resolution to resolve these structures.

The current observed $\deltagapmin$ span 10-1000, but we do not yet have a large enough sample to understand the selection effects. Below we will use $\deltagapmin=10$ as a conservative requirement for our planet-opened gaps and also assess our results for $\deltagapmin=100$ and 1000.

\subsection{The inner disk}\label{sec:inner_disk}

The gas depletion in the very inner part of the disk (i.e., $\sim$AU scales) is practically unconstrained based on current ALMA gas observations and chemical disk modeling. Interpreting mm gas emissions from the inner disk is difficult due to the challenge of chemical disk modeling with the high but uncertain temperatures close to the star. CO ro-vibrational emissions at 4.7$\micron$ indicate the existence of gas at $\sim$AU scale in at least some cases \citep[e.g.,][]{pontoppidan08,salyk09}; however the amount of gas is not well determined. In the analysis below, we will simply consider the gas surface density $\sigmagas$ in the inner $\sim$AU scale disk as unconstrained.

\subsection{Occurrence rate, $\ftdobs$}\label{sec:ftd}

Studies of young clusters have revealed that on the order of 10\% of protoplanetary disks are transitional disks with significant gap sizes. Based on analysis of Spitzer photometry, \citet{luhman10} found that the fraction of protoplanetary disks that are transitional disks, $\ftdobs$, is $\sim$13\% (15/113) in Taurus\footnote{\citet{luhman10} defined transitional disks as objects with weak or no excess at $\lambda<10\micron$, and large excess at longer wavelengths. SED modeling \citep[e.g.,][]{espaillat07,andrews11} has shown that this kind of SED is indicative of large gap sizes of typically about 30 AU.}, consistent with a number of other star forming regions at 2-10 Myr such as Chamaeleon I and IC 348. \citet{muzerolle10} found a similar $\ftdobs$ --- $\sim$12\% (51/417) --- for a sample of 7 clusters\footnote{We include both the  ``warm'' and ``weak excess'' transitional disks in \citet{muzerolle10} in the fraction, as both are likely to comply with our definition of transitional disks.}. Note that these fractions are computed for samples of host stars with a range of properties, such as spectral type and age.
\citet{muzerolle10} found $\ftdobs$ may depend on age and spectral type but the sample size is too small to robustly test correlations; \citet{andrews11} found $\ftdobs$ is higher around mm bright disks ($\sim25\%$ for the upper quartile of the total mm flux distribution). It is not yet known whether this fraction primarily reflects nature ($\sim 10\%$ disks develop wide gaps) or nurture (most disks spend $\sim 10\%$ of their lifetime in the transitional phase). We will further explore the interpretation $\ftdobs$ in Section~\ref{sec:statistics}.

\subsection{Summary}\label{sec:gapsummary}

\begin{figure}
\begin{center}
\includegraphics[trim=0 0 0 0, clip,width=\textwidth,angle=0]{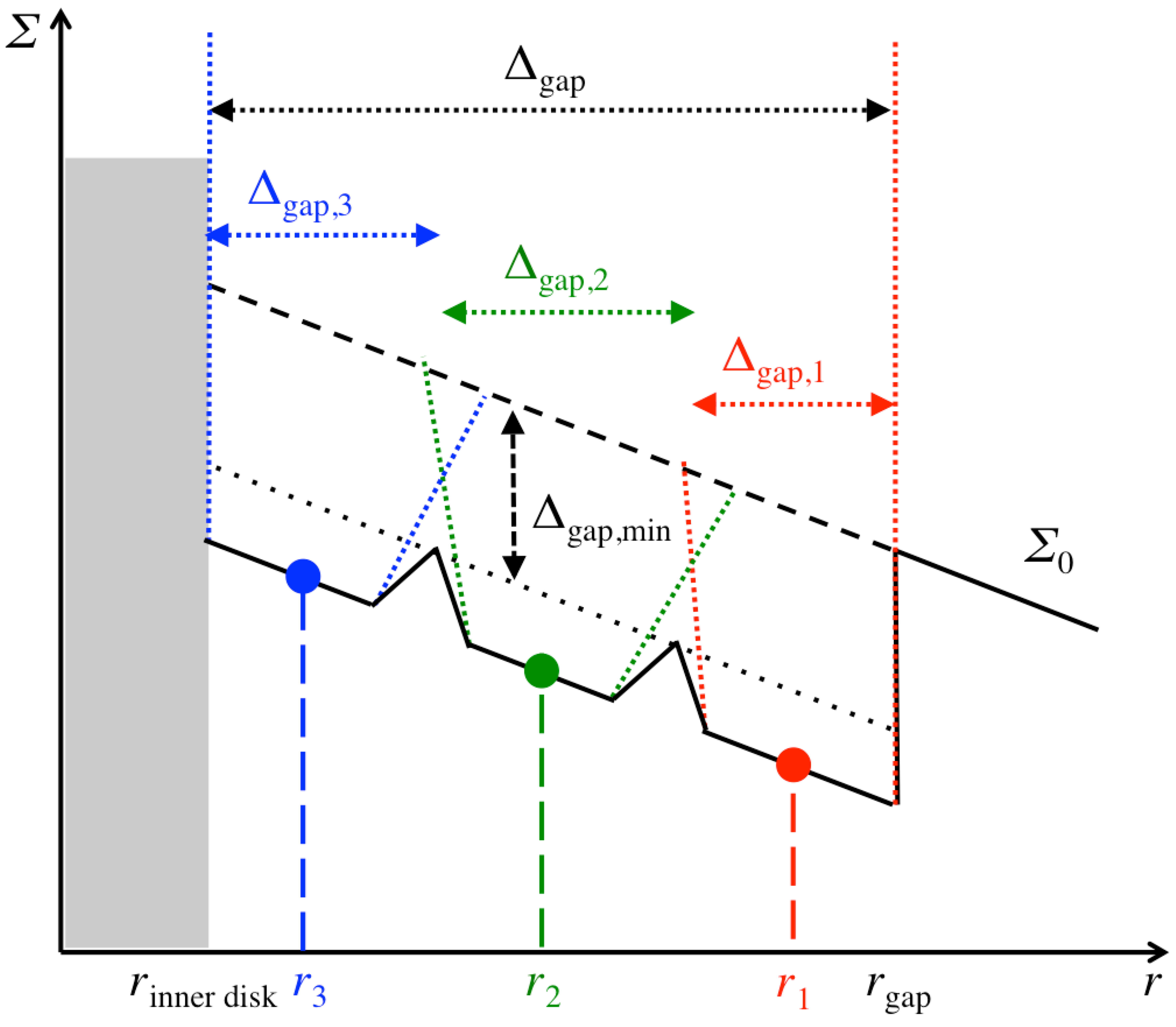}
\end{center}
\figcaption{Illustration of a common gap opened by 3 planets located at $r_1$, $r_2$, and $r_3$. $\Sigma_0$ (and its dashed line extension in the gap) indicates the original undepleted gas surface density. $\Delta_{\rm gap,1}$, $\Delta_{\rm gap,2}$, and $\Delta_{\rm gap,3}$ are the widths of the individual gaps opened by each planet if it were alone in the disk. $\deltagapmin$ is the minimum depletion factor inside the gap, which in the illustration is only reached at the joint edges of individual gaps, whereas most of the gap has a larger depletion factor. The shaded region marks the poorly constrained inner disk. See Section~\ref{sec:gapproperties} for details.
\label{fig:illustration_gap}}
\end{figure}

Based on the observed characteristics of transitional disks discussed above, we synthesize a set of properties for a typical transitional disk gap (illustrated in Figure~\ref{fig:illustration_gap}),
which we will use as fidicuial requirements in later explorations of planetary system configurations:
\begin{enumerate}
\item The gap is approximately continuous with no wide, interspersed un-depleted regions inside\footnote{Relaxing this requirement to a weaker one, that the total gas mass inside the gap is depleted by at least a factor of 10 (or 100, 1000) --- a more direct result from modeling ALMA observations --- makes little difference. If surface density scales with radius as $\Sigma_{\rm 0}\propto 1/r$ in the inner part of the disk as assumed in most models \citep[e.g.,][]{andrews11, vandermarel15}, a depletion factor of at least 10 in total gas mass means less than $10\%$ of radius inside the gap can be occupied by undepleted ring-like regions in between adjacent individual gaps opened by planets. On average, the distance between planets can increase by less than $10\%$ if the weaker condition is adopted, which has little effect on the configurations of planetary systems that can meet the condition.}.
\item The depletion factor, $\deltagap$, is at least 10 (or 100, 1000) everywhere in the common gap, except in the poorly constrained inner disk ($< \rinner$).
\item The typical size of the gas gap (i.e., its outer edge) is $\rgap=30$~AU.
\item The size of the inner disk $\rinner$, i.e., the inner edge of the gap, is no bigger than 3 AU, corresponding to 10\% of the gap size.
\end{enumerate}

\section{Requirements for multi-planet systems to open deep and wide gaps}\label{sec:planetrequirements}

We investigate the requirements for multi-planet systems to open deep and wide gaps consistent with transitional disk observations. Then we construct a series of such systems for use in the dynamical studies in Section~\ref{sec:stability}.

The profile of a gap opened by a single planet depends on the aspect ratio of the disk $h/r$; the parametrized disk viscosity $\alpha$ where the viscosity $\nu = \alpha h c_s$, $h$ is the scale height, and $c_s$ is the sound speed \citep{shakura73}; and the mass of the planet $q$ \citep[e.g.,][]{fung14, duffell15gap}. We assume that both $h/r$ and $\alpha$ are constant in time and location. The properties of a common gap opened by multiple planets also depend on the total number and the separation of the planets \citep[e.g.,][]{duffell15dong}. In general, given a set of disk properties $h/r$ and $\alpha$, a multi-planet system must  satisfy three requirements to open a common gap that resembles those observed:
\begin{enumerate}
\item The planets must be massive enough to carve out a deep gap.
\item The planets must be placed close enough together so that no wide, interspersed un-depleted regions exist.
\item The outermost planet has to be at the right location to account for the gap size, while the innermost planet has to be close enough to the star to extend the gap to $\sim$ 1 AU.\footnote{\citet[][see also \citealt{duffell14}]{durmann15} have shown that a gap opened by planets does not separate the inner and outer disk; instead, gas may cross the gap and reach the inner disk.}
\end{enumerate}

In the rest of this section, we will look at these requirements one by one. We assume that the planets' eccentricities do not significantly change the gap width (e.g., \citealt{duffell15e} find that the eccentricity damps rapidly once it begins large enough for the planet to collide with the gap wall). For simplicity, we assume equal mass planets; our results could be extended by mixing and matching planets of different masses and their gaps. For example, a configuration in which planet mass decreases with semi-major axis may avoid Type II migration in a flared disk \citep{crida09}.

\subsection{Planet masses}\label{sec:mp}

\citet{fung14} carried out extensive hydrodynamics simulations of gap opening in disks, and found that $\deltagapi$, the depth of a gap opened by a single planet\footnote{Throughout the paper, we reserve the subscript $_{\rm gap}$ for the common gap, and subscript $_{\rm gap,i}$ for the individual gap opened by a single planet. We use $(r)$ after a quantity (e.g., $\deltagap(r)$) to emphasize the radial dependence of the function. A quantity without $(r)$ represents the value of this quantity in the bulk part of the gap (e.g., $\deltagap$ represents the depletion factor in the core of a gap that is nearly independent of $r$).}, can be described by a simple power law with $h/r$ and $\alpha$ (see also \citealt{duffell13} and \citealt{kanagawa15}), segmented by the planet mass $q$. In the low mass planet regime $10^{-4} \lesssim \, q \lesssim \, 5 \times 10^{-3}$ (Eqn. 12 in \citealt{fung14})
\begin{align}
\deltagapi^{-1}=& \, 0.14 \left( \frac{q}{10^{-3}} \right)^{-2.16} \left( \frac{\alpha}{10^{-2}} \right)^{1.41} \left( \frac{h/r}{0.05} \right)^{6.61} \,,
\label{eqn:depthlowmp}
\end{align}
and for higher planet masses $5\times10^{-3} \lesssim \, q \lesssim \, 10^{-2}$ (Eqn. 14 in \citealt{fung14})
\begin{align}
\deltagapi^{-1} =& \, 4.7 \times 10^{-3} \left( \frac{q}{5 \times 10^{-3}} \right)^{-1.00} \left( \frac{\alpha}{10^{-2}} \right)^{1.26} \left( \frac{h/r}{0.05} \right)^{6.12} \,.
\label{eqn:depthhighmp}
\end{align}
\citet{duffell15dong} studied the profiles of gaps opened in multi-planet systems using hydrodynamics simulations. Low mass and/or widely separated planets open individual gaps well-characterized by the same scaling relation found in \citet{fung14}. At fixed separations, individual gaps formally merge at a critical planet mass $q=\qcrit$, when the high density ($\Sigma_0$) ring structure between each pair of adjacent gaps is squeezed out and a common gap forms with a flat bottom (i.e., constant $\deltagap(r)$ in the core of the common gap). When $q$ increases above $\qcrit$, $\deltagap$ is a constant in between $\qcrit$ and $\sim$2~$\qcrit$, before increasing again at $q\gtrsim2\qcrit$. Therefore Eqn.~\ref{eqn:depthlowmp} and \ref{eqn:depthhighmp} yield the lowest planet mass, $\qgap$, needed to open a common gap with multiple planets for a given $\deltagapmin$ at each [$\alpha$, $h/r$], which is same as the planet mass needed to open an individual gap with $\deltagapi=\deltagapmin$ and [$\alpha$, $h/r$]. A computed $\qgap$ is listed in Table~\ref{tab:mplimit} for various system parameters. All the planet masses exceed the viscous gap opening criteria for given $\alpha$ and $h/r$ \citep[e.g., Eqn. 25,][]{kratter10}. The viscosity, $\alpha$, has been estimated to be on the order of 0.01 based on the median accretion rate of T Tauri stars \citep{hartmann98}. At 30~AU, in a disk around a typical Herbig Ae/Be star with a stellar mass 2.5~$M_\odot$, a stellar radius  2~$R_\odot$, and a photosphere temperature $10,000$~K, the midplane temperature $T_{\rm mid}$ is about 65~K and $h/r$ is about 0.06; while in a disk around a typical T Tauri star with 0.5~$M_\odot$, 2~$R_\odot$, and $4000$~K, $T_{\rm mid}\sim30$~K and $h/r\sim0.08$. We will choose $\alpha$ of 0.1, 0.01 and 0.001 and $h/r$ of 0.05 and 0.1 as representative values. Finally, we note that the gap opening timescale considered here is generally much shorter than the age of transitional disks. \citet[][Fig. 5]{fung14} showed that with $\alpha=0.001$, the gap opened by a $1\mj$ planet reaches roughly its final state in a few thousand orbits, which is on the order of 0.1 Myr for the orbital periods considered here. The gap opening timescale is shorter for more massive planets and more viscous disks.

Observed lower limits on $\deltagap$ do not provide an upper limit on planet mass. We therefore turn to the outcomes of companion searches in transitional disks to set a meaningful upper limit. Recent near-IR interferometry \citep{pott10}, adaptive optics imaging \citep{cieza12}, and aperture masking observations \citep{kraus11} have shown that the observed inner holes and gaps in transitional disks are rarely due to close stellar companions or brown dwarfs \citep{alexander14}, as typically these objects would have been detected. In contrast, planetary mass companions (i.e., objects with a mass lower than the deuterium burning limit $\sim$13$\mj$) at tens of AU can be well hidden in current observations. Given these considerations, we limit our analysis to the planetary mass regime, $\mplanet\lesssim13\mj$.

\subsection{Planet separation}

\begin{figure}
\begin{center}
\includegraphics[trim=0 0 0 0, clip,width=0.5\textwidth,angle=0]{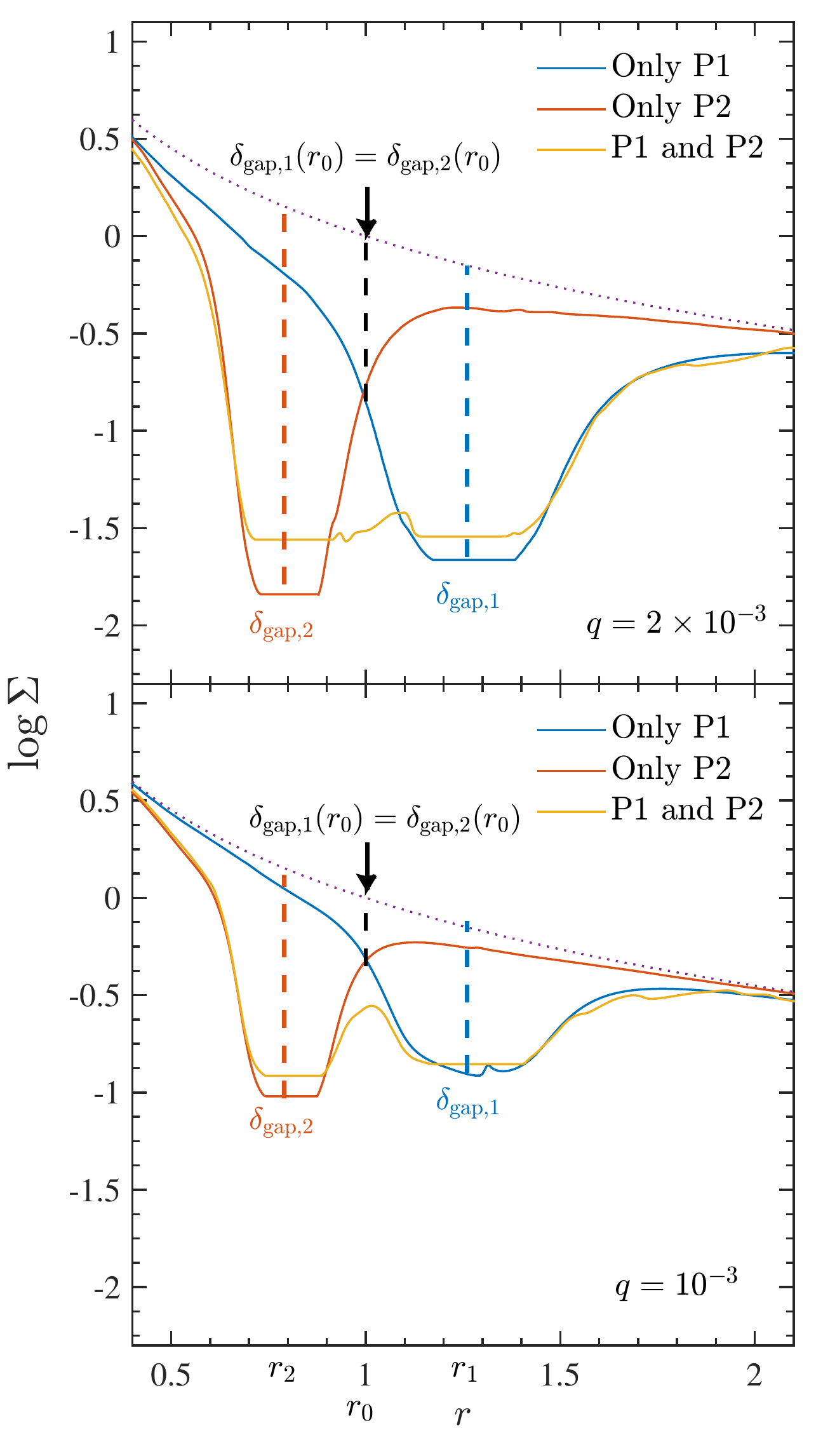}
\end{center}
\figcaption{The profiles of the individual gaps opened by a planet at $r_{\rm 1}$ (blue curves) or $r_{\rm 2}$ (red curves), and the profile of the common gap opened by both planets (yellow curves). Upper panel: individual gaps opened by two $q=2\times10^{-3}$ planets just merge to form a (complete) common gap ($\delta_{\rm gap,1}(r_0)=\delta_{\rm gap,2}(r_0)\approx\sqrt{\delta_{\rm gap,1}}\approx\sqrt{\delta_{\rm gap,2}}$; $r_0\approx1$). Bottom: two less massive planets with $q=10^{-3}$ at the same locations are unable to open a complete common gap ($\delta_{\rm gap,1}(r_0)=\delta_{\rm gap,2}(r_0)<\sqrt{\delta_{\rm gap,1}}\approx\sqrt{\delta_{\rm gap,2}}$). The black dotted line is the initial surface density profile (i.e., $\Sigma_0(r)$). The colored text boxes on the curves indicate the depletion factors of various gaps at certain locations, illustrated by the vertical dash lines of the same color.
\label{fig:sigma_rp}}
\end{figure}

At a given mass, planets have to be located close enough together to open a common gap instead of individual gaps. \citet{duffell15dong} found that when two neighboring gaps start to overlap, the overlapping region has an effective depletion factor 
\begin{align}
\delta_{\rm overlapping}(r)=\delta_{\rm gap,1}(r)\times\delta_{\rm gap,2}(r),
\label{eqn:deltaoverlapping}
\end{align}
where $\delta_{\rm gap,1}(r)$ ($\delta_{\rm gap,2}(r)$) is the depletion factor at $r$ for an individual gap opened by planet 1 (2). Figure~\ref{fig:sigma_rp} shows two examples of the profile of a gap opened by two planets. This relation holds until $\delta_{\rm overlapping}(r)\approx\delta_{\rm gap,1}\approx\delta_{\rm gap,2}$, at which point the two gaps formally merge and are replaced by a common gap. Therefore, in order for two planets to open a common gap with $\deltagapmin$ throughout, the two gaps must join at $r_0$ where $\delta_{\rm gap,1}(r_0)\approx\delta_{\rm gap,2}(r_0)\approx\sqrt{\deltagapmin}$. The location $r_0$ defines the maximum separation between the two planets. A precondition $\delta_{\rm gap,1} (\delta_{\rm gap,2})\gtrsim\deltagapmin$ is also required, i.e., that each planet is able to open a gap with a depletion factor at least $\deltagapmin$ (i.e., the condition in Section~\ref{sec:mp}).

\subsection{Planet locations}

Similar to $\deltagapi$, the width of a gap induced by an individual planet, $\widthgapi$, 
depends on $\alpha$, $h/r$, $q$, and the definition of the gap edge. As described in Section~\ref{sec:gapdepletion},
we focus on three illustrative values of $\deltagapmin$: 100, 100, and 1000. Respectively for the three values of $\deltagapmin$, we use planets with $q\geq q_{\rm gap}(10|\alpha,h/r)$, $q\geq q_{\rm gap}(100|\alpha,h/r)$, and $q\geq q_{\rm gap}(10|\alpha,h/r)$. Applying Eqn. \ref{eqn:deltaoverlapping}, we define the edges of individual gaps as the location where $\delta_{\rm gap,i}(r) =$ 3, 10, and 30. The gap width, $\widthgapi$, is the distance between the two edges. We obtain $\widthgapi$ numerically using the gap profiles calculated by \citet{fung14} using the code {\tt PENGUIN} \citep{fung15thesis}. In general, $\widthgapi$ is a monotonically decreasing function of $\alpha$ and $h/r$, and a monotonically increasing function of $\mplanet$ \citep[see also][]{crida06}. 

Finally, we construct maximally-spaced planetary systems that can open common gaps satisfying the conditions listed in Section~\ref{sec:gapsummary} for a given set of $\deltagapmin$, $\alpha$, and $h/r$. Planets are placed so their gap edges just touch; therefore $\deltagap(r)$ reaches 10 (or 100, 1000) at the joint point between each pair of adjacent individual gaps, and $\deltagap(r)\geq10$ (or 100, 1000) everywhere else in the common gap. The outermost planet, which maintains the outer edge of the common gap at $\rgap=30$~AU, is designated as planet 1. Inner planets are designated with increasing numbers. We use a minimum number of planets in the sense that the inner gap edge maintained by the innermost planet ($\rinner$) is inside 3~AU. Table~\ref{tab:planetarysystems} lists the parameters of all systems.

Opening a deep and wide common gap requires a system of 3--6 massive and closely-spaced planets. The separation between a pair of adjacent planets $i$ and $o$, $r_o-r_i$, in mutual Hill radius $\rmh=\frac{r_i+r_o}{2}\sqrt{\frac{2q}{3}}$ (the last column in Table~\ref{tab:planetarysystems}) is generally in the range of 3--6. This kind of planetary systems may be dynamically unstable on time scale shorter than the typical lifetime of disks \citep[e.g.,][]{smith09}, which is the subject to be addressed in the next section.

\section{Stability analysis}\label{sec:stability}

In this section, we will show that without gas damping, many of the systems of planets closely-spaced and massive enough to open deep and wide gaps are unstable. However, damping forces from residual gap gas can stabilize the systems and drive them into mean motion resonances, which can further enhance the stability. We demonstrate these conclusions through $N$-body simulations that take into account both planet-planet interactions (Section~\ref{sec:nogasdamping}) and various forms of gas damping forces (Section~\ref{sec:damping}).\footnote{Our 
approach -- synthesizing planet-opened gaps based on hydrodynamical simulations and following with $N$-body calculations of planet-planet interactions that incorporate prescriptions for gas damping --  allows us to carry out thousands of simulations to assess the statistics of the stability and resonant behavior of these systems. Our two stage approach is adequate to study whether the systems remain stable and compact because the gas damping depends on the bulk properties of the gaps (e.g., order of magnitude of the gas surface density inside), which are insensitive to small variations in the planets' orbits (e.g, gaps remain largely the same for $\lesssim10\%$ changes to the semi-major axes and eccentricities of the sculpting planets).}

\subsection{No gas damping}\label{sec:nogasdamping}

To explore the stability of the planetary systems constructed in Section~\ref{sec:planetrequirements}, we carry out $N$-body simulations using {\tt mercury6} with the hybrid symplectic integrator \citep{Cham96}. In the first set of simulations presented in this subsection, we use purely $N$-body gravitational forces and neglect the effects of gas (i.e., planetary migration and eccentricity/inclination damping caused by planet-disk interactions are not included). For each multi-planet system in Table~\ref{tab:planetarysystems}, we perform 10 $N$-body simulations for an initial assessment. All simulations are run for 27 Myr, much longer than the typical disk lifetime. In each trial, planets are assigned an initial eccentricity $e=0$; a mean anomaly, and longitude of ascending node drawn randomly between 0 to 360$^\circ$; a semi-major axis drawn from a normal distribution with a median from Table~\ref{tab:planetarysystems} and a standard deviation 10\% the gap width; and an inclination $i$ drawn from a normal distribution with median $0$ and standard deviation $0.01 c_s/v_{\rm kep}$ rad, where $c_s = 1.29 {\rm km/s} \left(\frac{a}{\rm AU}\right)^{-1/4}$ is the sound speed and $v_{\rm kep}$ is the Keplerian velocity. If planets in any of the ten initial runs undergo instability, an additional set of 40 trials are carried out to more precisely determine the median unstable time scale $\tu$ and the 1$^{\rm st}$ quartile unstable time scale $\tuq$ (i.e., when 25\% of the runs become unstable). A system is considered unstable if a planet is ejected $(a > 1000 {\rm AU})$, a planet's semi-major axis changes by more than 20\% due to scattering \footnote{For changes in a planet's semi-major axis greater than 20\% but less than 50\%, we visually inspect the orbital elements to ensure that scattering occurred rather than high amplitude oscillations.}, or two planets collide.

\begin{figure}
\begin{center}
\includegraphics[trim=0 0 0 0, clip,width=0.49\textwidth,angle=0]{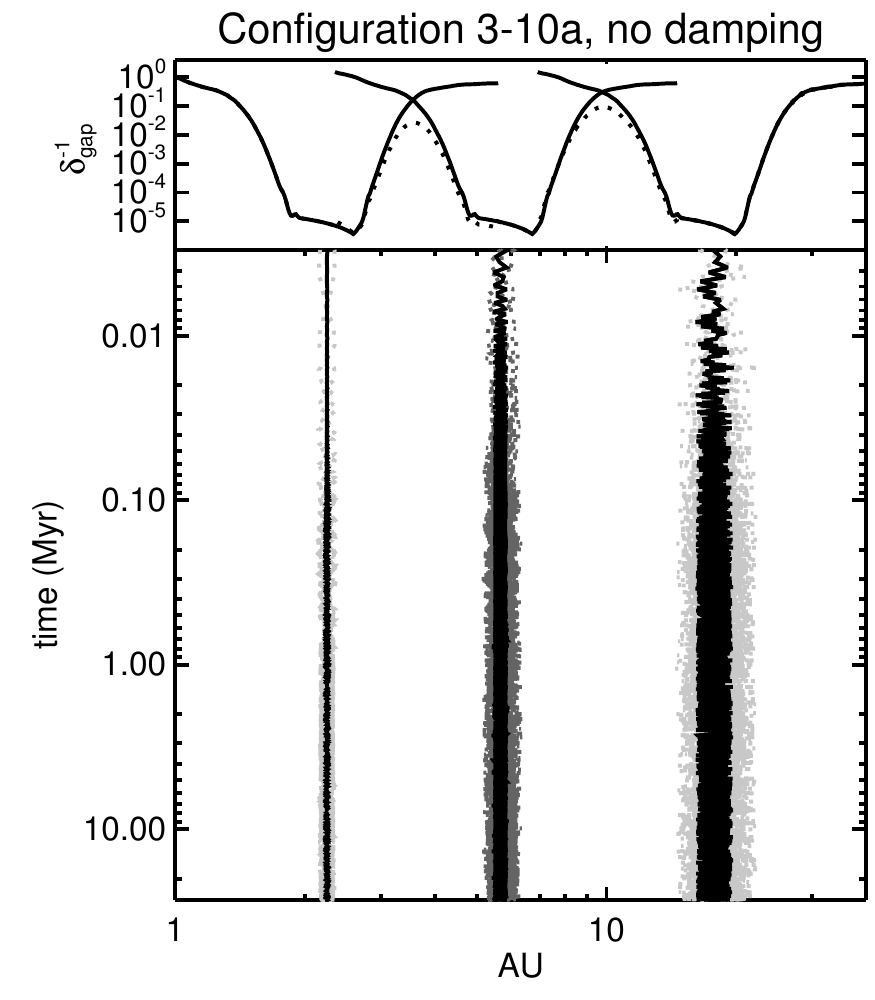}
\end{center}
\figcaption{The orbital evolution of system {\tt 3-10a} in Table~\ref{tab:planetarysystems} without gas damping forces. The system is stable for at least 27 Myr. The top panel shows the depletion caused by planets' gaps. The solid curves indicate individual gaps opened by each planet, while the dotted curve is the profile of the common gap calculated based on Equation~\ref{eqn:deltaoverlapping}. The bottom panel shows the temporal evolution of the semi-major axis $a$ (black) and periapse and apoapse (gray) of each planet. 
\label{fig:orbitalevolution_stable}}
\end{figure}

\begin{figure}
\begin{center}
\includegraphics[trim=0 0 0 0, clip,width=0.49\textwidth,angle=0]{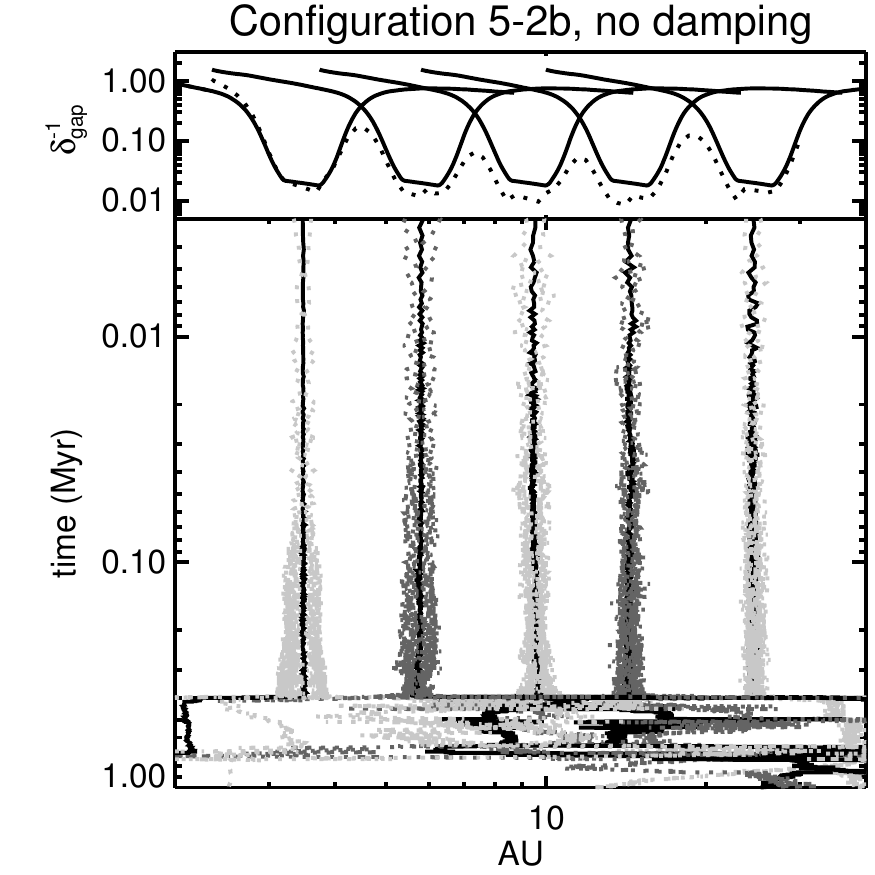}
\includegraphics[trim=0 0 0 0, clip,width=0.49\textwidth,angle=0]{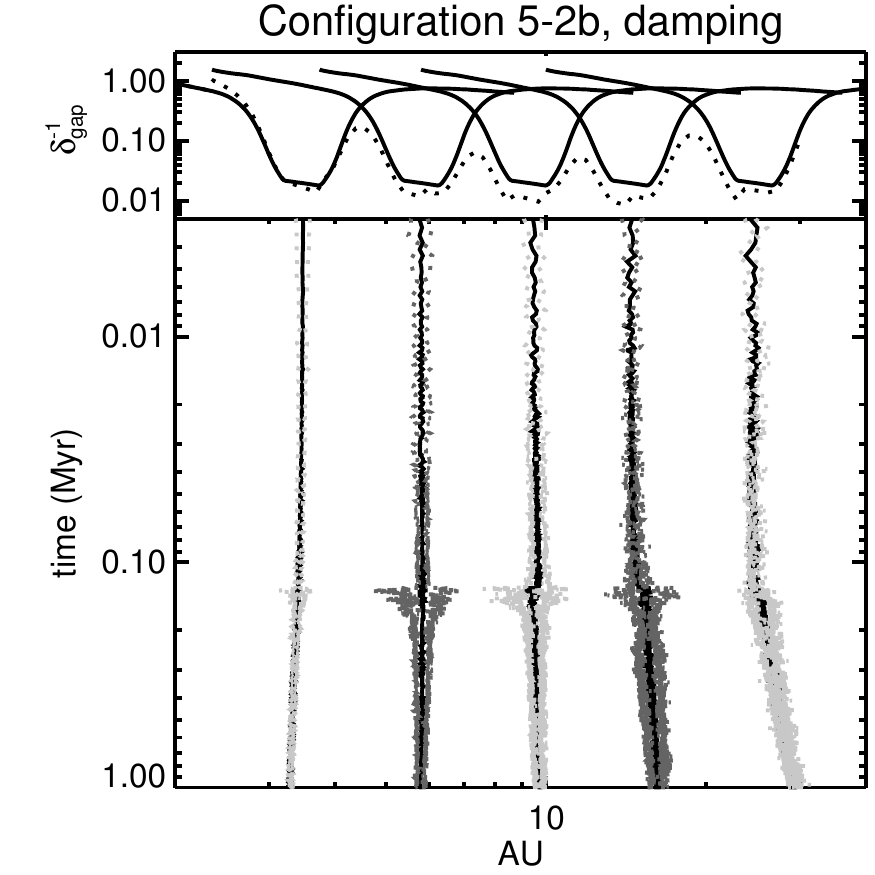}
\end{center}
\figcaption{The same as Figure~\ref{fig:orbitalevolution_stable}, but for system {\tt 5-2b} in Table~\ref{tab:planetarysystems}, showing the cases when gas damping forces are not included (left; unstable) and included (right; stable). The gas gap density normalization on the right is $\Sigma_{30} = 0.1$ g cm$^{-2}$.
\label{fig:orbitalevolution_unstable}}
\end{figure}

The outcomes of the $N$-body simulations are listed in Table~\ref{tab:stability}. Less than half of systems are stable for at least 27 Myr without the aid of gas damping forces. Figure~\ref{fig:orbitalevolution_stable} shows an example of a system that remains stable for $>$ 27 Myr (one trial of system {\tt 3-10a} in Table~\ref{tab:planetarysystems}) and contains three $q=0.01$ planets. Figure~\ref{fig:orbitalevolution_unstable} (left panel) shows a trial of system {\tt 5-2b} (containing five $q=0.002$ planets) that goes unstable shortly after 0.1 Myr. The planets scatter and two are ejected within 1 Myr. As expected from studies of orbit crossing timescale vs. spacing (e.g., \citealt{Cham96,Yosh99,Zhou07,Smit09}), systems with closer spacings (in units of $\rmh$) between planets or more planets tend to be more unstable. About 1/3 to 1/2 of unstable systems feature collisions between planets and most feature ejections of one or more planets. Many systems are left with only one or two giant planets, even those that started out with six.

In terms of gap characteristics, systems constructed for disks with larger $\deltagapmin$, $\alpha$, or $h/r$ tend to be less stable. This is expected, because for a given $\mplanet$, larger $\deltagapmin$, $\alpha$, or $h/r$ produces a smaller $\widthgapi$. Therefore to maintain a given common gap size a system requires more planets with tighter spacings and is thus less stable. Additionally, for a given $\deltagapmin$, $\alpha$, and $h/r$, systems composed of less massive planets tend to be less stable. This is because $\widthgapi$ often decreases faster with decreasing $\mplanet$ than $\rmh$; therefore the distance between adjacent planet pair shrinks in units of $\rmh$ as $\mplanet$ goes down (e.g., decreasing $rmh$ with $\mplanet$ for configurations {\tt 3-10b, 4-5c, 5-2b}, which all have $\deltagapmin=10$, $h/r=0.05$, $\alpha=0.01$). This trend is only marginal and does not hold in all cases. In addition, a system may need more planets to maintain the same common gap size as $\widthgapi$ decreases.

\subsection{Effects of gas on stability}\label{sec:damping}

Although the gas inside the gap is depleted, the remaining gas can still affect the dynamics and stability of the multi-planet systems through disk-planet interactions. Disk-planet interactions can alter the planet's eccentricity $e$ and inclination $i$ (typically damping them, e.g., \citealt{dunhill13}, but see also \citealt{tsang14,tsang14b,duffell15e}), semi-major axis $a$ (migration), or orbital orientation angles --- the longitude of periapse, $\varpi$, and the longitude of ascending node, $\Omega$  (precession). Gas damping and planetary migration can capture planets into mean motion resonance, and precession can alter the locations of the resonances. The possibility of resonant capture further motivates additional simulations that include the effects of gas because a resonant configuration can increase the stability timescale of a system, as explored for HR 8799 \citep{fabrycky10}, HL Tau \citep{tamayo15}, and LkCa~15 \citep{sallum15}.

We will first examine the effect of a non-zero $\dot{e}$ on the stability (Section~\ref{sec:dote}) and resonance capture (Section~\ref{sec:mmr}), followed by the effects of non-zero $\dot{a}$, $\dot{\varpi}$ and $\dot{\Omega}$ (Section~\ref{sec:dota}). We will conclude that if planets are responsible for the gaps, we expect to see many young planets in mean motion resonant configurations at the beginning of their post-gas-disk era evolution, and these configurations may persist for the older systems observable by non-direct imaging techniques (which we will discuss in Section~\ref{sec:disrupt} and Appendix~\ref{subsec:resage}).

\subsubsection{Non-zero $\dot{e}$ only}\label{sec:dote}

The strength of these effects depends on $\sigmagas$ inside the gap, which is not well known. As discussed in Section~\ref{sec:gapdepletion}, in many cases current gas observations can only put a lower limit of $\deltagapmin\sim10$. Under this limit, the total gas mass within the gap is usually on the order of $\mj$ or less \citep{vandermarel15}, which is small compared with the combined mass of the planets (usually $\sim3-40\mj$, Table~\ref{tab:planetarysystems}). If the planets' self-stirring is dominated by close encounters (e.g., \citealt{Gold04}), gas damping is expected to shut off for $\Sigma_{\rm gas} < \Sigma_{\rm planets}$, but gas damping can still be important for widely-spaced planets on low-eccentricity orbits. (See e.g., Section~2 of \citealt{Daws15} for a discussion and \citealt{dawson15b} for examples.) This motivates us to explore the effects of gas eccentricity and inclination damping using simulations. In contrast, migration of planets that have opened gaps tends to be slow: type II migration may occur on a timescale comparable to the disk viscous timescale \citep[][but see also \citealt{duffell14}]{ward97}, and the type I migration timescale is longer than the eccentricity damping timescale  by a factor of $(v_{\rm kep}/c_s)^2$. We do not include migration in the simulations in this subsection.

To explore the effects of gas damping, we use a customized version of {\tt mercury6} containing gas-damping forces (\citealt{dawson15b}, which makes use user-defined velocities and accelerations described in Appendix A of \citealt{Wolf12}). We adopt a gas surface density profile inside the gap of $\sigmagas=\Sigma_{30} \left(\frac{a}{30\rm AU}\right)^{-3/2}$, with a fiducial normalization of $\Sigma_{30} = $~1 g cm$^{-2}$ at the outer edge of the gap, the median value based on current ALMA transitional disk observations \citep{vandermarel15} and corresponding to a depletion factor of 10 relative to the minimum mass solar nebula. We run additional sets of simulations with $\Sigma_{30} = \{10, 0.1, 0.01, 0.001, 0.0001\}$ g cm$^{-2}$. Following \citet{dawson15b}, we use three damping timescales, based on the regimes described in \citet{Papa00,Komi02,Ford07} and \cite{Rein12}:
\begin{eqnarray}
\label{eqn:damp}
\tau = 0.029 \frac{{\rm g\ cm}^{-2}}{\Sigma_{30}} \left(\frac{a}{\rm AU}\right)^2 \frac{M_\odot}{M_{\rm p}} {\rm yr} \times & \nonumber\\
& 1&, v < c_s \nonumber \\
&\left(\frac{v}{c_s}\right)^3&, v > c_s, i < c_s/v_{\rm kep} \nonumber  \\
&\left(\frac{v}{c_s}\right)^4&, i > c_s/v_{\rm kep} \nonumber \\
\end{eqnarray}
\noindent where $v = \sqrt{e^2+i^2} v_{\rm kep}$ and $v_{\rm kep}$ is the Keplerian velocity. We impose $\dot{e}/e = -1/\tau$ and $\dot{i}/i = -2/\tau$ \citep{Komi02}.

\begin{figure}
\begin{center}
\includegraphics[trim=0 0 0 0, clip,width=0.49\textwidth,angle=0]{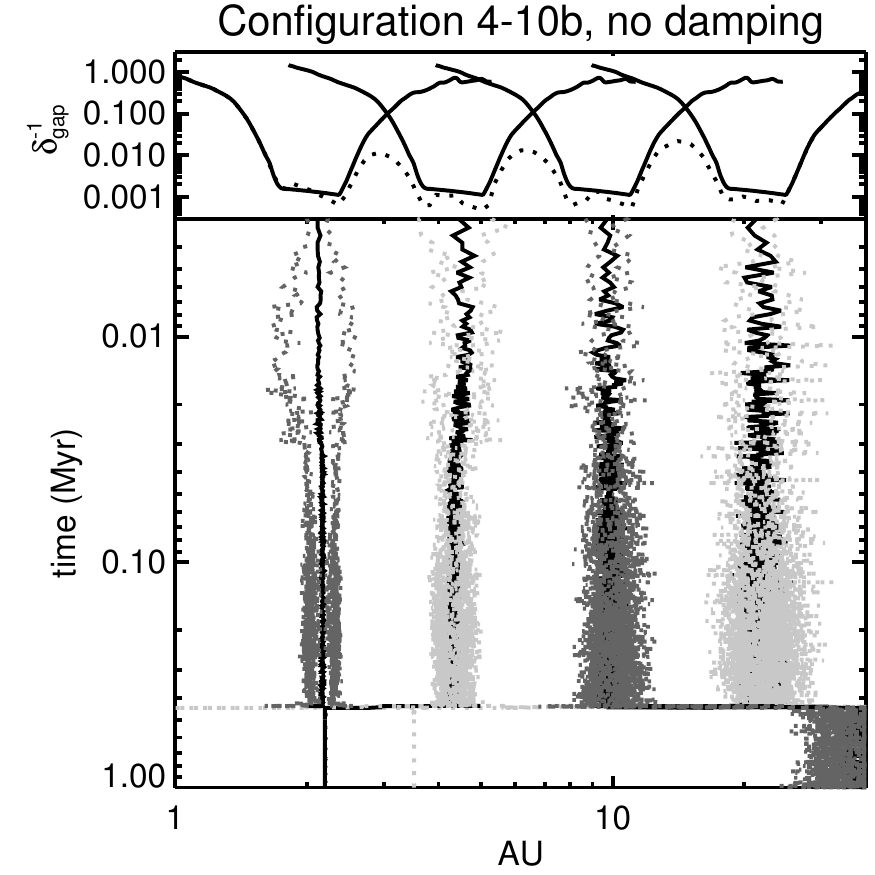}
\includegraphics[trim=0 0 0 0, clip,width=0.49\textwidth,angle=0]{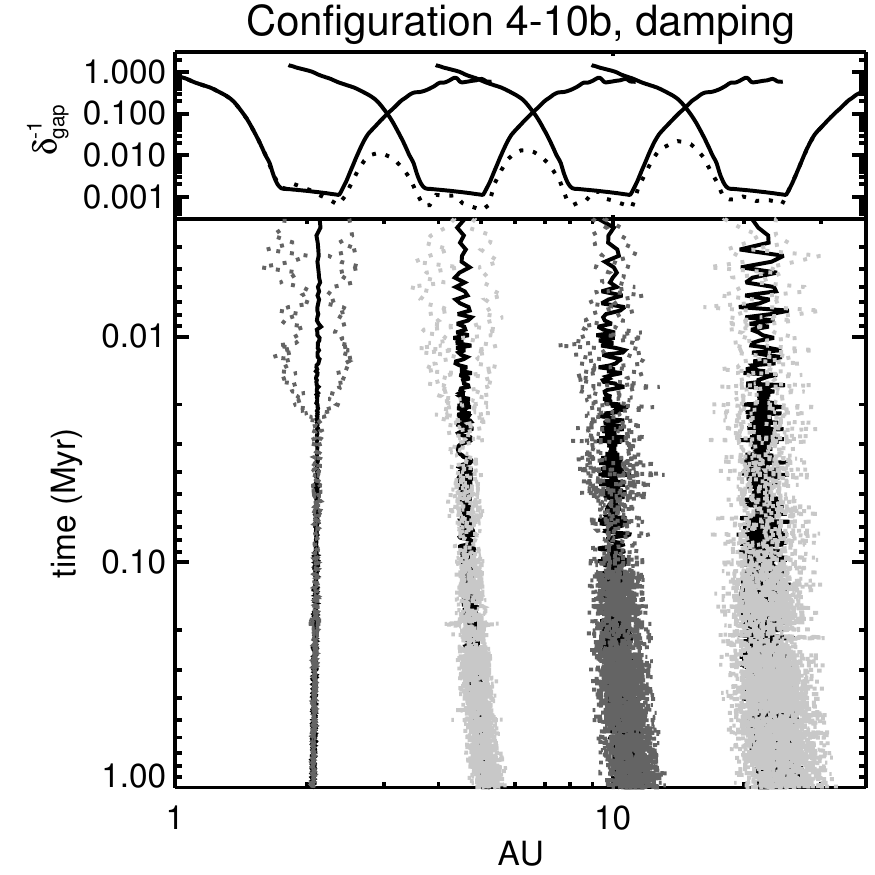}
\end{center}
\figcaption{The same as Figure~\ref{fig:orbitalevolution_unstable}, but for system {\tt 4-10b} in Table~\ref{tab:planetarysystems}. The gas gap density normalization on the right is $\Sigma_{30} = 0.01$ g cm$^{-2}$.
\label{fig:orbitalevolution_unstable1}}
\end{figure}

\begin{figure}
\begin{center}
\includegraphics[trim=0 0 0 0, clip,width=0.49\textwidth,angle=0]{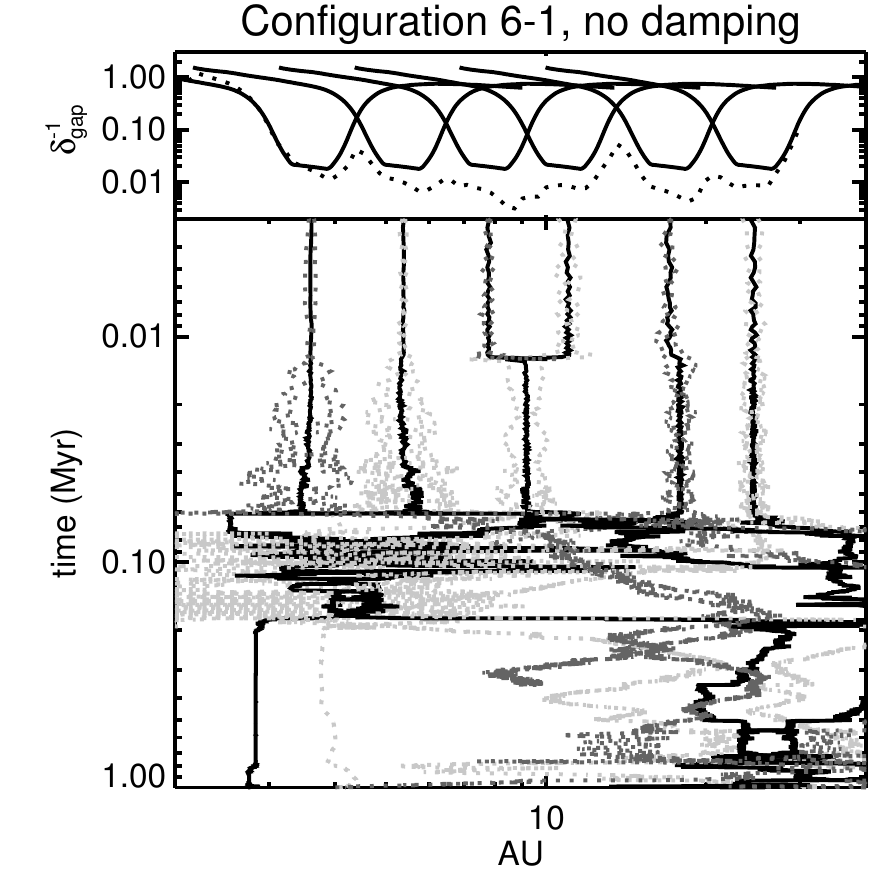}
\includegraphics[trim=0 0 0 0, clip,width=0.49\textwidth,angle=0]{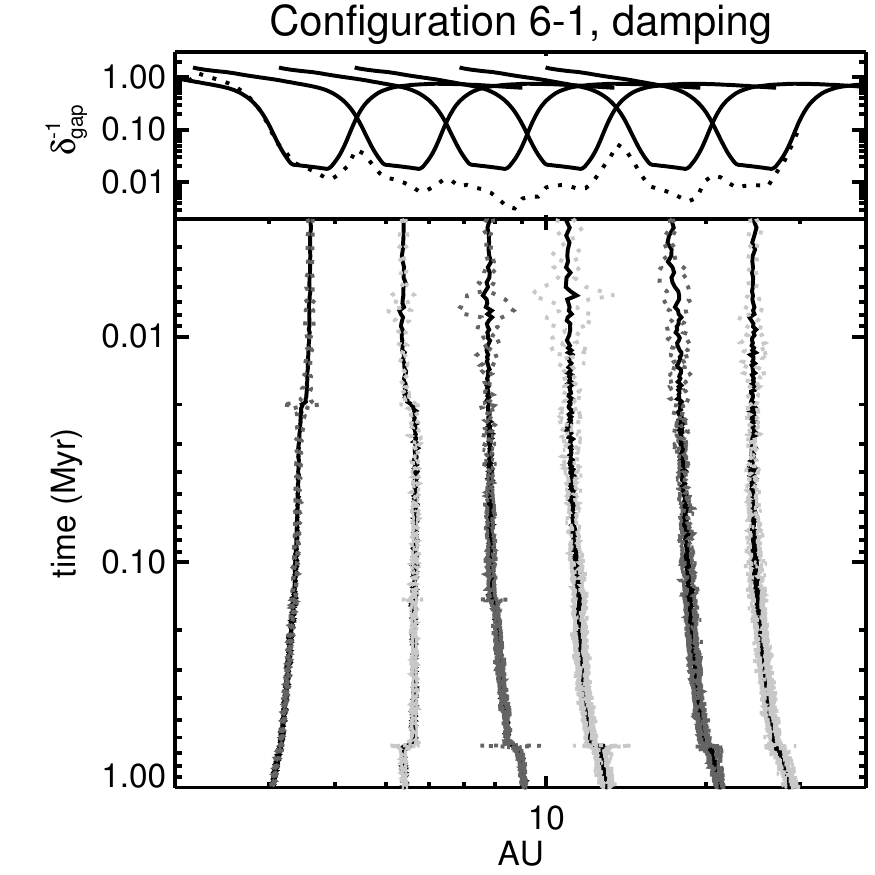}
\end{center}
\figcaption{The same as Figure~\ref{fig:orbitalevolution_unstable}, but for system {\tt 6-1} in Table~\ref{tab:planetarysystems}. The gas gap density normalization on the right is $\Sigma_{30} = 1$ g cm$^{-2}$.
\label{fig:orbitalevolution_unstable2}}
\end{figure}

Damping stabilizes planetary systems. The left panel in Figure~\ref{fig:orbitalevolution_unstable} shows the evolution of a five-planet system that was unstable without gas damping. With sufficient gas damping (right panel), eccentricities remain small and no scattering, collisions, or ejections occur. We show examples for four and six planet systems in Figures \ref{fig:orbitalevolution_unstable1} and \ref{fig:orbitalevolution_unstable2} respectively.

Next we assess the stability requirements for all the configurations developed in Section~\ref{sec:planetrequirements}. First we consider the criterion that at least 50\% of trials are stable, which we parametrize in terms of the minimum gas surface density normalization for 1 Myr stability $\Sigma_{30s,50\%}$. We list $\Sigma_{30s,50\%}$ for each system in Table \ref{tab:stability}. About half of configurations, including all the three planet configurations, are stable without gas damping. For other configurations, $\Sigma_{30s,50\%}$ ranges from 0.0001 g cm$^{-2}$ to 10 g cm$^{-2}$. Next we consider the stricter criterion that at least 90\% of trials are stable (parametrized as $\Sigma_{30s,90\%}$). About 1/3 of systems meet this criterion without gas damping. For other configurations, $\Sigma_{30s,90\%}$ ranges from 0.0001 g cm$^{-2}$ to 10 g cm$^{-2}$.

Even without explicit migration forces imposed (by default we set $\dot{a} = 0$ in our implementation of \citealt{Wolf12}, Appendix A), planets' semi-major axes can change as they planets repel each other. Figure~\ref{fig:orbitalevolution_unstable2}, right panel shows an example of repulsion in a six planet system. The planets are originally spaced from 3.5 to 26 AU. After 1 Myr with gas damping imposed, they are spaced from 3.0 to 30 AU. The repulsion may be a manifestation of the resonant-repulsion proposed for driving \kep super-Earths to period ratios wide of commensurability (e.g., \citealt{lithwick12}). Here the dissipative force comes from gas instead of tides. The mechanism operates even when no explicit $\dot{a}$ is imposed (e.g., \citealt{lithwick12}, Eqn. 22-23).

For many configurations, the repulsion is so strong with the fiducial value of $\Sigma_{30} = 1 $g cm$^{-2}$ 
that the planets cannot maintain a common gap for 1 Myr. These systems require a lower surface density, which we parametrize as $\Sigma_{30c}$, to  avoid excessive repulsion (Table \ref{tab:stability}). About 1/3 of configurations remain compact (which we define as moving no further than one gap width apart) over 1 Myr with  $\Sigma_{30c}=1$g cm$^{-2}$. Others require $\Sigma_{30c}\le 0.1$g cm$^{-2}$ or $\Sigma_{30c}\le 0.01$g cm$^{-2}$; these configurations tend to contain more massive planets.

For a few configurations ({\tt 6-2, 4-10cd}), too little gas means the system quickly goes unstable and too much gas means the planets repel each other excessively. The former value is $\Sigma_{30}=\{1, 0.01,0.01\}$ g cm$^{-2}$ for configurations \{{\tt 6-2, 4-10c, 4-10d}\} respectively and the latter value respectively is $\Sigma_{30}=\{10,0.1,0.1\}$ g cm$^{-2}$. We did not explore whether an intermediate value would allow for stability without excessive repulsion but if so, the conditions would require fine-tuning.

Overall, the range of $\Sigma_{30}$ that stabilizes the system without excessively repelling the planets is consistent with the plausible range of gas densities in the gap. The undepleted gas surface density at 30 AU ranges from about 10--100 g cm$^{-2}$ in observed transitional disks, interpolating from the gas density in the outer region (\citealt{vandermarel15}, Fig. 5). We multiply the 10--100 g cm$^{-2}$ range by the depletion factor at the bottom of the gap from the \citet{fung14} models (the last column in Table~\ref{tab:stability}) and compare to various $\Sigma_{30}$. We caution $\Sigma_{30}$ from our simulations is correct only to an order of magnitude due to approximations in the coefficients of Eqn. \ref{eqn:damp}. A handful of systems exhibit potential inconsistencies. For configurations {\tt 4-10a, 5-5ac}, the $\Sigma_{30} \le 0.01$ g cm$^{-2}$ required to avoid excessive repulsion is an order of magnitude or more lower than the expected density inside the gap. For configurations {\tt 6-2} and {\tt 6-1}, the large amount gas necessary to stabilize the system is several orders of magnitude larger than the expected gas density inside the gap.

In addition, an alternative form of eccentricity damping can further solve several of the inconsistencies described above. In theory planet-disk interactions can excite eccentricities instead of damping them, but the excitation is limited. For example, \citet{duffell15e} found that torques from a gas disk can excite a giant planet's eccentricity up to $\sim c_s/v_{\rm kep}$, above which eccentricity damping dominates. We ran a new set of simulations for Configuration {\tt 5-2b} in which we set the eccentricity damping to 0 when $v<c_s$. We found this damping to be insufficient to stabilize the system for $\Sigma_{30} = 0.1$ g cm$^{-2}$ but sufficient for $\Sigma_{30} = 1$ g cm$^{-2}$, which is still consistent with $\Sigma_{\rm gap}$ in Table \ref{tab:stability}. This damping still results in capture in mean motion resonances, which are discussed further below. For configurations {\tt 4-10a, 5-5ac}, $\Sigma_{30} = 0.1, 1, 1$ g cm$^{-2}$ respectively stabilizes the system while avoiding excess repulsion; this higher surface density is consistent with the expected surface density inside the gap.

Potentially gas outside the common gap could resist the resonant repulsion, but we do not expect this mechanism to be effective for configurations explored here. We expect the timescale to clear a new gap to be much less than the repulsion timescale so that the gap effectively moves with the planet, rather than the gas pushing the planet back toward the gap center. Furthermore, only a thin ring of gas with width on the order of the distance moved by the planet, typically a few AU, is expected to strongly interacts with the planet. The total mass in that gas ring is expected to be much smaller than the combined mass of the planets responsible for the repulsion. However, hydrodynamical simulations are necessary to fully explore the effect of the gas outside the common gap on resonant repulsion and the dependence on disk parameters. Because resonant repulsion only causes potential inconsistencies for a few configurations and the inconsistencies are solved by the alternative form of eccentricity damping, we leave such an exploration for future studies.

\subsubsection{Eccentricity damping driven capture into mean motion resonance}\label{sec:mmr}

Eccentricity damping drives the planets into mean motion resonances, defined as the libration of a resonant argument, $j_1 \lambda_o + j_2 \lambda_i + j_3 \varpi_o +j_4\varpi_i+j_5\Omega_o+j_6\Omega_i$ where $\{j_1,j_2,j_3,j_4,j_5,j_5\}$  is a set of integers that sums to 0. In the context of wide common gaps opened by multi-giant-planets, capture into resonances has been preliminarily shown in restricted parameter spaces with a small number of hydrodynamical simulations that included planetary sculpting of the gap, gravitational interactions among planets, and back reaction from the gas disk to the planets \citep[e.g.,][]{zhu11}.

\begin{figure}
\begin{center}
\includegraphics[trim=0 0 0 0, clip,width=0.75\textwidth,angle=0]{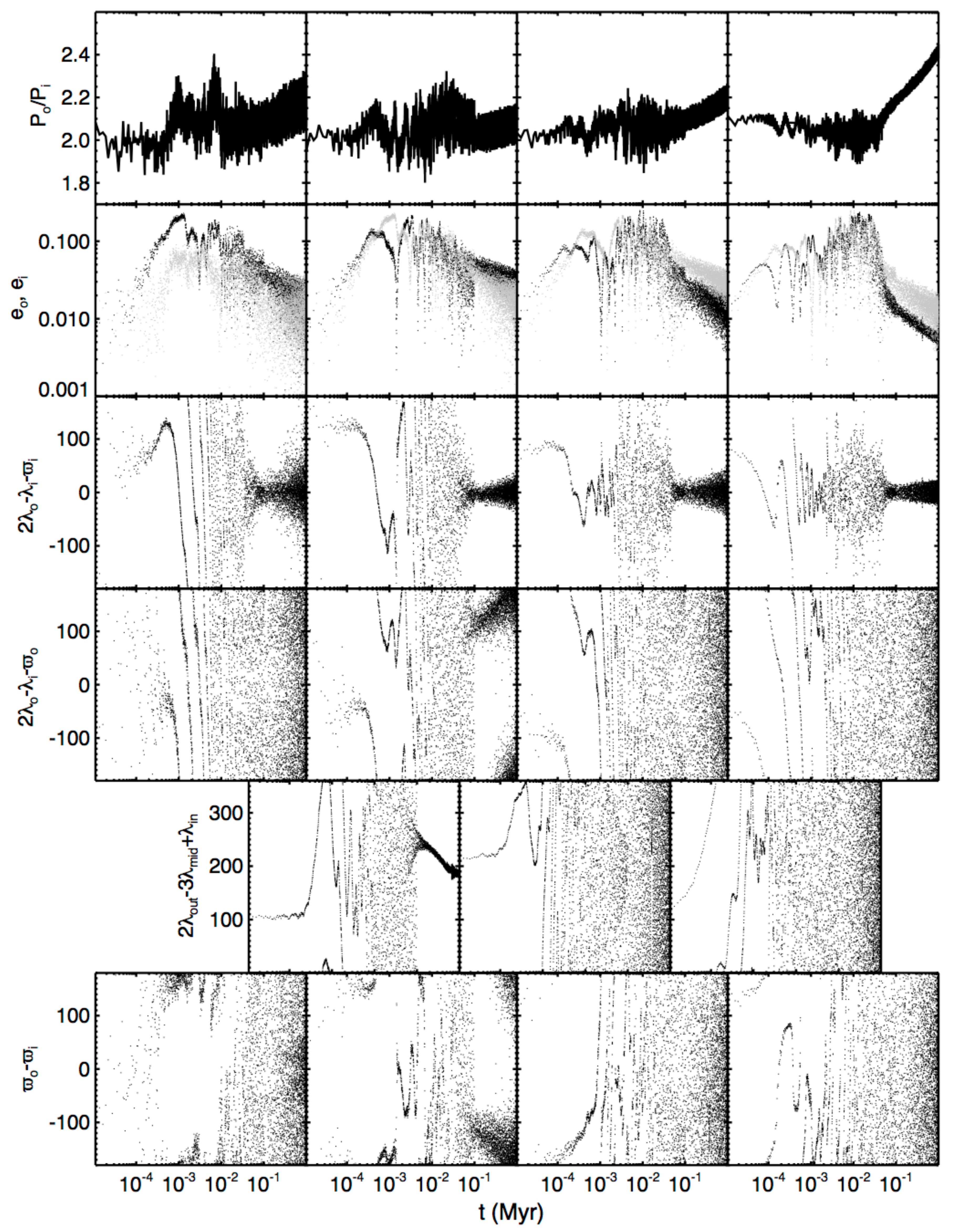}
\end{center}
\figcaption{Evolution of orbital elements for System {\tt 5-2b} with gas damping included ($\Sigma_{30} = 0.1$ g cm$^{-2}$; right panel in Figure~\ref{fig:orbitalevolution_unstable}). Row 1: period ratio of each adjacent pair of planets, from outer to inner. Row 2: eccentricity of outer planet (gray) and inner planet (black). Row 3: 2:1 mean motion resonant argument involving the longitude of periapse of the inner planet. Row 4: 2:1 mean motion resonant argument involving the longitude of periapse of the outer planet. Row 5: Three-body resonant argument (Laplace resonance). Row 6: Separation of adjacent planets' longitude of periapses.
\label{fig:damping}}
\end{figure}

In many of our simulations that include gas damping, the planetary systems end up configured in a chain of mean motion where each pair of adjacent planets are in mean motion resonance. As posited for \kep super-Earths, the period ratios are wide of resonance due to dissipation. In Figure~\ref{fig:damping}, we plot the resonant arguments of pairs of planets for the same system featured in Figure~\ref{fig:orbitalevolution_unstable}. The planets are captured into resonance on a $\sim$ 0.1 Myr timescale and then the libration amplitude grows. In column 4, row 1, the period ratio approaches 2.4 at 1 Myr yet the libration of the resonant argument persists. Libration occurs when the time derivative of the resonant argument, e.g., $2n_o + \dot{\xi_o} - n_i - \dot{\xi_i} - \dot{\varpi_i}$ where $n$ is the mean motion and $\xi$ is the mean longitude at epoch, is 0 (or oscillates about 0). Eccentricity damping produces a $\dot{\varpi_i, \xi_o, \xi_i}$ that allows $2n_o - n_i$ to deviate from 0, leading to non-commensurate period ratios. 

We show more examples of resonant behavior in Appendix A and summarize the types of behavior in Table \ref{tab:res}. For each configuration, we show an example where $\Sigma_{30}$ allows for stability, avoids excess resonant repulsion, and is consistent (to within an order of magnitude) with the expected gas density in the gap (Table \ref{tab:stability}). Here we will use the term libration to refer to both true libration where the amplitude is bounded and angles that spend most of their time bounded but technically circulate through all values. In Fig. \ref{fig:app13} row 3, column 4 is an example of the former and column 1 of the latter. For every configuration (Table \ref{tab:res}, column 3) gas damping drives one or more pairs into libration of the 2:1 resonance. Higher multiplicity configurations are more likely to have the 2:1 resonance librating for all pairs. Other resonances, such as the 3:2 mean motion resonance, librate for a subset of configurations (e.g., Fig. \ref{fig:app19}). In many of the higher multiplicity configurations, one or more three-body resonant argument librates, including the Laplace resonance (e.g., Fig. \ref{fig:app14}) and others (e.g., the 3:4:1 angle in Fig. \ref{fig:app10} and the 3:5:2 angle in Fig. \ref{fig:app18}). However, we caution that even when an example configuration shown here includes libration of a three-body angle, other random versions of that configuration do not necessary include libration of the angle, so capture into a three-body resonance is not guaranteed. We find that the mean motion and three-body resonances generally persist after gas damping shuts off (Table \ref{tab:res}, final column).

Many pairs exhibit libration of the separation of periapses $\varpi_o-\varpi_i$. Unlike libration of the 2:1 resonance, libration of $\varpi_o-\varpi_i$ is not necessarily a signature of dissipation, particularly for the more widely-spaced planets, because much of the parameter space can librate for hierarchical systems (e.g., \citealt{michtchenko04}).

Finally, we note that although all configurations feature libration of a resonant argument, not all feature period commensurabilities ({\tt 3-10cd, 4-2}) and for many, a different resonant argument is librating than the commensurability. For example, the inner pair in {\tt 3-5} (Fig. \ref{fig:app1}) have a 4:1 period commensurability, but the 4:1 resonant arguments do not librate; instead, a 2:1 resonant argument librates. Many systems featuring libration of the 2:1 argument do not feature 2:1 commensurability. This poses a challenge for comparing to observations because the damped eccentricities are small (a few percent or less), making it difficult to measure $\varpi$ and hence determine whether resonant arguments are librating\footnote{Resonance libration timescales are quite long so rather than directly measuring the libration, we would need to constrain the orbital elements precisely enough to constrain the libration amplitude by forward integration.}.

\subsubsection{Non-zero $\dot{a}$, $\dot{\varpi}$ and $\dot{\Omega}$ in addition to $\dot{e}$}\label{sec:dota}

\begin{figure}
\begin{center}
\includegraphics{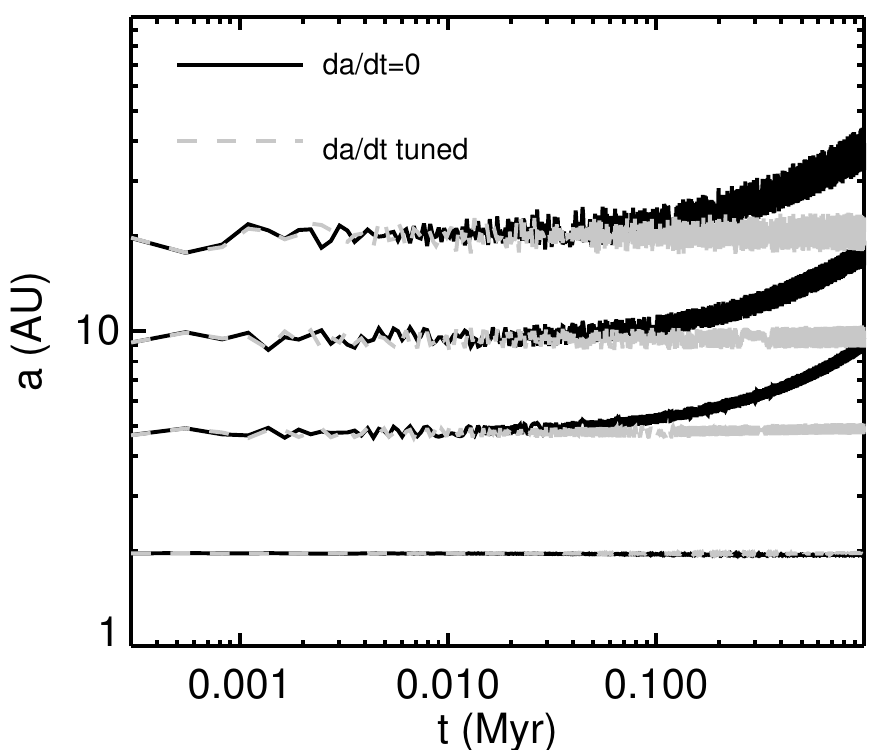}
\end{center}
\figcaption{The application of an $\dot{a}$ fine-tuned to each planet (gray) can counteract the effects of resonant repulsion (black,$\dot{a}$). The example shown is for Configuration {\tt 4-10a}.
\label{fig:mig}}
\end{figure}

We run additional simulations with $\dot{a}$ (migration) and $\dot{\varpi}$ and $\dot{\Omega}$ (precession) imposed in addition to $\dot{e}$ but find the same qualitative behavior. Migration can enhance or undo resonant repulsion but requires fine tuning. For example, recall that in Configuration {\tt 4-10a}, the system went unstable with too little gas but repelled each other excessively with too much gas. Migration can counteract the repulsion (Fig. \ref{fig:mig}). However, the migration rate ($\dot{a}/a$) must be positive for the innermost planet and negative for the other planets. We fine-tuned the migration timescale to 0.9 Myr for the outer three planets and 9 Myr for the inner planet. For either Type I or Type II migration, we generally expect the migration timescale to increase with semi-major axis, so these rates would require fine-tuning of the disk conditions to produce the required magnitude and direction of migration. 

\begin{figure}
\begin{center}
\includegraphics{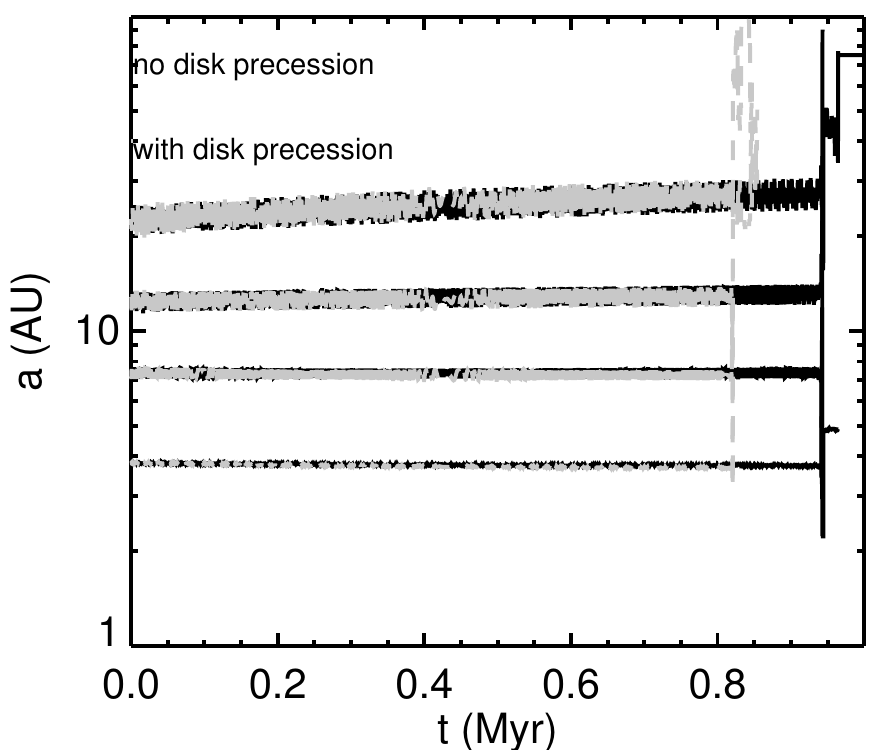}
\end{center}
\figcaption{Precession caused by an outer disk (approximately as a 10 Jupiter mass ring at 50 AU) does not significantly change the stability timescale. The example shown is for Configuration {\tt 4-10d} with $\Sigma_{30}=0.01$g cm$^{-2}$.
\label{fig:precess}}
\end{figure}

In principle, precession ($\dot{\varpi}$, $\dot{\Omega}$) caused by the undepleted outer disk can affect the stability of system. Primarily the outer disk would cause the planetary orbits to precess, which can stabilize them against their own planet-planet secular interactions. However, in practice the precession timescale is slow to make a difference: on a 1 Myr timescale, the systems are destabilized by resonant and synodic planet-planet interactions rather than secular planet-planet interactions. For example, without precession caused by an outer disk, a simulation in the {\tt 4-10d} configuration with $\Sigma_{30} = 0.01 $ gcm$^{-2}$ went unstable on a timescale of 0.95 Myr and thus barely failed our stability criterion. Next we approximated the outer disk as a 10 Jupiter mass ring at 50 AU and added the precession to each orbit caused by this ring\footnote{We apply precession by assigning an equivalent stellar $J_2$ to each planet.}. The system goes unstable on a similar timescale (Fig. \ref{fig:precess}). On a much longer timescale than considered here, a remnant planetesimal disk could affect the system's stability (e.g., \citealt{thommes08,moore13}), but this does not affect the requirements for the gas disk stage.

\section{Comparison with the statistics of giant planets at large separations}\label{sec:statistics}

In this section, we consider the occurrence rates of giant planets at large separations, i.e., those that could be the survivors of young planetary systems that sculpted transitional disks in the gas disk era. We compare their occurrence rates (Section~\ref{sec:fplanet}) to the occurrence rate of transitional disks to determine whether there are enough giant planets to account for the occurrence rate of transitional disks (Section~\ref{sec:fplanetvsftd}). Finally, we connect the initial conditions of the multi-planet systems at the beginning of the post-gas-disk era to the observed properties of these systems at older stages (Section~\ref{sec:disrupt}).

As shown in Section~\ref{sec:planetrequirements} and \ref{sec:stability}, multiple giant planets with $q\sim10^{-4}-10^{-2}$ separated by $\sim$3-6 $\rmh$ are necessary to open a common gap as seen in transitional disks and usually get locked into mean motion resonance and are stable for as long as the gas disk is around (a few Myr). Such configurations may go unstable sometime after the gas disk dissipates so we cannot assume that the resonant chains persist until today. Instead, we conservatively assume that systems with giant planets in a wide range of configurations today may have produced transitional disks early in their history.

Thoroughly following the evolution of systems similar to those we simulated in Section~\ref{sec:stability} for Gyr after the dissipation of the gas disk is beyond the scope of this paper. We explore the post gas-disk evolution briefly in Section~\ref{sec:disrupt} but mostly draw conclusions from the outcomes of the generically spaced multi-giant systems simulated by \citet{juric08}. \citet{juric08} find that the final stable configurations for planets initially spaced by $\lesssim8 \rmh$ typically contain 2--3 planets (see also \citealt{chatterjee08}) and have final semi-major axes similar to their initial. We therefore synthesize the following necessary (but not sufficient) conditions for judging whether a planet could have been part of a common-gap-opening multi-planet system at its infancy:  $q\geq\qgap$, and 1~AU$\leq a\leq$50~AU. Below we will call the occurrence rate of such planets $\fplanet(\deltagapmin|\alpha,h/r)$ (the fraction of stars with planets). Note that the specific choices of the minimum and maximum semi-major axis are not important, because the occurrence rate of giant planets outside the above semi-major axis range is low.

\subsection{The occurrence rate of planets at large separations, $\fplanet$}\label{sec:fplanet}

The occurrence rate of giant planets at tens of of AU is not well constrained by current observations. The best method to detect such planets is direct imaging. However, direct imaging observations are challenging due to the low planet/star contrast ratio at these separations, and the detection limit has only recently reached the planetary mass regime. So far, only a handful of planet candidates have been discovered roughly in the parameter space explored here \citep[e.g.,][]{marois08, lagrange10, marois10, kuzuhara13, rameau13, macintosh15}. Combining non-detections and a handful of detections (including both planets and brown dwarfs) from several direct imaging surveys containing a few hundred stars, \citet{brandt14} proposed a single power-law distribution as a function of $\mplanet$ and $a$,
\begin{align}
dN=0.057\%\times\mplanet^{-0.65\pm0.60} a^{-0.85\pm0.39} d\mplanet da,
\label{eqn:brandt}
\end{align}
\noindent to account for the data, and concluded that 1.7\% (maximum likelihood) of stars host a 5-70~$\mj$ companion between 10-100 AU, the most applicable range of the statistics. 

On the other hand, 
$\fplanet$ at smaller separations of up to a few AU has been well constrained based on the results of radial velocity (RV) surveys. Using 8 years of RV data, \citet{cumming08} fit a power law distribution of the planet occurrence rate as a function of $\mplanet$ and $a$,
\begin{align}
dN=0.74\%\times\mplanet^{-1.31\pm0.2} a^{-0.61\pm0.15} d\mplanet da,
\label{eqn:cumming}
\end{align}
\noindent and concluded that 7.5\%\footnote{Note that the actual number quoted in \citet{cumming08} is 10.5\%. However this number is inconsistent with the power law indexes and normalization constant given in the paper, as well as the quoted planet assurance rates in other parameter spaces, e.g., their Table 1. The 7.5\% number quoted here is calculated directly using the power law indexes and normalization constant given by \citet{cumming08}.} of solar type stars have a $0.3<\mplanet<10\mj$ planet at $0.03<a<3$~AU. Long term radial velocity surveys are pushing to longer orbital periods. For example, \citet{wittenmyer16} find that the giant planet occurrence rate from 3--7 AU is consistent with an extrapolation of \citet{cumming08}. \citet{bryan16} found that occurrence rate of long period, giant planetary companions to known planets declines beyond $\sim 3-10$ AU, but it is unclear whether the companions of shorter period planets have the same period distribution as the general population. Since these recent studies do not have a sufficiently large sample to estimate a power law that we can extrapolate to 30 AU, we use the power law from \citet{cumming08}.

It is unclear to how large separations the $\fplanet$ at small separations probed by the RV surveys can be extrapolated. This largely depends on whether widely separated giant planets are formed in the same way as giant planets as $\sim1-5$ AU, in which case a smooth and homogeneous distribution across the disk is expected. Although there is a general consensus that giant planets in the RV sample are probably formed through the core accretion scenario \citep{pollack96}, the conditions in a protoplanetary disk far from the star may not support core accretion. Both observational \citep[e.g.,][]{brandt14} and theoretical studies \citep[e.g.,][]{dodsonrobinson09, rice15} have proposed that massive companions (including both giant planets and brown dwarfs) beyond $\sim30$ AU may not be the long-period tail of the RV/core accretion samples. Instead, they may be formed through disk fragmentation \citep{boss98, rafikov05, kratter10}. Theoretical expectations for how $\fplanet$ changes over a wide range of $a$ are uncertain at the moment. 

Table~\ref{tab:mplimit} shows the expected percentage of (solar type) stars with planets,
$\fplanet$, using the \citet{cumming08} power law  (Eqn.~\ref{eqn:cumming}) for planets with $\qgap<q<0.013$ at 1~AU$<a<50$~AU for each disk property combination [$\alpha,h/r$] and a minimum gap depletion factor $\deltagapmin$ (i.e., for the range of planet masses that would open deep enough caps for each set of disk properties). For comparison, we also list the corresponding $\fplanet$ using the \citet{brandt14} power law (Equation \ref{eqn:brandt}). In general, $\fplanet$ based on the direct imaging statistics is 5--10 times smaller than based on the RV statistics. We caution that neither Eqn.~\ref{eqn:brandt} nor \ref{eqn:cumming} fully applies to our parameter space, which is too distant compared to the RV sample and too low mass compared to the direct imaging sample. 

\subsection{$\fplanet$ $vs$ $\ftdobs$}\label{sec:fplanetvsftd}

In this section we compare the percentage of stars with giant planets that can open gaps, $\fplanet$ (Section~\ref{sec:fplanet}), to the percentage of proto-planetary disks that are transitional disks, $\ftdobs$ (Section~\ref{sec:ftd}). 

The conventional interpretation of transitional disks, as suggested by the class name, is that they represent a transient inside-out disk clearing phase at the end of the primordial/full disk stage that almost all protoplanetary disks undergo. Under this interpretation, the occurrence rate of transitional disks $\ftdobs$ in a homogeneous disk sample across all ages is simply the ratio of the disk dispersal timescale $\tau_{\rm dispersal}$ to the disk life timescale $\tau_{\rm life}$: $\ftdobs\approx\tau_{\rm dispersal}/\tau_{\rm life}$. Thus, a small $\ftdobs$ ($\approx$10\%) suggests that the clearing process is rapid. This interpretation dates back to \citet{skrutskie90}, who concluded that $\tau_{\rm dispersal}\sim$0.3 Myr, 10\% of $\tau_{\rm life}$ ($\sim$3 Myr), and has been adopted in the literature (e.g.,  \citealt{luhman10,koepferl13}).

However, a more generalized interpretation of $\ftdobs$ is as follows. Only a fraction of disks, $\ftdin$, ever go through a transitional disk phase, which lasts for $\tau_{\rm TD}$. The occurrence rate of transitional disks is determined by both factors, $\ftdobs\approx\ftdin\times\tau_{\rm TD}/\tau_{\rm life}$. The conventional interpretation represents one extreme: $\ftdin\approx1$ and $\tau_{\rm TD}\approx\tau_{\rm dispersal}\approx \ftdobs\times\tau_{\rm life}$. 
In the other extreme, $\tau_{\rm TD}$ can be comparable to $\tau_{\rm life}$, and $\ftdobs\approx\ftdin$, meaning only on the order of 10\% of disks ever go through a long-lasting transitional disk phase. This possibility was raised by \citet[][see also \citealt{owen12clarke}]{muzerolle10}, who noted that if transitional disks do not represent a universal phase of disk evolution, then the conventional clearing timescale estimate $\tau_{\rm dispersal}=\ftdobs\times\tau_{\rm life}$ may be an underestimate.

If we assume that the majority transitional disks are disks with deep and wide gaps opened by planetary mass companions, we require $\fplanet \gtrsim \ftdin$ (and by definition $\ftdin\geq\ftdobs$), because otherwise there are not enough planets to open gaps. Comparing $\ftdobs$ (Section~\ref{sec:ftd}) with $\fplanet$ (Section~\ref{sec:fplanet}), the condition $\fplanet\geq\ftdobs$ can only be satisfied under the most favorable interpretations:
\begin{enumerate}
\item The occurrence rate of giant planets throughout most of the 1--50 AU radius range must be greater or equal to an extrapolation of the RV planet distribution to wider separations, because the direct imaging planet distribution extrapolated to lower masses predicts far too few planets in this range.
\item The gap region must have disk conditions $h/r<0.1$ and $\alpha\lesssim0.01$ ( $\alpha\lesssim0.001$ if future observations determine $\deltagapmin$ is typically $\sim 1000$), because $\fplanet$ with $h/r=0.1$ or $\alpha\gtrsim0.01$ is too small (due to the large planet masses required).
\item The conventional interpretation, that $\ftdin\approx1$ and $\tau_{\rm TD}\approx\ftdobs\times\tau_{\rm life}$, cannot hold, because giant planets at large separations are rare. $\fplanet$ is only 23\% or $\sim2\times\ftdobs$ under the most favorable conditions in the disk parameter space explored here: $\deltagapmin=10$, $\alpha=0.001$, and $h/r=0.05$. Instead, only a small fraction of disks, $\ftdin$, can go through the transitional disk phase, which must last for a significant fraction of the disk lifetime, $\tau_{\rm TD}\approx\tau_{\rm life}(\ftdobs/\ftdin)$, while $\ftdobs/\ftdin$ is close to unity. Meanwhile most, if not all, giant planets must participate in transitional disk sculpting process when they are in gaseous disks.
\end{enumerate}

A prediction can be made based the above conclusions. Since most giant planets need to be at work in transitional disk sculpting, stars with more (fewer) giant planets should show a higher (lower) $\ftdin$ and $\ftdobs$ when they are in the gas disk era. Current data tentatively support this prediction in the case of stellar mass. \citet{andrews11} found while $\ftdobs\sim10\%$ for their entire sample of 91 disks in Taurus and Ophiuchus star-forming regions, big holes preferentially exist in mm bright disks, indicating higher disk mass and higher mass of the central stars \citep{andrews13}. \citet{montet14} found the occurrence rate of giant planets ($1-13\mj$) at $a<20$ AU around lower-mass stars ($M$ dwarfs) is lower $(6.5\pm0.5\%)$ than around their higher mass counterparts, and \citet{shvartzvald16} found an occurrence rate of $5.0^{+4.0}_{-2.4}\%$ for Jupiters orbiting M-dwarfs in the $\sim$ 1.5--6 AU range. (See also consistent results for giant planets in much closer in orbits, \citealt{bowler10, johnson10}.) \citet{lada06} and \citet{downes15} find the fraction of stars with full (non-transitional) disks decreases with stellar mass, possibly because more massive stars are more likely to have giant planets to clear large cavities.

If future ALMA observations ascertain that the typical $\deltagapmin$ for transitional disk is 1000 or even higher, a tension will emerge (i.e., planets massive enough to open such deep gaps are not sufficiently common), as $\fplanet$ decreases with increasing $\deltagapmin$. We will come back to a potential mitigation to this issue in Section~\ref{sec:observations}. A final caveat is that the transitional disk statistics from which \citet{luhman10} estimate $\ftdobs$ are for stars with a median stellar mass of $\sim 0.7 M_\odot$, whereas here we do not consider the stellar mass dependence of the giant planet occurrence rate. 
\subsection{Transitional disks as constraints on initial conditions}
\label{sec:disrupt}

Here we discuss how transitional disks constrain the initial conditions for the post-gas evolution that establishes the architecture of planetary systems, and explore the 10 Gyr evolution of a single configuration as a case study. 
We also investigate the implications of the initial conditions established during the transitional disk stage for \citet{koriski11}'s statistical study of resonances vs. stellar age in Appendix~\ref{subsec:resage}.

We have argued that the relative prevalence of transitional disks and  rarity of giant planets means that most giant planet systems must have carved a transitional disk in their youth. Therefore the planet configurations that account for the observed characteristics of transitional disks (Section~\ref{sec:planetrequirements}) must comprise the initial conditions for the subsequent, gas-free dynamical evolution. This subsequent evolution is thought to play a major role in establishing the observed eccentricity distribution (e.g., \citealt{juric08}) and spawning hot and warm Jupiters through subsequent tidal evolution from highly elliptical orbits (e.g., \citealt{rasio96,nagasawa11,beauge12,dawson13}). The initial conditions for this evolution have been a major source of uncertainty. For example, \citet{juric08} state that the ``the theory is still too crude to allow'' initial conditions ``based on the predictions of planetary formation theory'' and draw planets' initial semi-major axes randomly from a log uniform distribution. \citet{beauge12} take an  ``uncertain leap of faith'' and place planets in chains of first order mean motion resonances, inspired by migration simulations that aim to account for resonances observed in exoplanet systems (e.g., \citealt{snellgrove01}) or necessary in the early Solar System (e.g., \citealt{morbidelli07}) in the Nice model \footnote{A popular hypothesis to account for the dynamical structure of the Kuiper belt, e.g., \citep{tsiganis05}.}. In the transitional disk sculpting scenario we have explored in this paper, the properties of transitional disks  serve as a helpful check-point for the ``initial'' (i.e., post gas-disk) conditions of giant planet systems. As shown in Section~\ref{sec:stability}, most of the suitable configurations feature a chain of planets in or near orbital resonances.

We showed in Section~\ref{sec:stability} that the resonant librations occurring in our simulations continue their libration if we remove gas damping and integrate for 1 Myr (Table \ref{tab:res}). To get a sense for how long the resonant configurations can survive after the gas disk's dissipation, we integrated ten versions of Configuration {\tt 5-2b} with $\Sigma_{30} = 0.1$ g/cm$^2$ for 1 Myr (Fig. \ref{fig:damping} is an example) and then for 10 Gyr without gas. Of those ten, three went unstable during the gas disk stage, three went unstable throughout the star's lifetime (at 63 Myr, 541 Myr, and 1.0 Gyr), and four remained stable for 10 Gyr. The seven systems that did not go unstable during the gas disk stage each feature between one and three (out of four) pairs within 10\% of 2:1 commensurability (with the other pairs within 20\% of commensurability). A detailed study of the post-gas stability timescales established by the configuration during gas disk stage is beyond the scope of this paper, but we preliminarily conclude that the resonant configurations sometimes survive to the present day but sometimes are disrupted on a wide range of timescales. Future studies could account for the range of gas disk lifetimes and the gradual dissipation of the gas disk, thoroughly determine the distribution of resonant disruption timescales, and focus on configurations that fulfill the more stringent constraints on $\deltagapmin$ expected from ongoing ALMA observations.

\section{Summary and discussion}\label{sec:summary}

We investigated the hypothesis that transitional disks are common gaps opened by multiple giant planets. We first synthesized the properties of the extended gaps in transitional disks based on ALMA gas observations (Section~\ref{sec:gapproperties}) and defined the multi-planet systems that can account for these properties  (Section~\ref{sec:planetrequirements}). A series of minimally-packed planetary systems were constructed (Table~\ref{tab:planetarysystems}), and their stability was explored using $N$-body simulations (Section~\ref{sec:stability}, Table~\ref{tab:stability}). We then compared the occurrence rate of transitional disks, $\ftdobs$, with the extrapolated occurrence of giant planets at large separations, $\fplanet$ and considered the subsequent dynamical evolution of the systems beyond the gas disk era (Section~\ref{sec:statistics}). Our main conclusions are:
\begin{enumerate}
\item To open a wide and deep gas gap around a solar type star consistent with observations, a system of 3--6 giant planets is indeed. The minimum planet mass increases with increasing disk scale height $h/r$ and viscosity $\alpha$ (as gaps become more difficult open), and increasing gap depth (Table~\ref{tab:mplimit}). The total number of planets inside a gap depends on the gap size: the planets must be placed close enough together to open a common gap. For given set of gap and disk properties, the less massive the planets, the larger the number and the closer the placement (Table~\ref{tab:planetarysystems}).
\item In general, without the aid from the gas damping, systems with a smaller number of more massive planets tend to be dynamically stable for the typical disk lifetime, while systems with a larger number of less massive planets may be unstable. Eccentricity damping from the residual gas inside the gaps can help stabilize systems by locking planets into mean motion resonances, establishing a chain of pairs resembling the older HR 8799 system \citep{fabrycky10}. However, in some cases eccentricity damping induces resonant repulsion that can drive planets away from commensurate period ratios, resulting in planets that are technically in resonance but not easily identifiable as such.
 \label{con:stability}
\item The giant planet occurrence rate at wide separations $\fplanet$ must equal or exceed the occurrence rate of transitional disks $\ftdobs\sim10\%$ for the planet-opening-common-gap scenario to remain viable.
This can only be satisfied under some of the most favorable conditions explored here, namely, $h/r<0.1$ and $\alpha<0.01$ in the gap regions; more importantly, the occurrence rate of giant planets at $\sim$1--50 AU has to largely follow the radial velocity statistics \citep{cumming08}, not the direct imaging statistics \citep{brandt14}. This situation may be significantly mitigated if the disk viscosity is lower than $10^{-3}$, enabling low mass planets to open gaps through non-linear wave damping processes (see discussion in Section~\ref{sec:alpha}).
\item The fact that $\fplanet\sim25\%\sim2\times\ftdobs$ under the most favorable conditions here ($h/r=0.05$, $\alpha=0.001$, $\deltagapmin=10$) implies that transitional disks are not a universal, fast disk dispersal phase at the end of protoplanetary disks' lifetimes, as assumed by most previous work dating back to \citet{skrutskie90}. Instead, the rarity of giant planets at large separations requires (1) most (if not all) giant planets to be at work in transitional disk sculpting in gaseous disks, (2) the transitional disk phase to be long-lasting, with a timescale comparable to typical disk lifetime, and (3) that only a small fraction ($\sim10\%$) of protoplanetary disks undergo this phase. If $\ftdobs$ increases with cluster age, the dependence reflects the time scale of giant planet formation instead of disk clearing.\label{con:fpvsftd}
\item The formation time scale for giant planets must be short (i.e., compared to the gas disk lifetime) so that they carve and maintain gaps for most of the disk lifetime. HL Tau, if its gaps are opened by giant planets, would be an example of fast planet formation; the system is believed to be $\lesssim 1$ Myr based on the youth of the Taurus cluster \citep{briceno02}.
\item As a consequence of the dynamical evolution of the multi-giant-planet systems inside the gaps, and the fact that most giant planets at large separations must participate in sculpting transitional disks in the gas disk era, most multi-giant systems have to be in mean motion resonances at the end of the gas disk era. The fraction of systems in resonance should decrease with time in the post gas disk era.
\item The properties of transitional disks serve as a helpful check-point for the ``initial'' (i.e., post gas-disk) conditions of giant planet systems. This subsequent evolution, for which the initial conditions have been a major source of uncertainty, is thought to play a major role in establishing the observed eccentricity distribution and spawning hot and warm Jupiters through subsequent tidal evolution from highly elliptical orbits (e.g., \citealt{rasio96,nagasawa11,beauge12,dawson13}).
\end{enumerate}
We re-emphasize that these conclusions only hold if the planetary sculpting hypothesis is responsible for most transitional disks, a mechanism with footings in both theory and observations (Section~\ref{sec:intro}), and in particular, supported by the recent detection of three companions inside LkCa~15's wide gap \citep{sallum15}. 

In drawing these conclusions, we made a couple assumptions that simplified the analysis without losing much generality. We assumed a single planet mass for each configuration and when comparing $\fplanet$ to $\ftdobs$, we did not take into account the different distribution of stellar masses in the samples. The current limited sample sizes of transitional disks and giant planets at large separations do not permit a detailed $\fplanet$--$\ftdobs$ comparison by subgroups.

\subsection{Disk viscosity and low mass planets}\label{sec:alpha}
\label{sec:visc}

The tension between $\fplanet$ and $\ftdobs$ can be significantly mitigated if the disk viscosity is lower than $10^{-3}$. For a given $\deltagapmin$ and $h/r$, lower $\alpha$ leads to a lower gap opening planet mass limit and thus more planets potentially capable of opening gaps. At least at close separations, the planet occurrence rate increases with decreasing planet mass \citep{howard10,mayor11}. Limited by the availability of the large systematic sets of gap opening simulations used to constructed our multi-planet systems \citep{fung14}, we did not go below $\alpha=10^{-3}$ and $h/r=0.05$. However, while $h/r$ can hardly go any lower than 0.05, the tension between $\fplanet$ and $\ftdobs$ suggests that $\alpha$, a quantity that is hard to measure observationally and whose nature is not well understood theoretically, may span lower values in reality. We will argue below that a lower $\alpha$ can be consistent with recent observational and theoretical work.

As discussed in Section~\ref{sec:intro}, ALMA observations have found that the mm continuum ring in a few transitional disks is asymmetric (lopsided), with most of the emission coming from only one side of the disk \citep[e.g.,][]{vandermarel13, casassus13, perez14}. One interpretation is that these asymetric features result from vortex formation at the edge of planet-induced gaps, generated by the Rossby wave instability \citep[e.g.,][]{li00, lin10, lin12}, and dust trapping in vortices \citep[e.g.,][]{lyra09, heng10, zhu14votices, zhu14stone, lyra13}. A key ingredient in this scenario is a low viscosity. As \citet{zhu14stone} pointed out, $\alpha\lesssim10^{-3}$ is required to form vortices at the gap edge.

Other observational evidence comes from ALMA gas disk observations. First, non-detections of non-thermal motions induced by disk turbulence in a few systems suggest low viscosity. For example, based on ALMA CO observations of HD~163296, \citet{flaherty15} determined that the level of turbulence in this disk is lower than $3\%$ of the local sound speed, which implies $\alpha<10^{-3}$. Second, through detailed modeling of the gas gap structures and comparing with simulations of gap opening, \citet{vandermarel16} also proposed $\alpha\lesssim10^{-3}$ inside the gap of a few transitional disks.

On the theory side, low $\alpha$ in protoplanetary disks has recently gained some footing. The Magnetorotational instability (MRI) has been put forward as a prime candidate to provide disk viscosity. The operation of the MRI requires the disk to be sufficiently ionized and well coupled to the magnetic field. In the part of the disk that these conditions are not satisfied, a deadzone with no MRI and low viscosity is expected \citep{gammie96}. Magnetohydrodynamics (MHD) simulations and chemical disk modeling have shown than non-ideal MHD effects, in particular ambipolar diffusion, can significantly suppress  MRI in disks at tens of AU, resulting in very low viscosity equivalent to $\alpha<10^{-3}$ at the bulk of the disk \citep{bai11, bai11stone, bai11-grains,perezbecker11-td, perezbecker11, simon13,  turner14, bai15}.

The gap opening process in an extremely low viscosity environment with $\alpha<10^{-3}$ is different from the viscous disk case (i.e., the \citet{fung14} models used here), as the gap is now opened by nonlinear evolution, instead of the viscous damping, of waves. Nonlinear evolution of waves was explored by \citet{goodman01} and \citet{rafikov02}, who predicted that even  planets much less massive than the thermal mass, $M_{\rm th}=c_{\rm s}^3/G\Omega_{\rm p}$ (about Saturn mass at 15 AU), can open gaps. Their theory was later confirmed by numerical simulations of disk-planet interactions in (nearly) inviscid disks \citep{li09, muto10, dong11-linear, dong11-nonlinear, duffell13, zhu13}. We note that gap opening by low mass planets in nearly inviscid disks can be slow; it has not yet been explored whether the timescale is consistent with observed disk lifetimes. Finally, if future observational and numerical studies demonstrate that disk viscosity is generally very low (e.g., $\alpha\lesssim10^{-4}$), {\it and} this low viscosity enables Earth and super-Earth-like planets to participate in the transitional disk sculpting processes, {\it and} the abundance of such planets found by Kepler (e.g., \citealt{howard12}) extends to tens of AU, we caution that the conclusion made in Section~\ref{sec:fplanetvsftd} -- namely the $\sim10\%$ occurrence rate of transitional disks reflects nature not nurture -- may be altered.

\subsection{Connections to ongoing observations}\label{sec:observations}

In this paper we discussed gaps with $\deltagapmin=10$, a conservative limit set by most current observations, as well as $\deltagapmin=10$ and 100, more aggressive limits put forward by pioneering studies with the latest ALMA results \citep{vandermarel16}. As additional attenae are commissioned for ALMA,
future observations with finer angular resolution and better sensitivity may systematically push $\deltagap$ to as high as $\sim10^3$. If such a large depletion is confirmed, even with the most gap-friendly disk properties in our models ($\alpha=0.001$ and $h/r=0.05$) and RV planet statistics at large separations, there may not be enough giant planets capable of opening such deep gaps to account for $\ftdobs$ (Table~\ref{tab:mplimit}).
Therefore the planetary sculpting hypothesis would only be viable for the ultra-low viscosity $\alpha<10^{-3}$ discussed in Section~\ref{sec:visc} and not considered in the configurations explored in the paper.

As we concluded most giant planets have to be at work in gap opening, stars with more (fewer) giant planets should show a higher (lower) $\ftdin$ and $\ftdobs$. This expected correlation is tentatively supported by the data for different stellar masses (Section~\ref{sec:fplanetvsftd}). Future measurements of $\ftdobs$ and $\fplanet$ for subgroups of stars can further test this prediction.

We concluded that most giant planets had to participate in gap opening when they were in protoplanetary disks and gas damping established stable resonant configurations. For some configurations (particular those with four or more massive planets), the period ratios are commensurate and the system could be identifiable as a resonant configuration. The young HR~8799, with all four planets likely in a chain of 2:1 resonances \citep{fabrycky10}, is an excellent example. For other configurations (particularly those with only three planets), we found resonant repulsion drives the planets away from recognizable commensurability. We found that each of the configurations we studied features libration of the 2:1 resonant argument and many feature libration of three-body resonances (e.g., Laplace). The resonance librations persist after gas damping shuts off but may be difficult to measure from observations because the eccentricities are small. After the protoplanetary disk phase, the fraction of multi-giant systems in resonance is expected to decrease over time (Section~\ref{sec:disrupt}), a prediction that can be more robustly tested in the future with a sample planets whose host stars have well-determined ages that span multiple orders of magnitude.

The two major ongoing direct imaging surveys with VLT/SPHERE \citep{beuzit08} and Gemini/GPI \citep{macintosh08} are expected to establish much better statistics of giant planets at tens of AU from the star, resolving the question of whether giant planets are sufficiently common for planetary sculpting to be predominant cause of transitional disks.


\section*{Acknowledgments}

We thank the referee, John E. Chambers, for a constructive report that improved the quality and the clarity of the paper. R.D. thanks Xue-Ning Bai for teaching him about MRI in disks, and Ewine van Dishoeck and Nienke van der Marel for educating him on the topic of ALMA disk observations. We also thank Sean Andrews, Tim Brandt, Sourav Chatterjee, Eugene Chiang, Paul Duffell, Misato Fukagawa, Andrea Isella, John Johnson, Heather Knutson, Chalie Lada, Renu Malhotra, Rebecca Martin, and Ben Montet for insightful discussions. We particularly thank Jeffery Fung and Paul Duffell for kindly sharing the simulation data in \citet{fung14} and \citet{duffell15dong} with us. This project is partially supported by NASA through Hubble Fellowship grant HST-HF-51320.01-A (first R.D.) awarded by the Space Telescope Science Institute, which is operated by the Association of Universities for Research in Astronomy, Inc., for NASA, under contract NAS 5-26555, and by the Miller Institute for Basic Research in Science (second R.D.). Simulations were run on the SAVIO computational cluster provided by Berkeley Research Computing.


\begin{appendix}
\section{Plots of resonant behavior for additional configurations}
Here we plot orbital elements and resonant arguments for additional systems including gas damping. The figures here (Fig. \ref{fig:app1}, \ref{fig:app2}, \ref{fig:app3}, \ref{fig:app4}, \ref{fig:app5}, \ref{fig:app6}, \ref{fig:app7}, \ref{fig:app7b}, \ref{fig:app8}, \ref{fig:app9}, \ref{fig:app10}, \ref{fig:app11}, \ref{fig:app12}, \ref{fig:app13}, \ref{fig:app14}, \ref{fig:app15}, \ref{fig:app16}, \ref{fig:app17}, \ref{fig:app18}, \ref{fig:app19}) are analogous to Fig. \ref{fig:damping} in the main text.

\section{Implications for the statistical study of resonances vs. stellar age in Koriski \& Zucker (2011) }
\label{subsec:resage}

Several years ago, \citet{koriski11} reported a correlation between stellar chromospheric age (based on H and K emission lines) and orbital resonances. They found that systems containing detected giant planets with period ratios near 2:1 are younger (median age 4.08 Gyr) than those without (median age 6.23 Gyr). Using a permutation test, they found a 0.4\% probability that such a difference would occur by chance. In order to believe this result is not due to chance, we need a viable alternative hypothesis that yields a higher probability for the age difference. Below we provide two such alternative hypotheses and make connections to the outcome of the dynamical processes studied in Section~\ref{sec:stability}. 

\begin{figure}
\begin{center}
\includegraphics{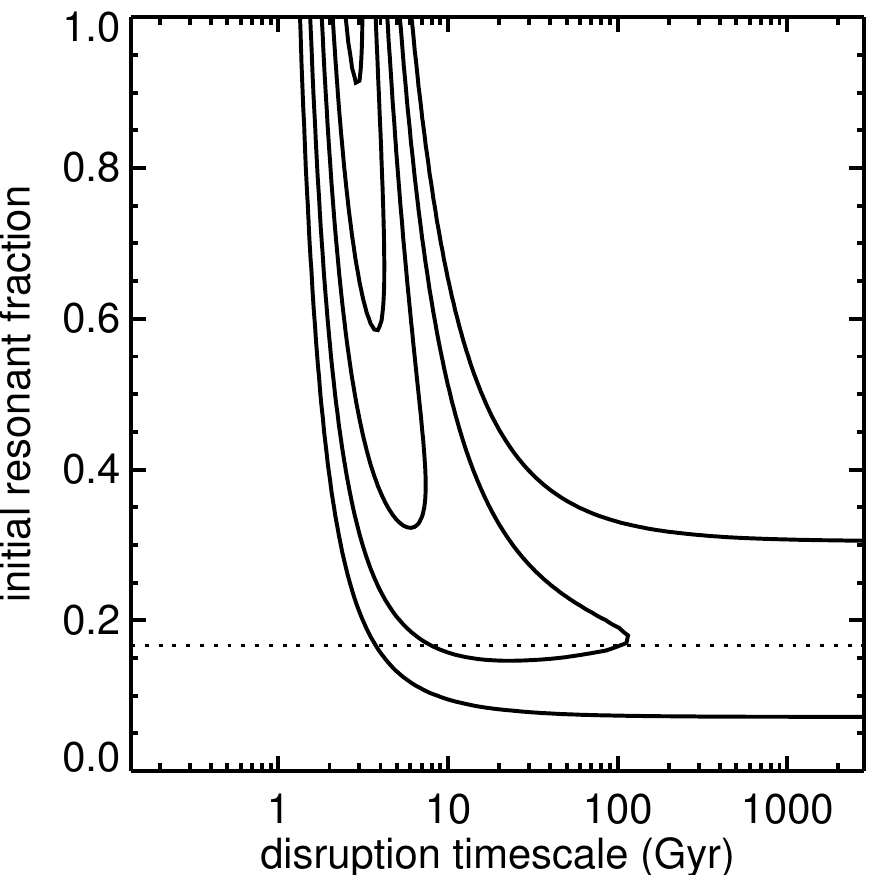}
\end{center}
\figcaption{Probability contour of initial resonant fraction vs. disruption timescale (Eqn. \ref{eqn:prob} and \ref{eqn:probres}) to account for the stellar ages and absence or presence of 2:1 commensurabilities in the \citet{koriski11} collection. The contours are for $[.01,.05,.2,.5,.9]$, normalized to the highest probability. The dashed line represents the present day 5/30 fraction of 2:1 commensurabilities in the collection. 
\label{fig:res}}
\end{figure}

First we consider the hypothesis that initially a fraction $f_{\rm res,0}$ of giant planet systems begin in resonance and exponentially ``decay" out of resonance on a characteristic timescale resonant disruption timescale $t_{\rm res}$. We compute the probability of stellar ages for resonant\footnote{Here we adopt the terminology ``resonant'' for period commensurabilities, although in practice it is unknown for most observed systems whether the resonant angle librates.} systems $\{t_{\star, \rm res}\}$, of which there are $N_{\rm res} = 5$ in \citet{koriski11} sample, and non-resonant systems $\{t_{\star, \rm non-res}\}$, of which there are $N_{\rm non-res} = 25$, as
\begin{eqnarray}
    {\rm prob} = \prod_{i}^{N_{\rm res}}f_{\rm res,0} {\rm prob_{res}} (t_{\star,i}) \prod_{j}^{N_{\rm non-res}} \left[1-f_{\rm res,0} {\rm prob_{res}}(t_{\star,j})\right] , \label{eqn:prob}\\
    {\rm prob_{res} } (t_\star)= \exp \left(-t_{\star, \rm res}/t_{\rm res}\right) \label{eqn:probres}.
\end{eqnarray}
We contour the probability in Fig. \ref{fig:res}. The maximum probability occurs for $f_{\rm res,0} = 1$ and $t_{\rm res} = 2.7$ Gyr and is about 22 times higher than the probability of $f_{\rm res,0} = 5/30$ and $t_{\rm res} = \infty$ (i.e., such that systems are non-resonant by nature, not nurture, and the difference in age is due to chance). 

To choose between the chance hypothesis as an explanation for the data and the hypothesis that all systems start out in resonance and are disrupted on a character 3 Gyr timescale, we also need to consider which is more likely a priori. Our Section~\ref{sec:stability} and ~\ref{sec:statistics} results allow us to weigh in. The \citet{koriski11} sample has smaller semi-major axes than the configurations studied here (e.g., their observed sample of 2:1 resonant pairs have a median $a=1.5$ AU for the outer planet, whereas our configuration {\tt 5-2b}, which features 2:1 commensurabilities, has its innermost planet at 2.4 AU). However, the processes we study here may apply to planets at smaller semi-major axes, extending in to the Period Valley (e.g., \citealt{jones03}) at 0.8 AU. Their sample of 2:1 resonant planets have masses $\lesssim 3 M_J$, similar to the masses in our configurations of planets with recognizable commensurabilities (Table \ref{tab:res}).

We favor a high $f_{\rm res,0}$ because nearly all giant planets must participate in the transitional disk stage (Section~\ref{sec:fplanetvsftd}) and the planetary systems that can account for observed transitional disk properties are often configured in resonant chains (Section~\ref{sec:stability}). However, in our case study of long-term evolution (Section~\ref{sec:disrupt}), we did not find a typical resonance disruption timescale of 3 Gyr; instead, we found the timescales spanned many orders of magnitude. Nonetheless, we do find that long resonance disruption timescales of $\sim$ Gyr are possible (Section~\ref{sec:disrupt}), in contrast to other types of processes that would primarily operate on shorter timescales (e.g., interactions with a remnant planetesimal disk, \citealt{thommes08,moore13}).

\begin{figure}
\begin{center}
 \includegraphics{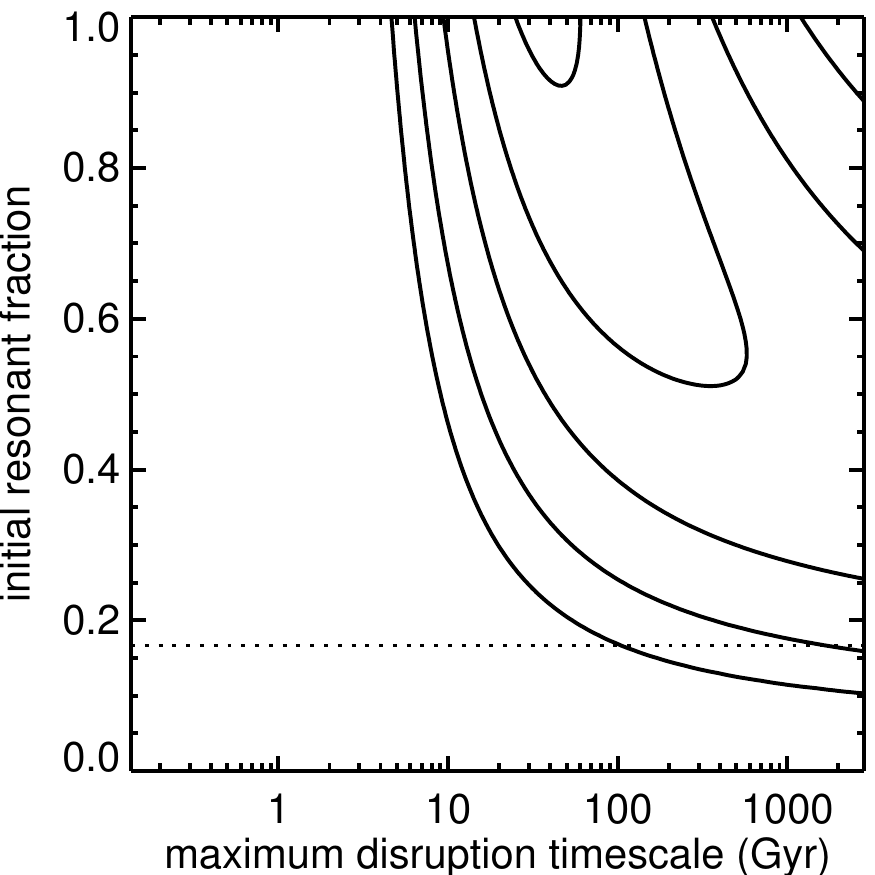}
\end{center}
\figcaption{Probability contour of initial resonant fraction vs. maximum disruption timescale (Eqn. \ref{eqn:prob} and \ref{eqn:probres2}; the maximum disruption timescale is an upper cut-off for a log uniform distribution of disruption timescales) to account for the stellar ages and absence or presence of 2:1 commensurabilities in the \citet{koriski11} collection. The contours are for $[.01, .05, .2, .5, .9]$, normalized to the highest probability. The dashed line represents the present day 5/30 fraction of 2:1 commensurabilities in the collection.
\label{fig:res2}}
\end{figure}

\begin{figure}
\begin{center}
\includegraphics[trim=0 0 0 0, clip,width=0.8\textwidth,angle=0]{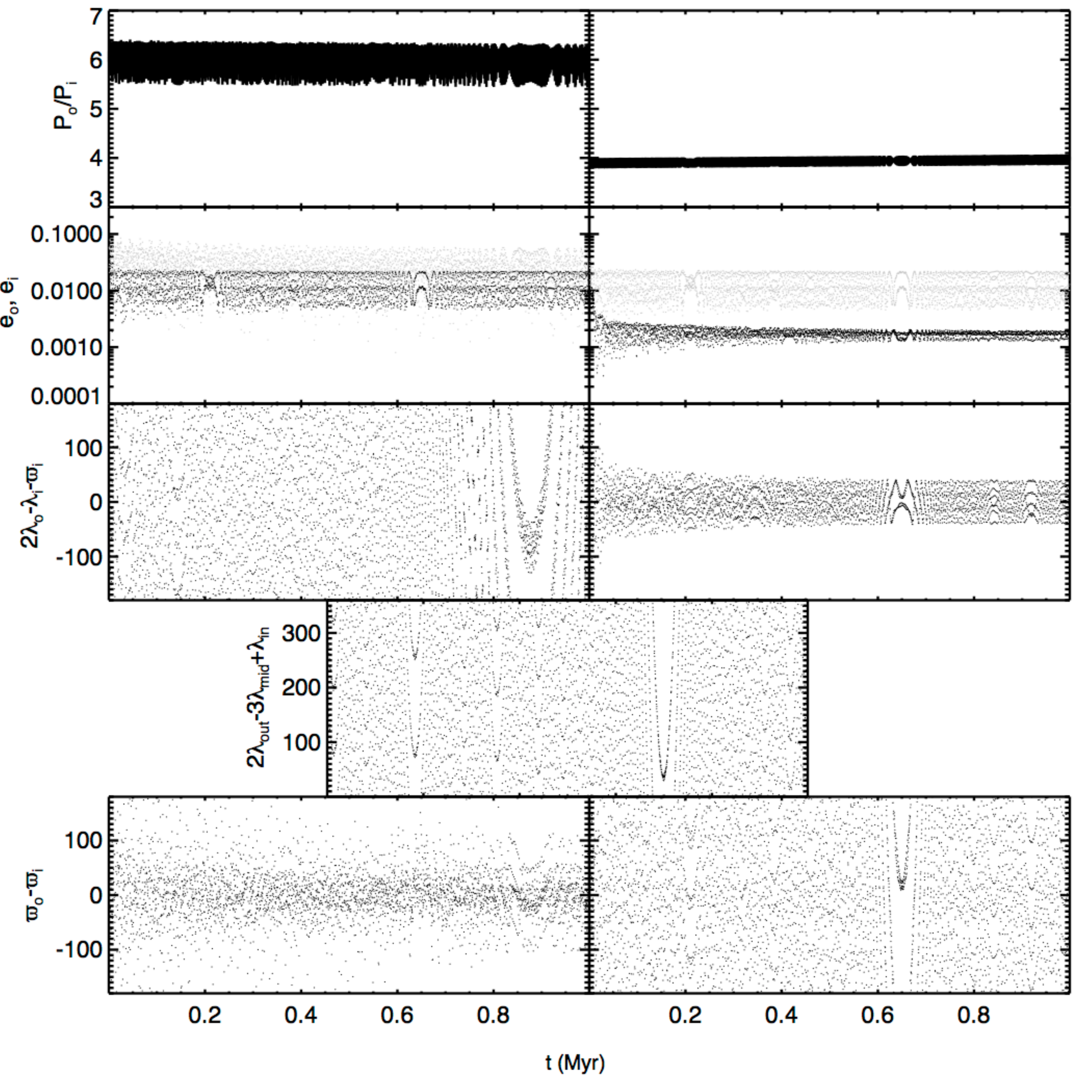}
\end{center}
\figcaption{Evolution of orbital elements for System {\tt 3-5} with gas damping included (Table~\ref{tab:planetarysystems}; $\Sigma_{30} = 0.01$ g cm$^{-2}$). Row 1: period ratio of each adjacent pair of planets, from outer to inner. Row 2: eccentricity of outer planet (gray) and inner planet (black). Row 3: 2:1 mean motion resonant argument involving the longitude of periapse of the inner planet. Row 4: Three-body resonant argument (Laplace resonance). Row 5: Separation of adjacent planets' longitude of periapses.
\label{fig:app1}}
\end{figure}

\begin{figure}
\begin{center}
\includegraphics[trim=0 0 0 0, clip,width=0.8\textwidth,angle=0]{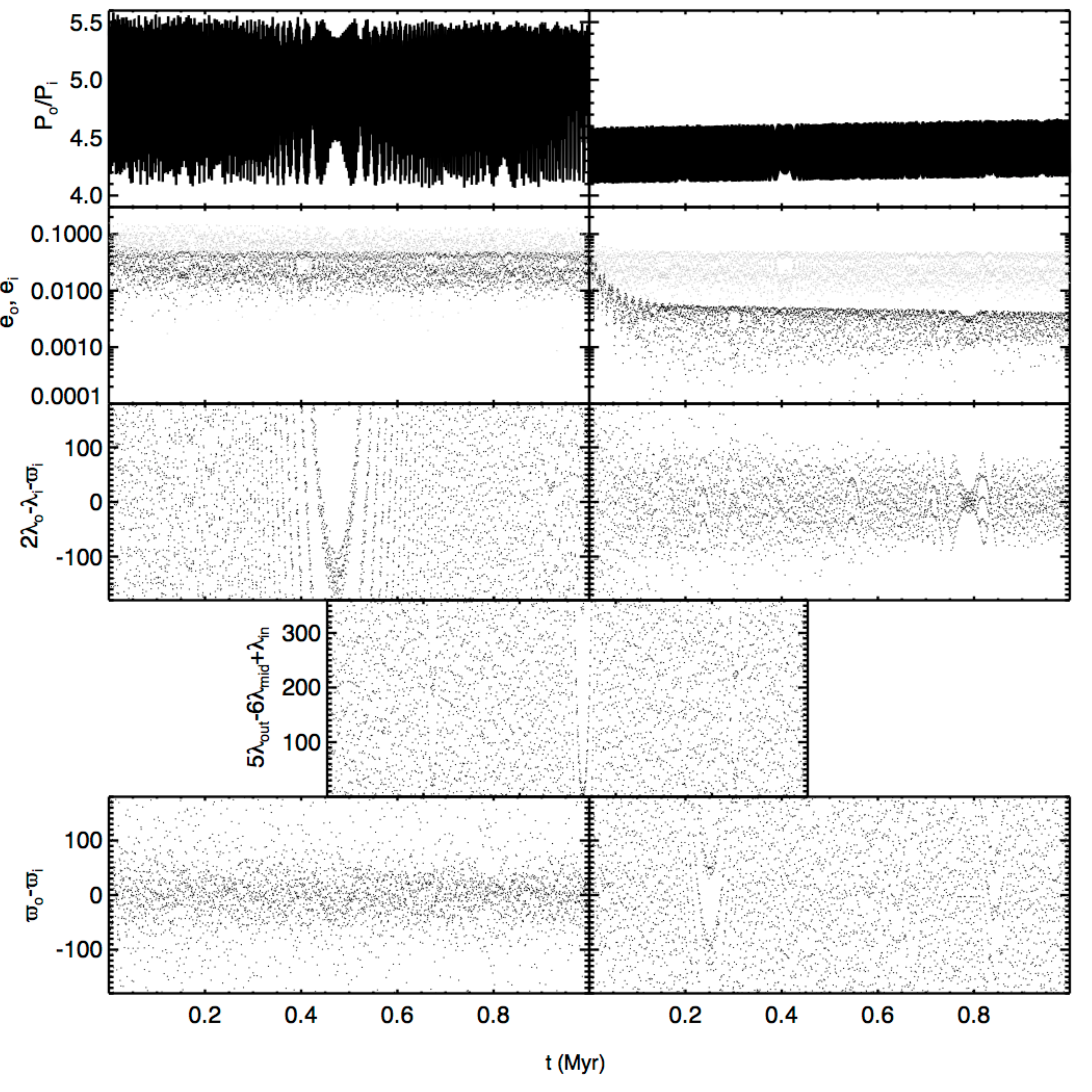}
\end{center}
\figcaption{Evolution of orbital elements for System {\tt 3-10a} with gas damping included (Table~\ref{tab:planetarysystems}; $\Sigma_{30} = 0.001$ g cm$^{-2}$). Row 1: period ratio of each adjacent pair of planets, from outer to inner. Row 2: eccentricity of outer planet (gray) and inner planet (black). Row 3: 2:1 mean motion resonant argument involving the longitude of periapse of the inner planet. Row 4: Three-body resonant argument. Row 5: Separation of adjacent planets' longitude of periapses.
\label{fig:app2}}
\end{figure}

\begin{figure}
\begin{center}
\includegraphics[trim=0 0 0 0, clip,width=0.8\textwidth,angle=0]{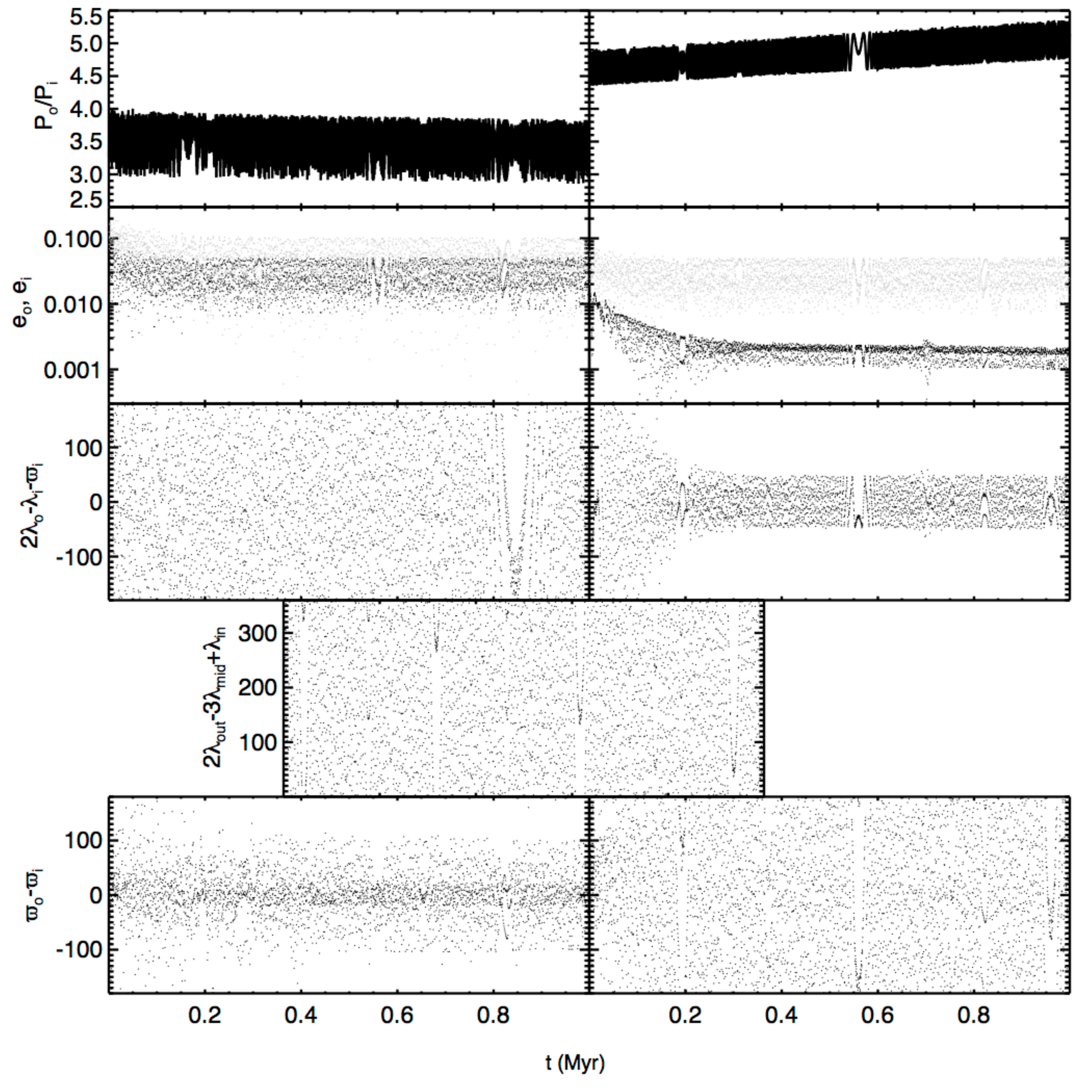}
\end{center}
\figcaption{Evolution of orbital elements for System {\tt 3-10b} with gas damping included (Table~\ref{tab:planetarysystems}; $\Sigma_{30} = 0.01$ g cm$^{-2}$). Row 1: period ratio of each adjacent pair of planets, from outer to inner. Row 2: eccentricity of outer planet (gray) and inner planet (black). Row 3: 2:1 mean motion resonant argument involving the longitude of periapse of the inner planet. Row 4: Three-body resonant argument (Laplace resonance). Row 5: Separation of adjacent planets' longitude of periapses.
\label{fig:app3}}
\end{figure}

\begin{figure}
\begin{center}
\includegraphics[trim=0 0 0 0, clip,width=0.8\textwidth,angle=0]{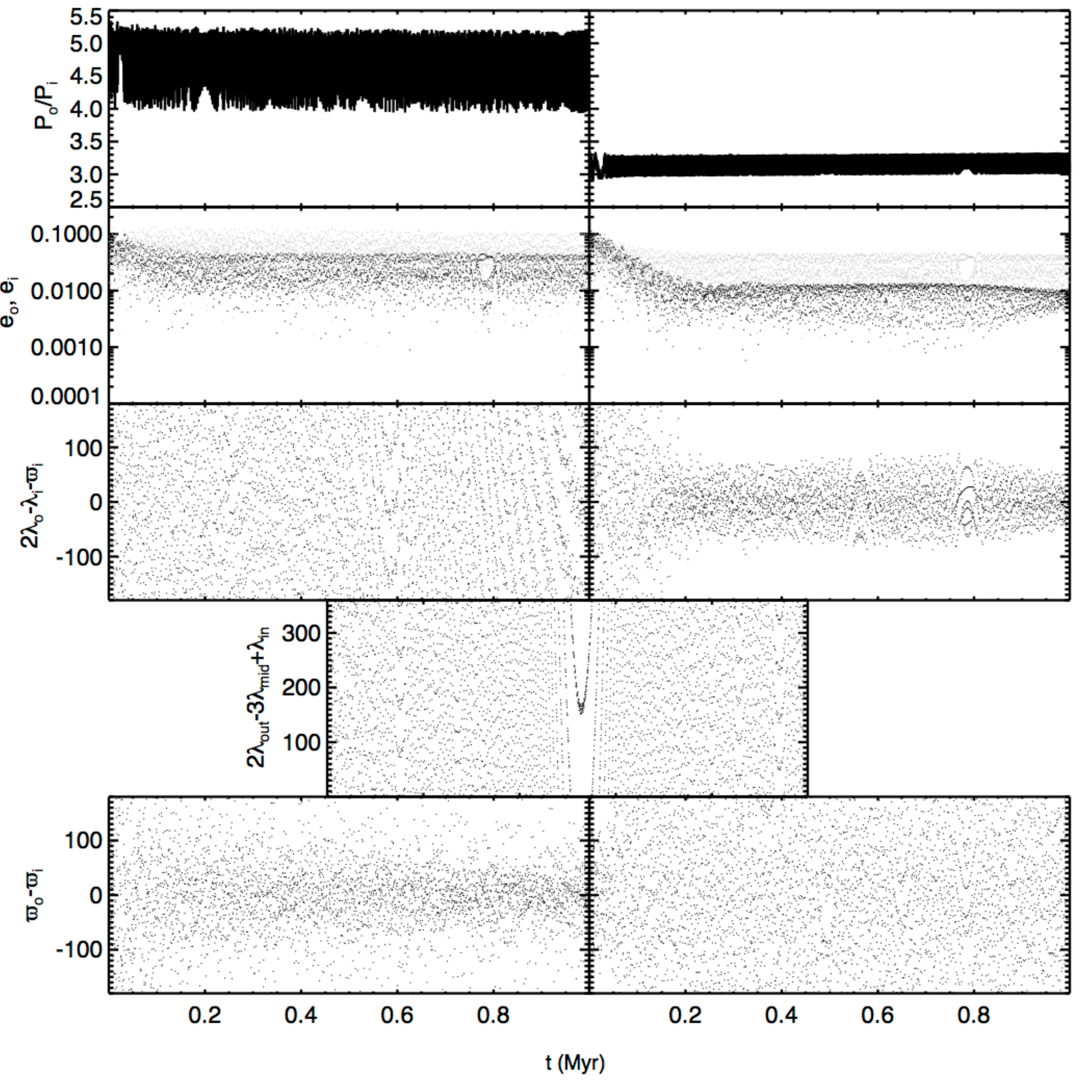}
\end{center}
\figcaption{Evolution of orbital elements for System {\tt 3-10c} with gas damping included (Table~\ref{tab:planetarysystems}; $\Sigma_{30} = 0.001$ g cm$^{-2}$). Row 1: period ratio of each adjacent pair of planets, from outer to inner. Row 2: eccentricity of outer planet (gray) and inner planet (black). Row 3: 2:1 mean motion resonant argument involving the longitude of periapse of the inner planet. Row 4: Three-body resonant argument (Laplace resonance). Row 5: Separation of adjacent planets' longitude of periapses.
\label{fig:app4}}
\end{figure}

\begin{figure}
\begin{center}
\includegraphics[trim=0 0 0 0, clip,width=0.8\textwidth,angle=0]{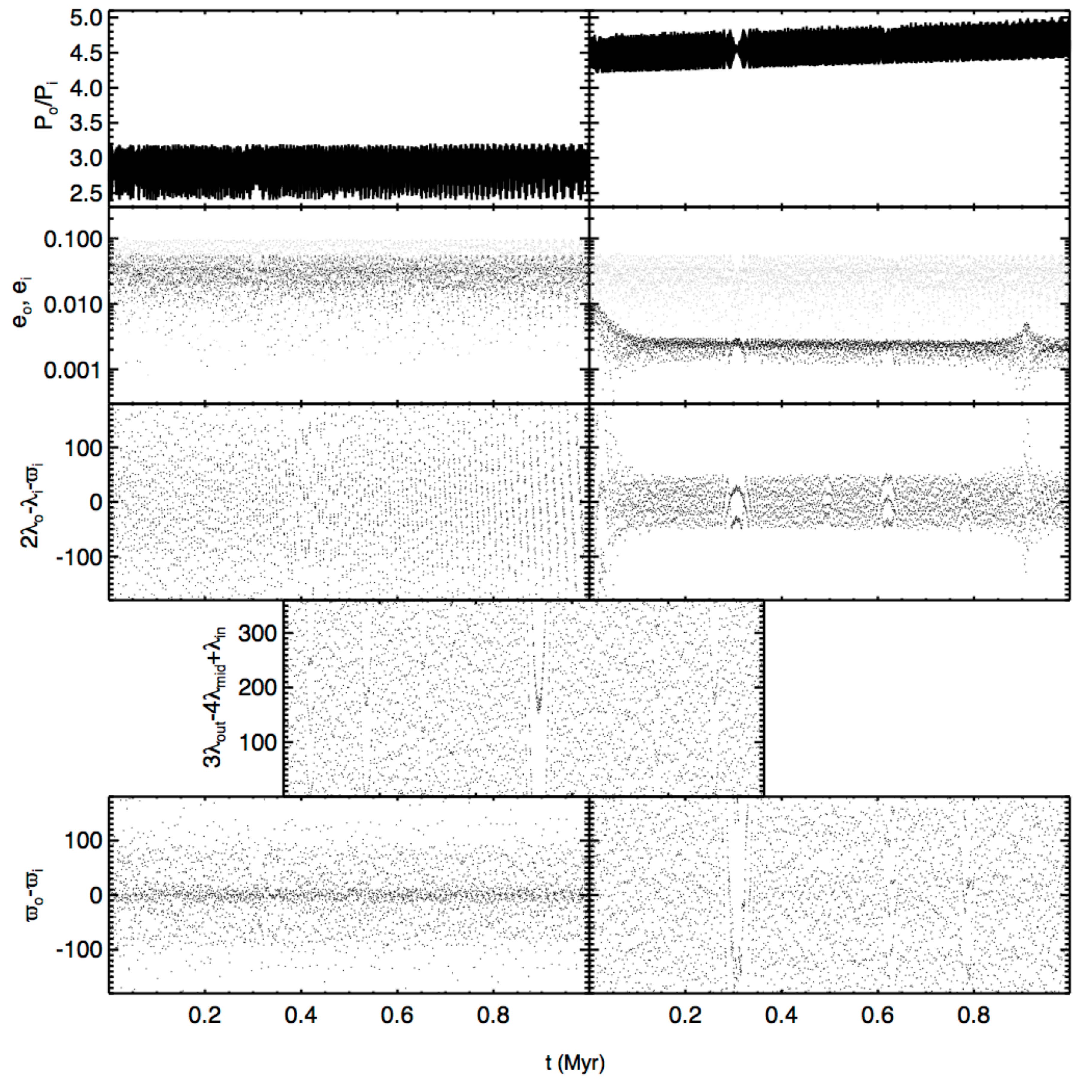}
\end{center}
\figcaption{Evolution of orbital elements for System {\tt 3-10d} with gas damping included (Table~\ref{tab:planetarysystems}; $\Sigma_{30} = 0.01$ g cm$^{-2}$). Row 1: period ratio of each adjacent pair of planets, from outer to inner. Row 2: eccentricity of outer planet (gray) and inner planet (black). Row 3: 2:1 mean motion resonant argument involving the longitude of periapse of the inner planet.Row 4: Three-body resonant argument. Row 5: Separation of adjacent planets' longitude of periapses.
\label{fig:app5}}
\end{figure}

\begin{figure}
\begin{center}
\includegraphics[trim=0 0 0 0, clip,width=0.8\textwidth,angle=0]{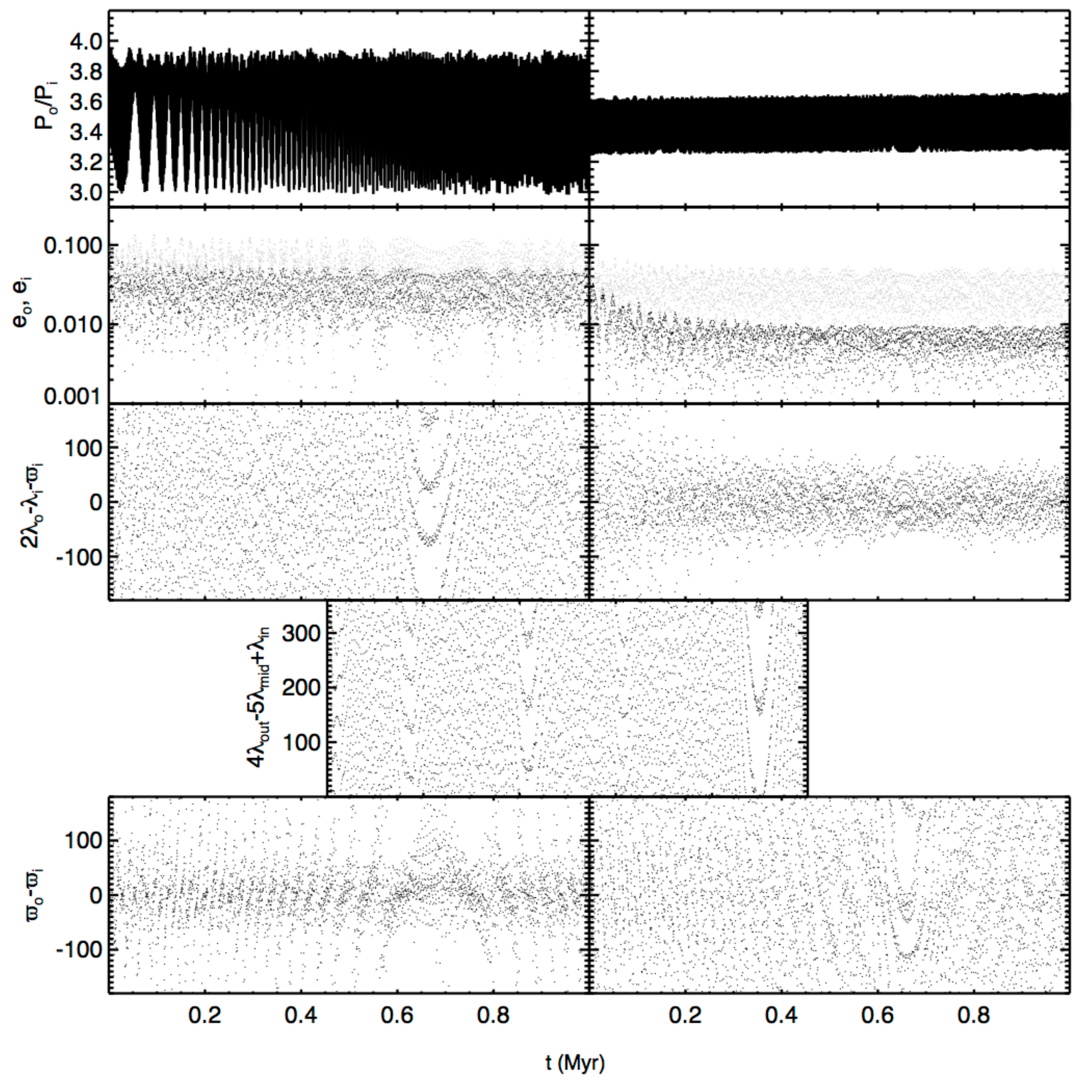}
\end{center}
\figcaption{Evolution of orbital elements for System {\tt 3-10e} with gas damping included (Table~\ref{tab:planetarysystems}; $\Sigma_{30} = 0.1$ g cm$^{-2}$). Row 1: period ratio of each adjacent pair of planets, from outer to inner. Row 2: eccentricity of outer planet (gray) and inner planet (black). Row 3: 2:1 mean motion resonant argument involving the longitude of periapse of the inner planet. Row 4: Three-body resonant argument. Row 5: Separation of adjacent planets' longitude of periapses.
\label{fig:app6}}
\end{figure}

\begin{figure}
\begin{center}
\includegraphics[trim=0 0 0 0, clip,width=0.8\textwidth,angle=0]{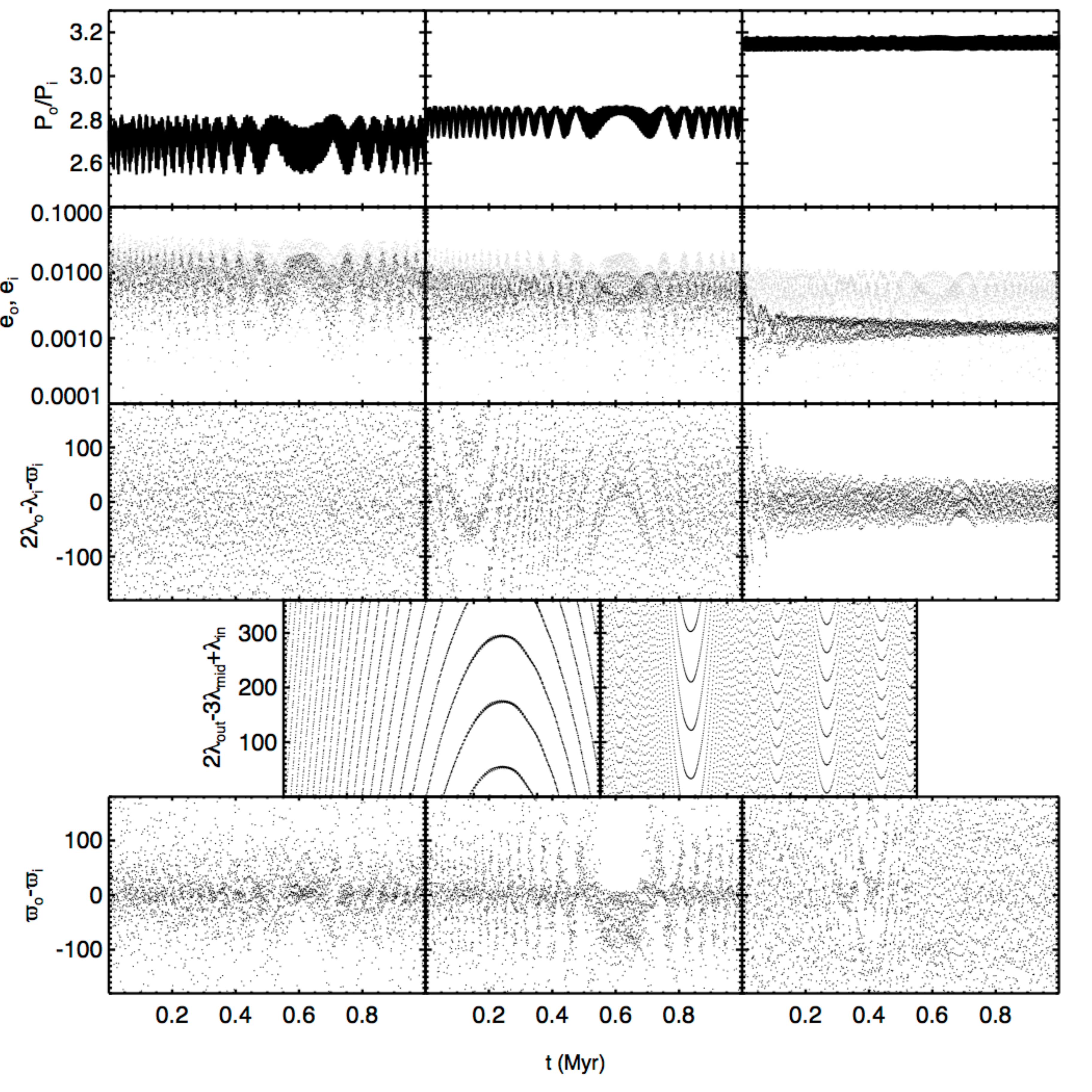}
\end{center}
\figcaption{Evolution of orbital elements for System {\tt 4-2} with gas damping included (Table~\ref{tab:planetarysystems}; $\Sigma_{30} = 0.01$ g cm$^{-2}$). Row 1: period ratio of each adjacent pair of planets, from outer to inner. Row 2: eccentricity of outer planet (gray) and inner planet (black). Row 3: 2:1 mean motion resonant argument involving the longitude of periapse of the inner planet. Row 4: Three-body resonant argument (Laplace resonance). Row 5: Separation of adjacent planets' longitude of periapses.
\label{fig:app7}}
\end{figure}

\begin{figure}
\begin{center}
\includegraphics[trim=0 0 0 0, clip,width=0.8\textwidth,angle=0]{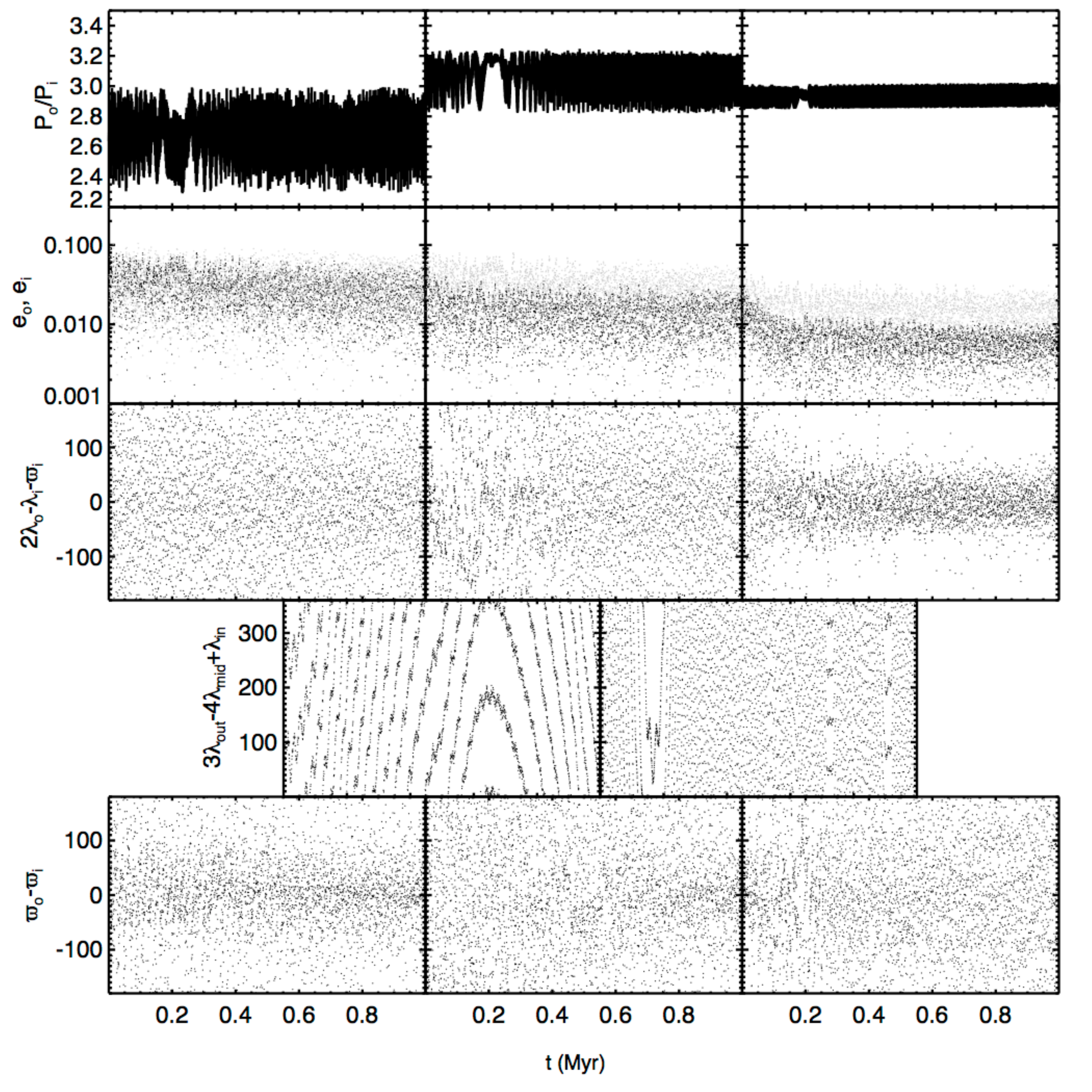}
\end{center}
\figcaption{Evolution of orbital elements for System {\tt 4-5a} with gas damping included (Table~\ref{tab:planetarysystems}; $\Sigma_{30} = 0.001$ g cm$^{-2}$). Row 1: period ratio of each adjacent pair of planets, from outer to inner. Row 2: eccentricity of outer planet (gray) and inner planet (black). Row 3: 2:1 mean motion resonant argument involving the longitude of periapse of the inner planet. Row 4: Three-body resonant argument. Row 5: Separation of adjacent planets' longitude of periapses.
\label{fig:app7b}}
\end{figure}

\begin{figure}
\begin{center}
\includegraphics[trim=0 0 0 0, clip,width=0.8\textwidth,angle=0]{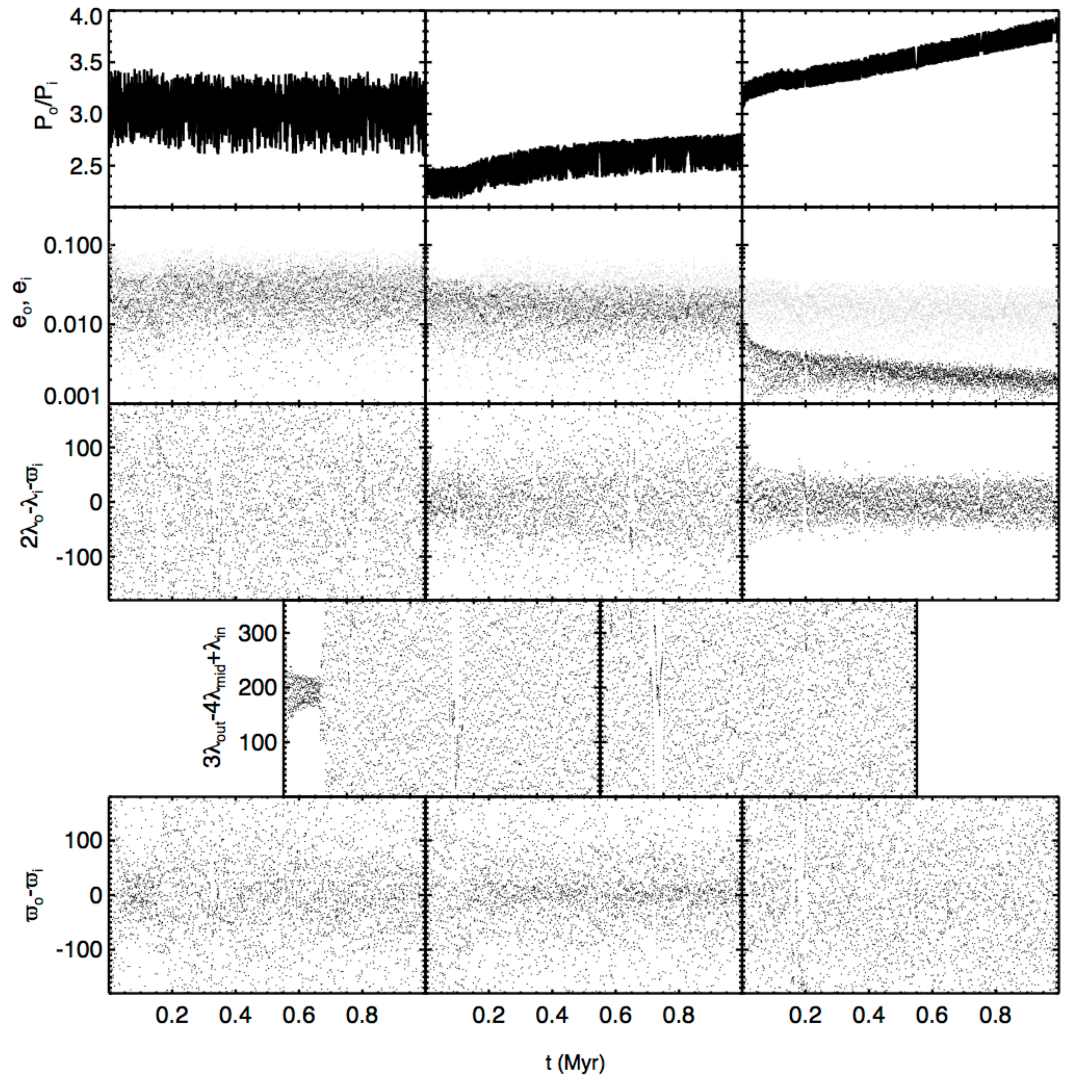}
\end{center}
\figcaption{Evolution of orbital elements for System {\tt 4-5b} with gas damping included (Table~\ref{tab:planetarysystems}; $\Sigma_{30} = 0.01$ g cm$^{-2}$). Row 1: period ratio of each adjacent pair of planets, from outer to inner. Row 2: eccentricity of outer planet (gray) and inner planet (black). Row 3: 2:1 mean motion resonant argument involving the longitude of periapse of the inner planet. Row 4: Three-body resonant argument. Row 5: Separation of adjacent planets' longitude of periapses.
\label{fig:app8}}
\end{figure}

\begin{figure}
\begin{center}
\includegraphics[trim=0 0 0 0, clip,width=0.8\textwidth,angle=0]{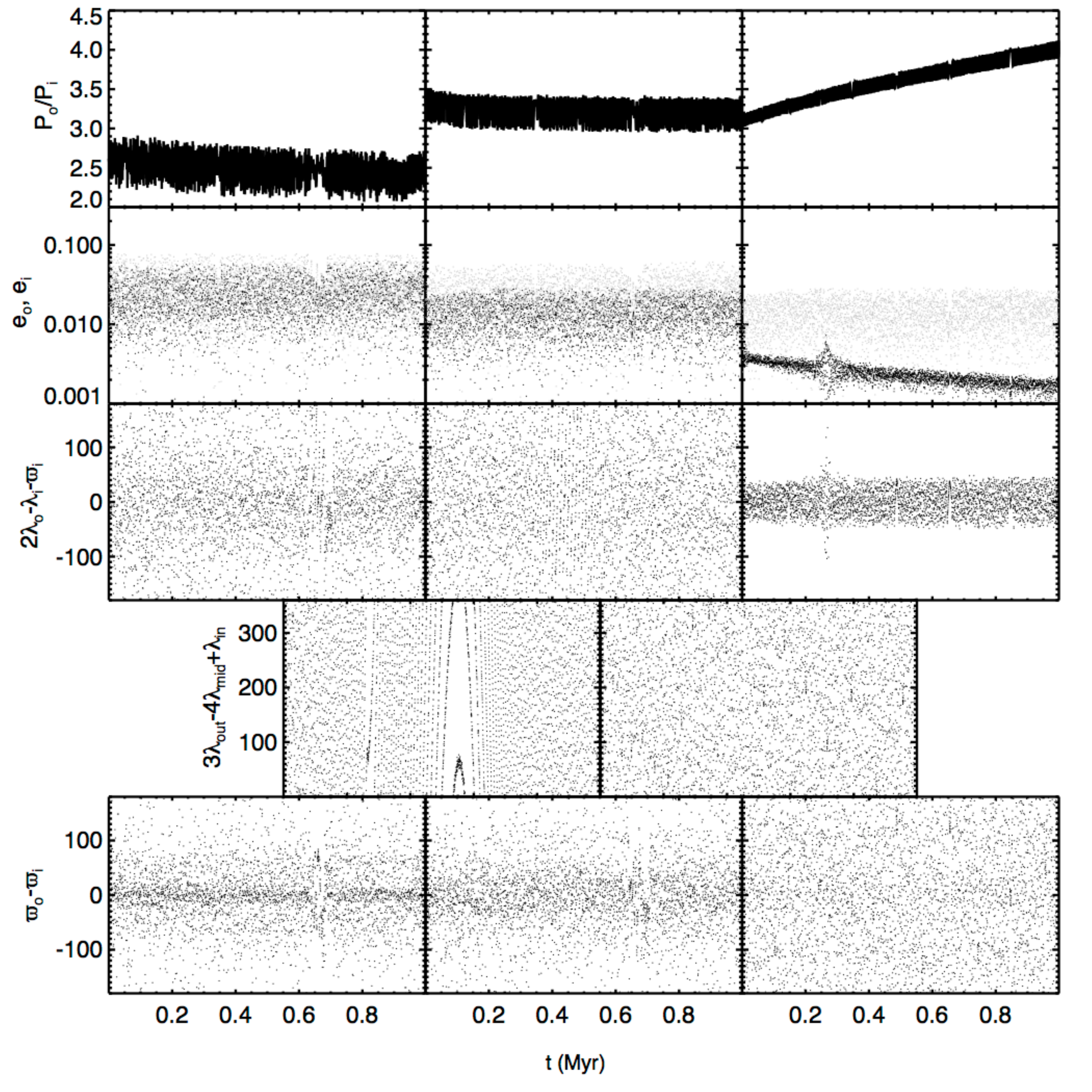}
\end{center}
\figcaption{Evolution of orbital elements for System {\tt 4-5c} with gas damping included (Table~\ref{tab:planetarysystems}; $\Sigma_{30} = 0.1$ g cm$^{-2}$). Row 1: period ratio of each adjacent pair of planets, from outer to inner. Row 2: eccentricity of outer planet (gray) and inner planet (black). Row 3: 2:1 mean motion resonant argument involving the longitude of periapse of the inner planet. Row 4: Three-body resonant argument. Row 5: Separation of adjacent planets' longitude of periapses.
\label{fig:app9}}
\end{figure}

\begin{figure}
\begin{center}
\includegraphics[trim=0 0 0 0, clip,width=0.8\textwidth,angle=0]{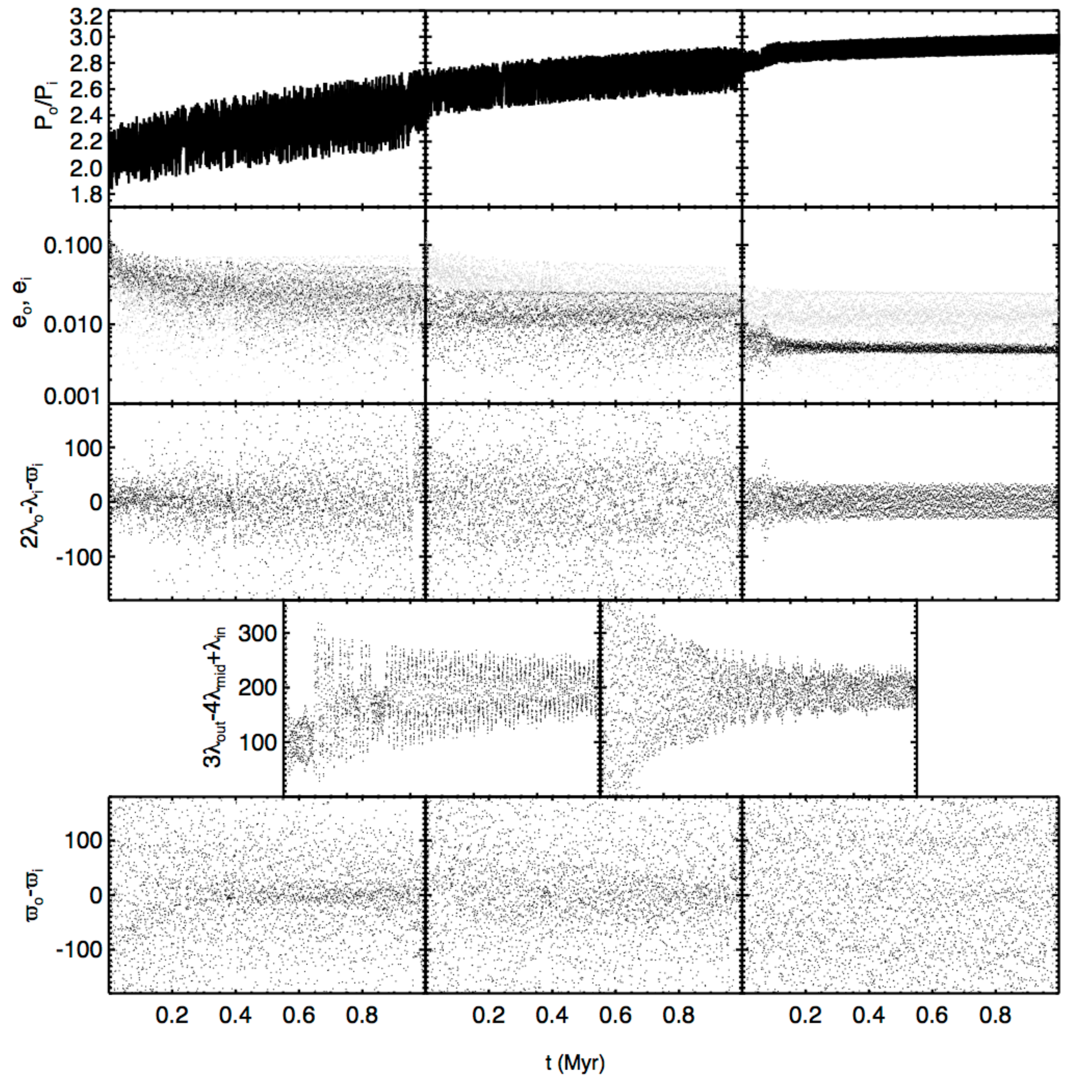}
\end{center}
\figcaption{Evolution of orbital elements for System {\tt 4-5d} with gas damping included (Table~\ref{tab:planetarysystems}; $\Sigma_{30} = 0.1$ g cm$^{-2}$). Row 1: period ratio of each adjacent pair of planets, from outer to inner. Row 2: eccentricity of outer planet (gray) and inner planet (black). Row 3: 2:1 mean motion resonant argument involving the longitude of periapse of the inner planet. Row 4: Three-body resonant argument. Row 5: Separation of adjacent planets' longitude of periapses.
\label{fig:app10}}
\end{figure}

\begin{figure}
\begin{center}
\includegraphics[trim=0 0 0 0, clip,width=0.8\textwidth,angle=0]{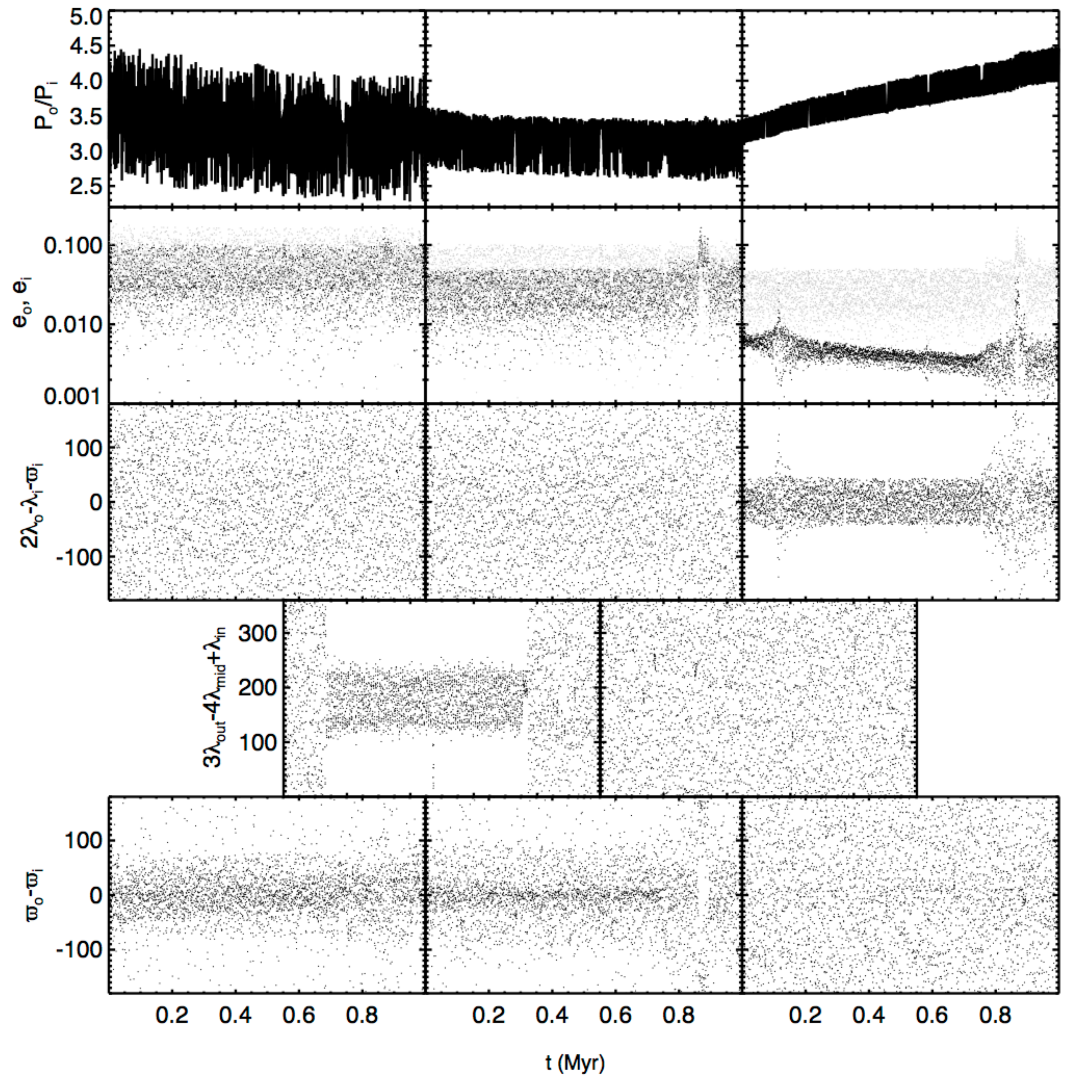}
\end{center}
\figcaption{Evolution of orbital elements for System {\tt 4-10a} with gas damping included (Table~\ref{tab:planetarysystems}; $\Sigma_{30} = 0.01$ g cm$^{-2}$). Row 1: period ratio of each adjacent pair of planets, from outer to inner. Row 2: eccentricity of outer planet (gray) and inner planet (black). Row 3: 2:1 mean motion resonant argument involving the longitude of periapse of the inner planet. Row 4: Three-body resonant argument. Row 5: Separation of adjacent planets' longitude of periapses.
\label{fig:app11}}
\end{figure}

\begin{figure}
\begin{center}
\includegraphics[trim=0 0 0 0, clip,width=0.8\textwidth,angle=0]{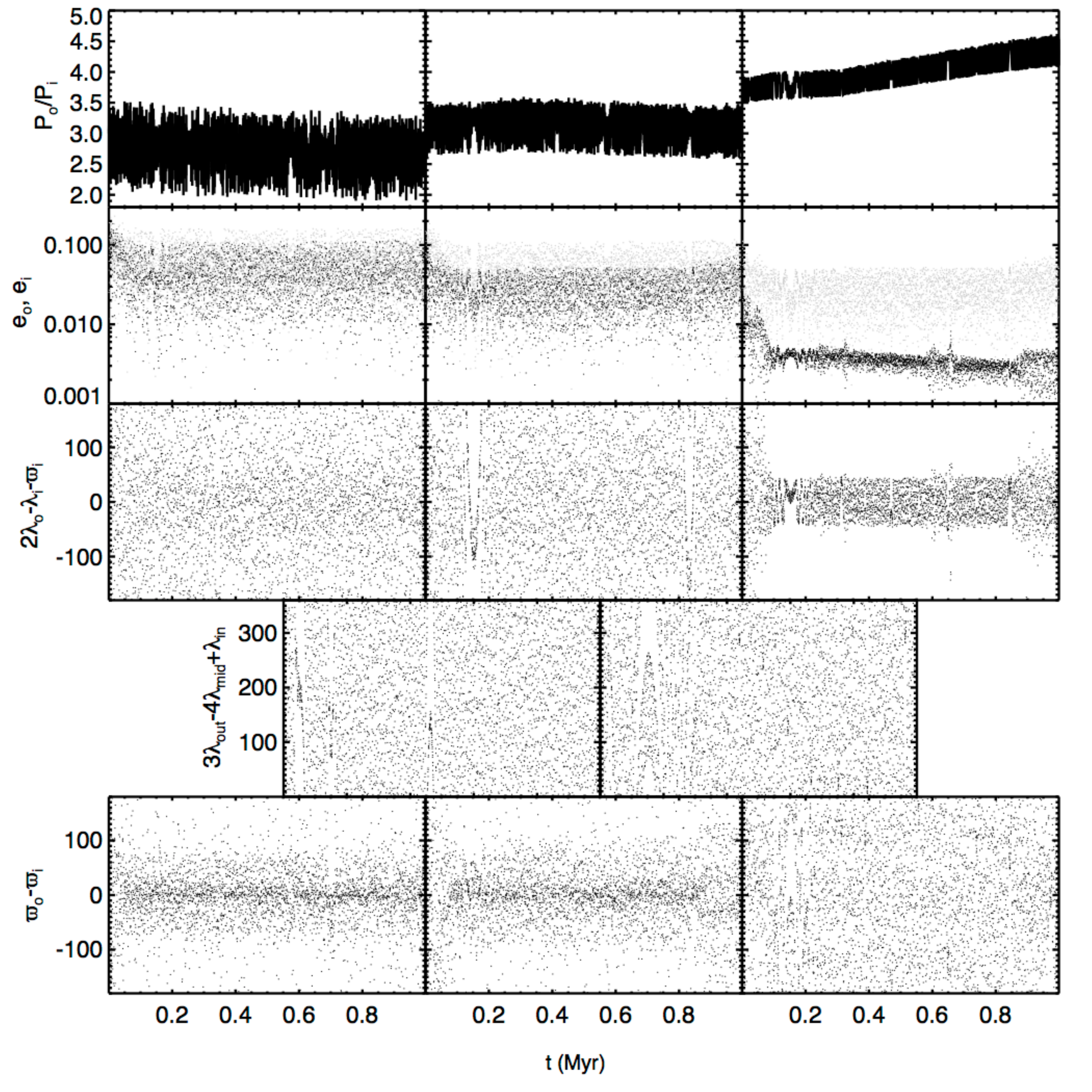}
\end{center}
\figcaption{Evolution of orbital elements for System {\tt 4-10b} with gas damping included (Table~\ref{tab:planetarysystems}; $\Sigma_{30} = 0.01$ g cm$^{-2}$). Row 1: period ratio of each adjacent pair of planets, from outer to inner. Row 2: eccentricity of outer planet (gray) and inner planet (black). Row 3: 2:1 mean motion resonant argument involving the longitude of periapse of the inner planet. Row 4: Three-body resonant argument. Row 5: Separation of adjacent planets' longitude of periapses.
\label{fig:app12}}
\end{figure}

\begin{figure}
\begin{center}
\includegraphics[trim=0 0 0 0, clip,width=0.8\textwidth,angle=0]{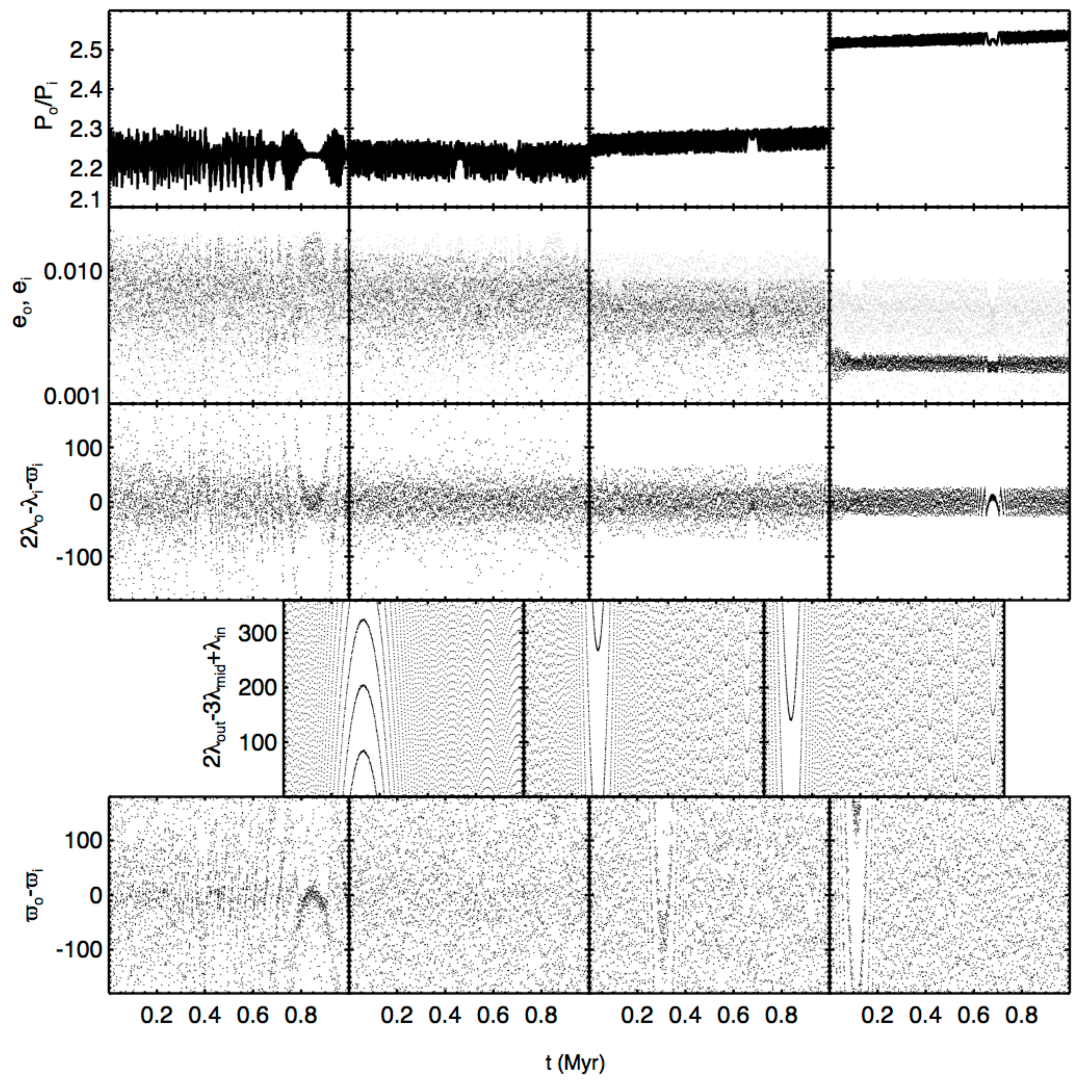}
\end{center}
\figcaption{Evolution of orbital elements for System {\tt 5-1} with gas damping included (Table~\ref{tab:planetarysystems}; $\Sigma_{30} = 0.1$ g cm$^{-2}$). Row 1: period ratio of each adjacent pair of planets, from outer to inner. Row 2: eccentricity of outer planet (gray) and inner planet (black). Row 3: 2:1 mean motion resonant argument involving the longitude of periapse of the inner planet. Row 4: Three-body resonant argument (Laplace resonance). Row 5: Separation of adjacent planets' longitude of periapses.
\label{fig:app13}}
\end{figure}

\begin{figure}
\begin{center}
\includegraphics[trim=0 0 0 0, clip,width=0.8\textwidth,angle=0]{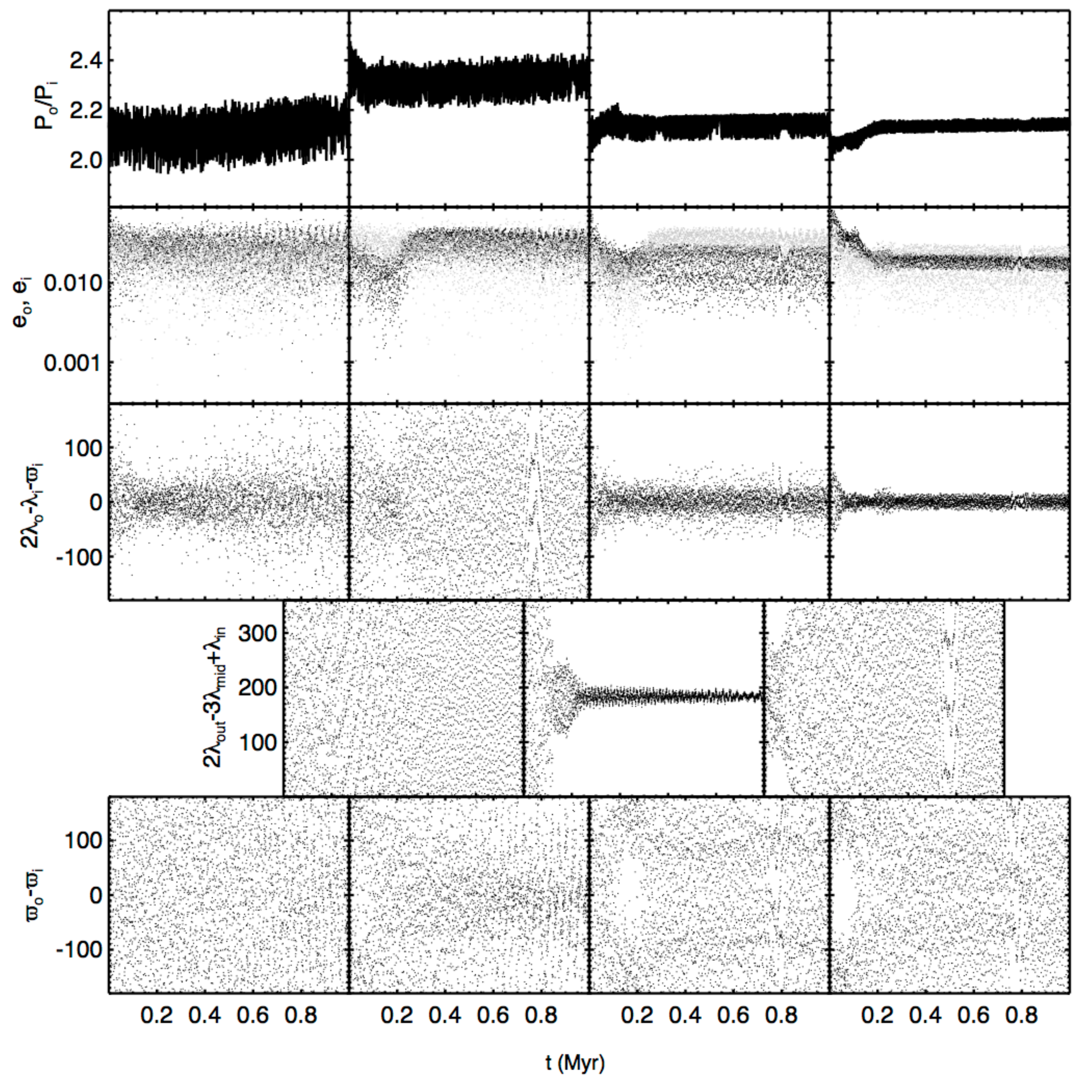}
\end{center}
\figcaption{Evolution of orbital elements for System {\tt 5-2a} with gas damping included (Table~\ref{tab:planetarysystems}; $\Sigma_{30} = 0.01$ g cm$^{-2}$). Row 1: period ratio of each adjacent pair of planets, from outer to inner. Row 2: eccentricity of outer planet (gray) and inner planet (black). Row 3: 2:1 mean motion resonant argument involving the longitude of periapse of the inner planet. Row 4: Three-body resonant argument (Laplace resonance). Row 5: Separation of adjacent planets' longitude of periapses.
\label{fig:app14}}
\end{figure}

\begin{figure}
\begin{center}
\includegraphics[trim=0 0 0 0, clip,width=0.8\textwidth,angle=0]{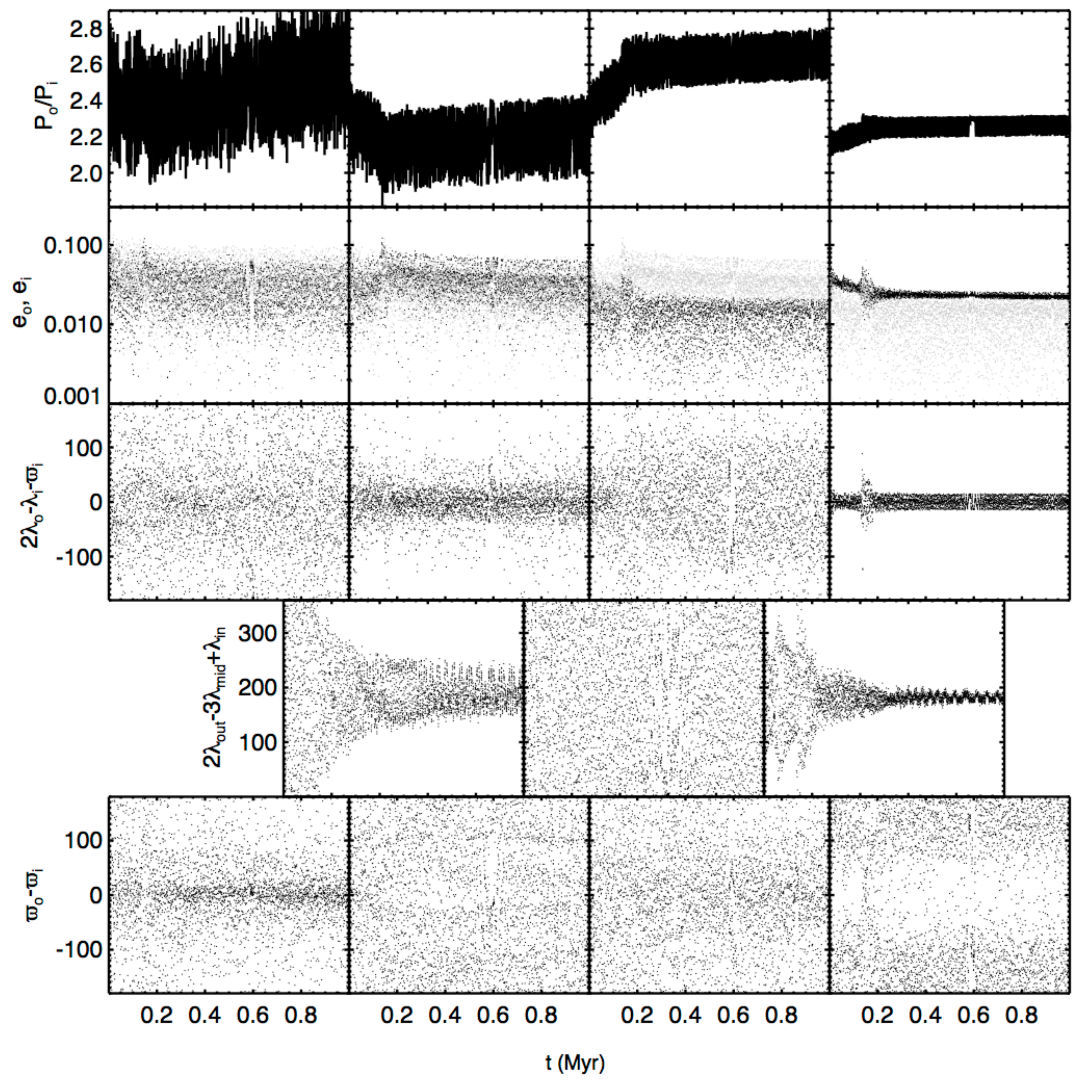}
\end{center}
\figcaption{Evolution of orbital elements for System {\tt 5-5a} with gas damping included (Table~\ref{tab:planetarysystems}; $\Sigma_{30} = 0.01$ g cm$^{-2}$). Row 1: period ratio of each adjacent pair of planets, from outer to inner. Row 2: eccentricity of outer planet (gray) and inner planet (black). Row 3: 2:1 mean motion resonant argument involving the longitude of periapse of the inner planet. Row 4: Three-body resonant argument (Laplace resonance). Row 5: Separation of adjacent planets' longitude of periapses.
\label{fig:app15}}
\end{figure}

\begin{figure}
\begin{center}
\includegraphics[trim=0 0 0 0, clip,width=0.8\textwidth,angle=0]{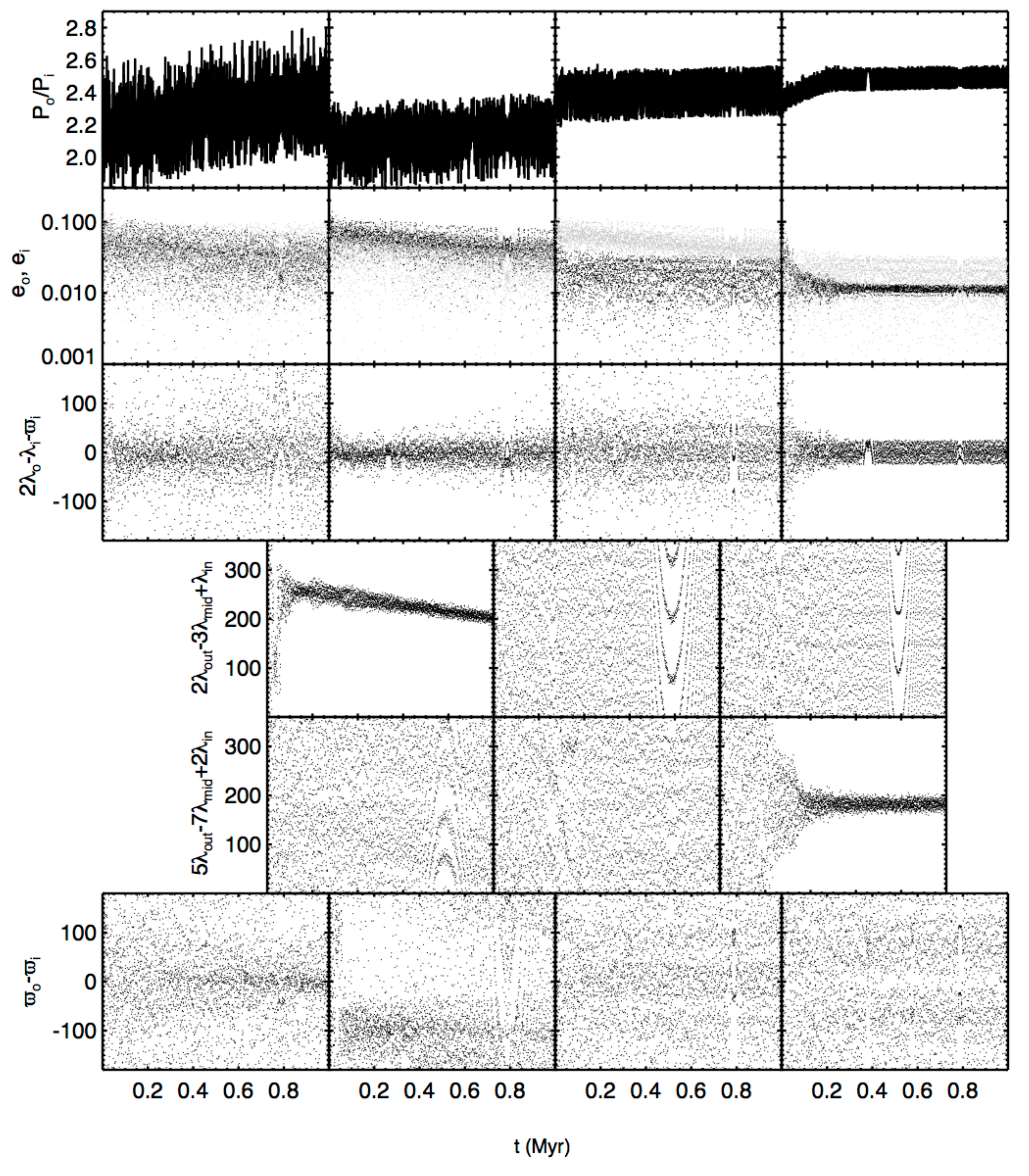}
\end{center}
\figcaption{Evolution of orbital elements for System {\tt 5-5b} with gas damping included (Table~\ref{tab:planetarysystems}; $\Sigma_{30} = 0.01$ g cm$^{-2}$). Row 1: period ratio of each adjacent pair of planets, from outer to inner. Row 2: eccentricity of outer planet (gray) and inner planet (black). Row 3: 2:1 mean motion resonant argument involving the longitude of periapse of the inner planet. Row 4: Three-body resonant argument (Laplace resonance). Row 5: Separation of adjacent planets' longitude of periapses.
\label{fig:app16}}
\end{figure}

\begin{figure}
\begin{center}
\includegraphics[trim=0 0 0 0, clip,width=0.8\textwidth,angle=0]{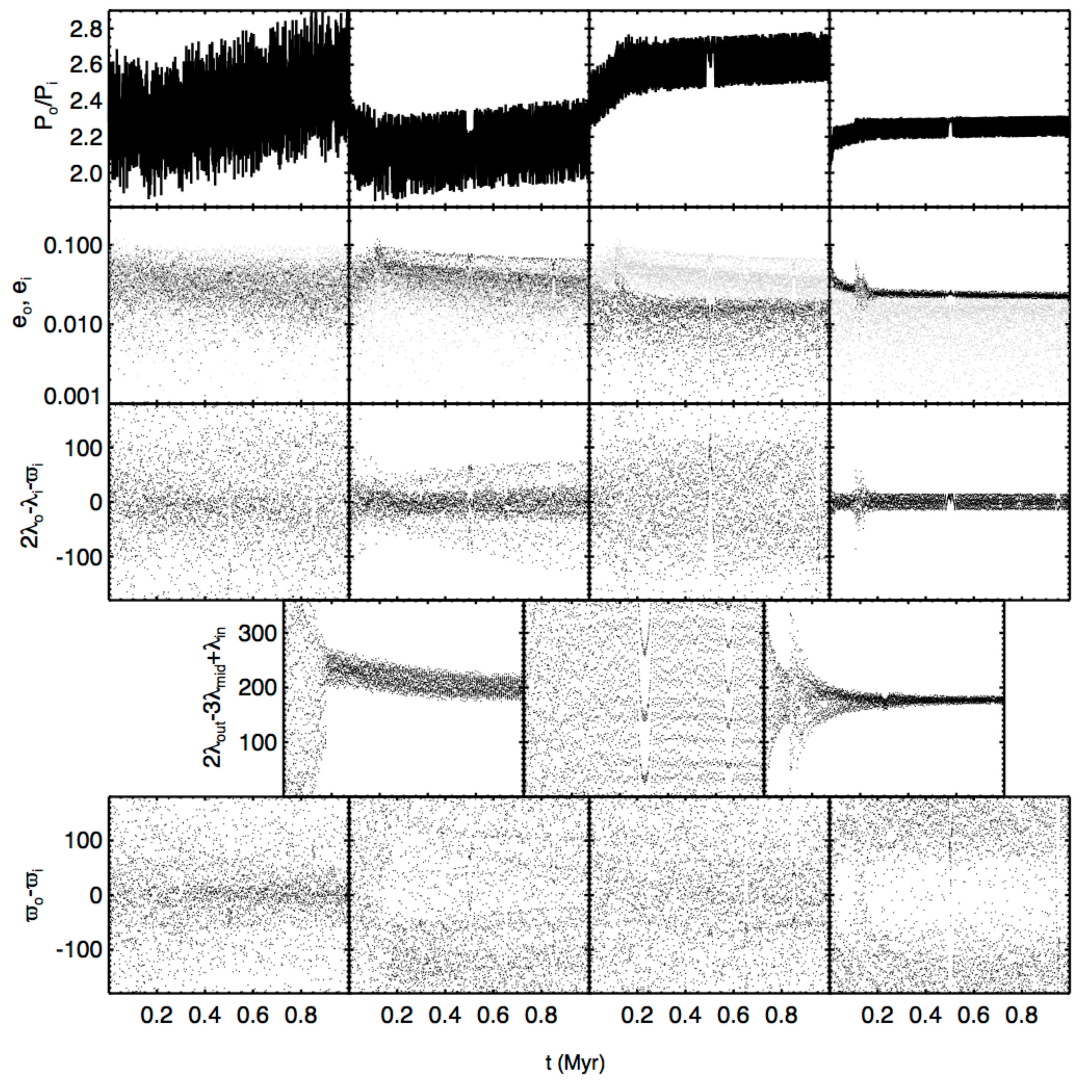}
\end{center}
\figcaption{Evolution of orbital elements for System {\tt 5-5c} with gas damping included (Table~\ref{tab:planetarysystems}; $\Sigma_{30} = 0.01$ g cm$^{-2}$). Row 1: period ratio of each adjacent pair of planets, from outer to inner. Row 2: eccentricity of outer planet (gray) and inner planet (black). Row 3: 2:1 mean motion resonant argument involving the longitude of periapse of the inner planet. Row 4: Three-body resonant argument (Laplace resonance). Row 5: Separation of adjacent planets' longitude of periapses.
\label{fig:app17}}
\end{figure}

\begin{figure}
\begin{center}
\includegraphics[trim=0 0 0 0, clip,width=0.8\textwidth,angle=0]{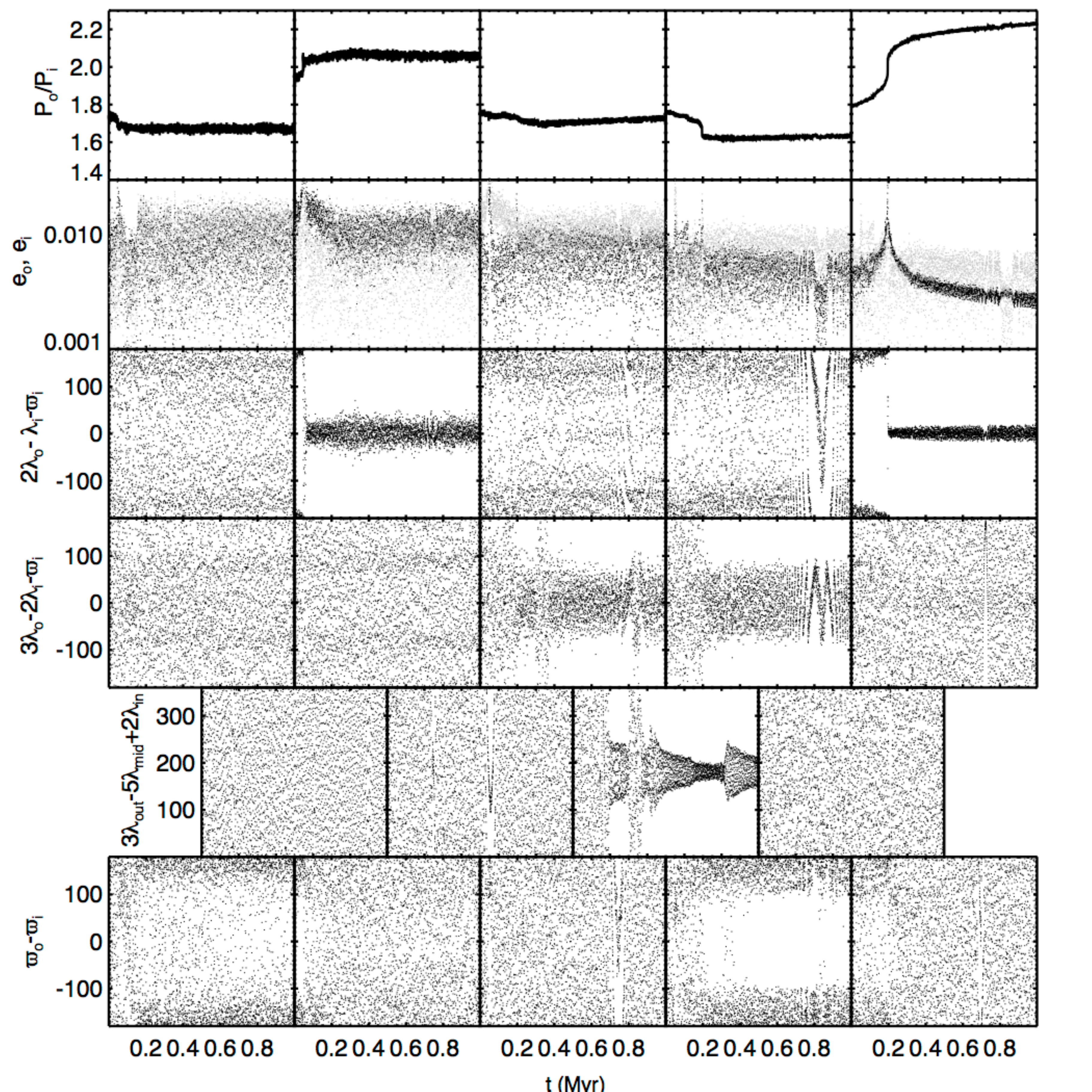}
\end{center}
\figcaption{Evolution of orbital elements for System {\tt 6-0.5} with gas damping included (Table~\ref{tab:planetarysystems}; $\Sigma_{30} = 1$ g cm$^{-2}$). Row 1: period ratio of each adjacent pair of planets, from outer to inner. Row 2: eccentricity of outer planet (gray) and inner planet (black). Row 3: 2:1 mean motion resonant argument involving the longitude of periapse of the inner planet. Row 4: 3:2 mean motion resonant argument involving the longitude of periapse of the inner planet. Row 5: Three-body resonant argument. Row 6: Separation of adjacent planets' longitude of periapses.
\label{fig:app18}}
\end{figure}

\begin{figure}
\begin{center}
\includegraphics[trim=0 0 0 0, clip,width=0.8\textwidth,angle=0]{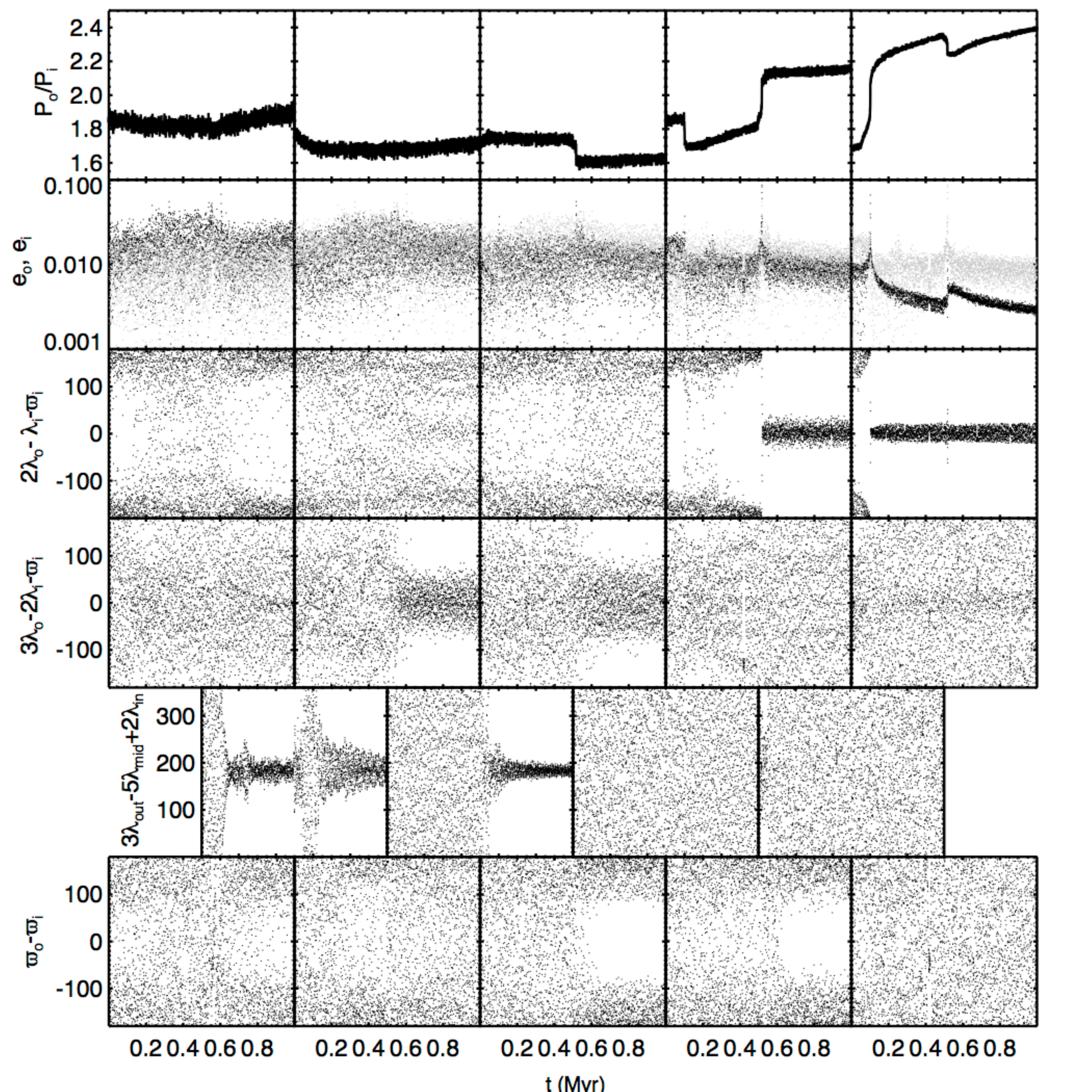}
\end{center}
\figcaption{Evolution of orbital elements for System {\tt 6-1} with gas damping included (Table~\ref{tab:planetarysystems}; $\Sigma_{30} = 1$ g cm$^{-2}$). Row 1: period ratio of each adjacent pair of planets, from outer to inner. Row 2: eccentricity of outer planet (gray) and inner planet (black). Row 3: 2:1 mean motion resonant argument involving the longitude of periapse of the inner planet. Row 4: 3:2 mean motion resonant argument involving the longitude of periapse of the inner planet. Row 5: Three-body resonant argument. Row 6:  Separation of adjacent planets' longitude of periapses.
\label{fig:app19}}
\end{figure}

Since our case studies leads us to expect a range of resonant disruption timescales spanning magnitude, we consider a third hypothesis that replaces Eqn. \ref{eqn:probres} with a truncated log-uniform distribution
\begin{eqnarray}
    {\rm prob_{res} } (t_\star)= \frac{1}{\log(\rm t_{res,max}) - \log(1 \rm Myr)} \int_{\log(1 \rm Myr)}^{\log(\rm t_{res,max})} \exp \left[-t_{\star, \rm res} \exp (-x) \right] dx \label{eqn:probres2}.
\end{eqnarray}
We plot the results in Fig. \ref{fig:res2}. The maximum probability occurs for $f_{\rm res,0} = 1$ and $t_{\rm res, max} = 38$ Gyr and is about four times higher than the probability of $f_{\rm res,0} = 5/30$ and $t_{\rm res, max} = \infty$ (i.e., such that the difference in age is due to chance).

Based on the assessment here, it is unclear whether the average younger ages stellar age of 2:1 resonant planets is due to chance  or a manifestation of resonance disruption. The hypothesis that all systems begin in resonance and are disrupted on a timescale that happens to fall within to the (single order of magnitude) range of ages in the sample is best at explaining the age difference but unlikely a priori. An alternative hypothesis -- that most planets start out in resonance but that resonances are disrupted on log uniform range of timescales -- is more likely a priori from the results of our study but not a clear winner over random chance. To better study resonance disruption empirically, we need a sample with a wide (multiple orders of magnitude) range of stellar ages and reliable stellar age uncertainties.

\end{appendix}
\clearpage

\begin{table}[h]
\centering
\caption{Gap opening mass $\qgap$ and planet extrapolated occurrence rates $\fplanet$}.
\footnotesize
\label{tab:mplimit}
\begin{tabular}{c|c|c|ccc}
\hline
$\deltagapmin$ & $h/r$ & $\alpha$ & $\qgap$ & $\fplanet(\deltagapmin|\alpha,h/r)$ & $\fplanet(\deltagapmin|\alpha,h/r)$ \\ \hline
 &  &  & ($10^{-3}$) & RV-based distribution & DI-based distribution  \\ \hline
\multirow{6}{*}{10} & \multirow{3}{*}{0.05} & 0.001 & 0.3 & 23.4\% & 1.6\% \\
 &  & 0.01 & 1.2 & 11.0\% & 1.2\% \\
 &  & 0.1 & 5.3 & 3.2\% & 0.6\% \\ \cline{2-6} 
 & \multirow{3}{*}{0.1} & 0.001 & 2.2 & 7.3\% & 1.0\% \\
 &  & 0.01 & $>13$ &  &  \\
 &  & 0.1 & $>13$ &  &  \\ \hline
\multirow{6}{*}{100} & \multirow{3}{*}{0.05} & 0.001 & 0.8 & 14.0\% & 1.4\% \\
 &  & 0.01 & 3.4 & 5.1\% & 0.8\% \\
 &  & 0.1 & $>13$ &  &  \\ \cline{2-6} 
 & \multirow{3}{*}{0.1} & 0.001 & 6.3 & 2.5\% & 0.5\% \\
 &  & 0.01 & $>13$ &  &  \\
 &  & 0.1 & $>13$ &  &  \\ \hline
\multirow{6}{*}{1000} & \multirow{3}{*}{0.05} & 0.001 & 2.2 & 7.3\% & 1.0\% \\
 &  & 0.01 & 9.9 & 0.9\% & 0.2\% \\
 &  & 0.1 & $>13$ &  &  \\ \cline{2-6} 
 & \multirow{3}{*}{0.1} & 0.001 & $>13$ &  &  \\
 &  & 0.01 & $>13$ &  &  \\
 &  & 0.1 & $>13$ &  &  \\ \hline
\end{tabular}
\tablecomments{The gap opening planet mass limit $\qgap$ for $\deltagapmin=10$, 100, 1000 under every combination of disk parameters $\alpha\in[0.1, 0.01, 0.001]$ and $h/r\in[0.05,0.1]$ (see Section~\ref{sec:mp} for details); and the fraction of solar type stars, $\fplanet(\deltagapmin|\alpha,h/r)$, hosting planets with $\qgap<q<13\times10^{-3}$ at 1~AU$<a<$50~AU per solar type star for each disk type, based on the RV planet distribution (Equation~\ref{eqn:cumming}, \citealt{cumming08}) and the direct-imaging (DI) planet distribution (Equation~\ref{eqn:brandt}, \citealt{brandt14}) (see Section~\ref{sec:fplanet} for details).}
\end{table}

\begin{table}
\centering
\caption{Planetary systems}
\scriptsize
\label{tab:planetarysystems}
\begin{tabular}{l|l|c|c|c|cccccccccc}
\hline
&Name 	&	 $\deltagapmin$                       	&	 $h/r$                  	&	 $\alpha$               	&	 $q$       	&	 $\rgap$ 	&	 $\rinner$ 	&	 $r_{p1}$ 	&	 $r_{p2}$ 	&	 $r_{p3}$ 	&	 $r_{p4}$ 	&	 $r_{p5}$ 	&	 $r_{p6}$ 	&	  $r_i-r_j$ 	\\ \hline
 &  	&	                                      	&	                        	&	                        	&	 ($10^{-3}$) 	&	 (AU)      	&	 (AU)        	&	 (AU)       	&	 (AU)       	&	 (AU)       	&	 (AU)       	&	 (AU)       	&	 (AU)       	&	($\rmh$)   	\\ \hline
1 &{\tt 3-10a}	&	 \multirow{12}{*}{10} 	&	 \multirow{10}{*}{0.05} 	&	 \multirow{5}{*}{0.001} 	&	10	&	30	&	1.2	&	18.1	&	6.2	&	2.1	&	          	&	          	&	          	&	5.2	\\ \cline{1-1}
2 &{\tt 3-5}	&	                                      	&	                        	&	                        	&	5	&	30	&	1.5	&	19.4	&	7.1	&	2.6	&	          	&	          	&	          	&	6.2	\\ \cline{1-1}
3 &{\tt 4-2	}&	                                      	&	                        	&	                        	&	2	&	30	&	1.8	&	21.6	&	10.7	&	5.3	&	2.7	&	          	&	          	&	6.1	\\ \cline{1-1}
4 &{\tt 5-1	}&	                                      	&	                        	&	                        	&	1	&	30	&	1.9	&	23.1	&	13.2	&	7.6	&	4.4	&	2.5	&	          	&	6.2	\\ \cline{1-1}
5 &{\tt 6-0.5}	&	                                      	&	                        	&	                        	&	0.5	&	30	&	2.8	&	24.9	&	16.8	&	11.3	&	7.6	&	5.2	&	3.5	&	5.6	\\ \cline{1-1} \cline{5-15} 
6 &{\tt 3-10b}	&	                                      	&	                        	&	 \multirow{3}{*}{0.01}  	&	10	&	30	&	1.6	&	18.5	&	6.9	&	2.6	&	          	&	          	&	          	&	4.8	\\ \cline{1-1}
7 &{\tt 4-5c}	&	                                      	&	                        	&	                        	&	5	&	30	&	1.7	&	21.3	&	10.4	&	5.1	&	2.5	&	          	&	          	&	4.6	\\ \cline{1-1}
8 &{\tt 5-2b}	&	                                      	&	                        	&	                        	&	2	&	30	&	2.9	&	23.9	&	15	&	9.4	&	5.9	&	3.7	&	          	&	4.2	\\ \cline{1-1} \cline{5-15} 
9 &{\tt 4-10a}	&	                                      	&	                        	&	 \multirow{2}{*}{0.1}   	&	10	&	30	&	1.4	&	21	&	9.8	&	4.5	&	2.1	&	          	&	          	&	3.9	\\ \cline{1-1}
10 &{\tt 5-5a}	&	                                      	&	                        	&	                        	&	5	&	30	&	1.9	&	22.9	&	13.2	&	7.5	&	4.3	&	2.5	&	          	&	3.6	\\ \cline{1-1} \cline{4-15} 
11 &{\tt 3-10d}	&	                                      	&	 \multirow{2}{*}{0.1}   	&	 \multirow{2}{*}{0.001} 	&	10	&	30	&	2.2	&	19	&	8	&	3.3	&	          	&	          	&	          	&	4.4	\\ \cline{1-1}
12 &{\tt 4-5b}	&	                                      	&	                        	&	                        	&	5	&	30	&	1.7	&	20.7	&	10.1	&	4.9	&	2.4	&	          	&	          	&	4.6	\\ \hline
13 &{\tt 3-10c}	&	 \multirow{8}{*}{100}   	&	 \multirow{6}{*}{0.05}  	&	 \multirow{4}{*}{0.001} 	&	10	&	30	&	2.1	&	19.6	&	8.1	&	3.3	&	          	&	          	&	          	&	4.4	\\ \cline{1-1}
14 &{\tt 4-5a}	&	                                      	&	                        	&	                        	&	5	&	30	&	1.6	&	21	&	10.1	&	4.8	&	2.3	&	          	&	          	&	4.7	\\ \cline{1-1}
15 &{\tt 5-2a}	&	                                      	&	                        	&	                        	&	2	&	30	&	2.2	&	23.3	&	13.8	&	8.2	&	4.8	&	2.9	&	          	&	4.6	\\ \cline{1-1}
16 &{\tt 6-1	}&	                                      	&	                        	&	                        	&	1	&	30	&	2.9	&	24.9	&	16.9	&	11.5	&	7.8	&	5.3	&	3.6	&	4.4	\\ \cline{1-1} \cline{5-15} 
17 &{\tt 4-10b}	&	                                      	&	                        	&	 \multirow{2}{*}{0.01}  	&	10	&	30	&	1.5	&	20.9	&	9.8	&	4.6	&	2.2	&	          	&	          	&	3.9	\\ \cline{1-1}
18 &{\tt 5-5b}	&	                                      	&	                        	&	                        	&	5	&	30	&	1.9	&	22.9	&	13.2	&	7.5	&	4.3	&	2.5	&	          	&	3.6	\\ \cline{1-1} \cline{4-15} 
19 &{\tt 4-10c}	&	                                      	&	 \multirow{2}{*}{0.1}   	&	 \multirow{2}{*}{0.001} 	&	10	&	30	&	1.8	&	20.6	&	10.2	&	5	&	2.5	&	          	&	          	&	3.6	\\ \cline{1-1}
20 &{\tt 5-5c}	&	                                      	&	                        	&	                        	&	5	&	30	&	2	&	22.5	&	13.1	&	7.6	&	4.4	&	2.6	&	          	&	3.5	\\ \hline
21 &{\tt 3-10e}	&	 \multirow{4}{*}{1000}  	&	 \multirow{4}{*}{0.05}  	&	 \multirow{3}{*}{0.001} 	&	10	&	30	&	2.9	&	20.6	&	9.4	&	4.3	&	          	&	          	&	          	&	4.0	\\ \cline{1-1}
22 &{\tt 4-5d}	&	                                      	&	                        	&	                        	&	5	&	30	&	2.5	&	22	&	11.8	&	6.3	&	3.4	&	          	&	          	&	4.0	\\ \cline{1-1}
23 &{\tt 6-2}	&	                                      	&	                        	&	                        	&	2	&	30	&	2.2	&	24.2	&	15.7	&	10.2	&	6.6	&	4.3	&	2.8	&	3.8	\\ \cline{1-1} \cline{5-15} 
24 &{\tt 4-10d}	&	                                      	&	                        	&	0.01	&	10	&	30	&	2.8	&	23.1	&	12.8	&	7.1	&	3.9	&	          	&	          	&	3.1		 \\ \hline
\end{tabular}
\tablecomments{Planetary systems constructed to open a common gap from $\rgap=30$~AU to $\rinner$ ($<0.1$~$\rgap$) around a solar type star, with a minimum gap depletion factor $\deltagapmin$ in a disk with $h/r$ and $\alpha$, using equal mass planets with $\qgap<q<13\times10^{-3}$. The last column is the separation between a pair of adjacent planets $i$ and $j$, $r_i-r_j$, in mutual Hill radii,  $\rmh=\frac{r_i+r_j}{2}\left(\frac{2q}{3}\right)^{1/3}$. Systems are named as [``number-of-planets''-``planet mass''(a, b, c... in case of multiples)] (the second column). See Section~\ref{sec:planetrequirements} for details.}
\end{table}

\begin{table}
\centering
\caption{Stability}
\footnotesize
\begin{tabular}{l|l|c|c|c|ccllllll}
\hline
&Name & $\deltagapmin$ & $h/r$ & $\alpha$ & $q$ & $r_i-r_j$ & $\tu$ & $\tuq$ & $\Sigma_{30s,50\%}$&$\Sigma_{30s,90\%}$&$\Sigma_{30c}$&$\Sigma_{\rm gap}$\\ \hline
 &&  &  &  & ($10^{-3}$) & ($\rmh$) & (Myr) & (Myr) & g cm$^{-2}$ & g cm$^{-2}$& g cm$^{-2}$& g cm$^{-2}$\\ \hline
1 & {\tt 3-10a} & \multirow{12}{*}{10} & \multirow{10}{*}{0.05} & \multirow{5}{*}{0.001} & 10 & 5.2 & S & S  &0&0&0.1&3--30 $\times10^{-5}$  \\ \cline{1-1}
2 & {\tt 3-5}       &&&& 5 & 6.2 & S & S &0&0&$>$1&4--40 $\times10^{-4}$  \\ \cline{1-1}
3 & {\tt 4-2}       &&&& 2 & 6.1 & S & S &0&0&$>$1&0.004--0.04 \\ \cline{1-1}
 4 & {\tt 5-1}      &&&& 1 & 6.2 & S & S &0&0&$>$1&0.02--0.2 \\ \cline{1-1}
 5 & {\tt 6-0.5}    &&&&  0.5 & 5.6 & S & 4  &0&0.0001&$>$1&0.2--2   \\ \cline{1-1} \cline{5-13} 
 6 & {\tt 3-10b}    &&& \multirow{3}{*}{0.01} & 10 & 4.8 & S & S  &0&0&0.1&0.011--0.11  \\ \cline{1-1}
 7 & {\tt 4-5c}     &&&& 5 & 4.6 & S & 4 &0&0.001&0.1&0.03--0.3   \\ \cline{1-1}
 8 & {\tt 5-2b}     &&&& 2 & 4.2 & 0.015 & 0.004 &0.1&1&0.1&0.18--1.8\\ \cline{1-1} \cline{5-13} 
 9 & {\tt 4-10a}    &&&   \multirow{2}{*}{0.1} & 10 & 3.9 & 0.05 & 0.006 &0.0001&0.01&0.01&0.2--2\\ \cline{1-1}
10 & {\tt 5-5a}     &&&& 5 & 3.6 & 0.07 & 0.02 &0.01&0.01&0.01&0.9--9 \\ \cline{1-1} \cline{4-13}
11 & {\tt 3-10d}    && \multirow{2}{*}{0.1} & \multirow{2}{*}{0.001} & 10 & 4.4 & S & S &0&0&0.1&0.03--0.3  \\ \cline{1-1}
12 & {\tt 4-5b}     &&&& 5 & 4.6 & S & S &0&0.0001&0.1&0.097--0.9 \\ \hline
13 &{\tt 3-10c} &\multirow{8}{*}{100} & \multirow{6}{*}{0.05} & \multirow{4}{*}{0.001} & 10 & 4.4 & S & S &0&0&0.1&4--40 $\times10^{-5}$  \\ \cline{1-1}
14 & {\tt 4-5a}     &&&&5 & 4.7 & S & S &0&0&0.1&4--40 $\times10^{-4}$  \\ \cline{1-1}
15 & {\tt 5-2a}     &&&& 2 & 4.6 & 3 & 0.8 &0&0.0001&$>$1&0.004--0.04  \\ \cline{1-1}
16 & {\tt 6-1}      &&&&1 & 4.4 & 0.04 & 0.008 &1&1&$>1$&0.02--0.2  \\ \cline{1-1} \cline{5-13} 
17 & {\tt 4-10b} & & & \multirow{2}{*}{0.01} & 10 & 3.9 & 0.04 & 0.006 &0.001&0.1&0.01&0.011--0.11  \\ \cline{1-1}
18 & {\tt 5-5b}         &&&& 5 & 3.6 & 0.08 & 0.03  &0.001&0.01&0.01&0.03--0.3 \\ \cline{1-1} \cline{4-13}
19 & {\tt 4-10c} && \multirow{2}{*}{0.1} & \multirow{2}{*}{0.001} & 10 & 3.6 & 0.02 & 0.003  &0.1&0.1&0.01&0.03--0.3\\ \cline{1-1}
20 & {\tt  5-5c}         &&&&5 & 3.5 & 0.05 & 0.010  &0.01&0.1&0.01&0.097--0.97 \\ \hline
 21 & {\tt3-10e}&\multirow{4}{*}{1000} & \multirow{4}{*}{0.05} & \multirow{3}{*}{0.001} & 10 & 4.0 & S & 0.6  &0&0.0001&0.1&4--40 $\times10^{-5}$  \\ \cline{1-1}
22 & {\tt 4-5d }         &&&&5 & 4.0 & 4 & 0.9&0&0.001&0.1&4--40 $\times10^{-4}$ \\ \cline{1-1}
23 & {\tt 6-2}         &&&&2 & 3.8 & 0.0013 & 0.0005 &10&10&1&0.004--0.04 \\ \cline{1-1} \cline{5-13} 
24 & {\tt 4-10d}         &&&0.01 & 10 & 3.1 & 0.004 & 0.002 &0.1&0.1&0.01&0.011--0.11 \\ \hline
\end{tabular}
\label{tab:stability}
\tablecomments{Outcomes of the $N$-body simulations for the multi-planet systems in Table~\ref{tab:planetarysystems}. All simulations are run for 27 Myr. The times $\tu$ and $\tuq$ are times at which 50\% and 25\% of the systems go unstable when gas damping forces is not included, respectively. A value of S for $\tu$ or $\tuq$ indicates that the system is stable for at least 27 Myr. $\Sigma_{30s,50\%}$ and $\Sigma_{30s,90\%}$ are the minimum gas gap surface density normalizations at 30 AU for at least 50\% and 90\% systems to be stable for at least 1~Myr. $\Sigma_{30c}$ is the upper limit of $\Sigma_{30}$ for the systems to stay compact (i.e., to avoid excessive repulsion). The last column is $\Sigma_{30}$ calculated as $\Sigma_{30,0}=$10--100 g cm$^{-2}$ (the undepleted value) divided by the depletion factor at the bottom of the gap from the \citet{fung14} models. See Section~\ref{sec:stability} for details.}
\end{table}

\begin{table}
\centering
\caption{Resonances}
\footnotesize
\begin{tabular}{l|l|lllll|l|l}
\\
\hline
Name & $\Sigma_{30}$ & \multicolumn{5}{c}{Libration}&	Commensurate& Fig.\\
	&	&	2:1 $\varpi_i$	&	Other MMR & Three-body	&	$\Delta \varpi$	& Post-gas&		 \\
& g cm$^{-2}$&&&&&&\\
\hline
{\tt 3-5}	&	0.01	&	Some	&	None	&	None & Some &     All&Some	&	\ref{fig:app1}   	\\
{\tt 3-10a}	&	0.001	&	Some	&	None	&	None & Some   & All&Some	&	\ref{fig:app2}     	\\
{\tt 3-10b}	&	0.01	&	Some	&	None	&	None & Some   & All&None	&	\ref{fig:app3}     	\\
{\tt 3-10c}	&	0.001	&	Some	&	None	&	None & Some&  All&None	&	\ref{fig:app4}	   	\\
{\tt 3-10d}	&	0.01	&	Some	&	None	&	None & Some& All & Some	&	\ref{fig:app5}	    	\\
{\tt 3-10e}	&	0.1	    &	Some	&	None	&	None & Some &  All& Some	&	\ref{fig:app6}	   	\\
{\tt 4-2}	&	0.01	&	Some	&	None	&	None & Some  & All & None	&	\ref{fig:app7}	   	\\
{\tt 4-5a}	&	0.001	&	Some	&	None	&	None & Some & All&Some	&	\ref{fig:app7b}	    	\\
{\tt 4-5b}	&	0.01	&	Some	&	None	&	None & Some & None$^{\rm a}$& Some	&	\ref{fig:app8}	    	\\ 
{\tt 4-5c}	&	0.1	     & 	Some	&	None	&	None & Some  & All & Some	&	\ref{fig:app9}	  	\\
{\tt 4-5d}	&	0.1	    &	Some	&	None	&	All & Some & All& Some	&	\ref{fig:app10}	    	\\
{\tt 4-10a}	&	0.01	&	Some	&	None	&	Some & Some&Some$^{\rm b}$ & None	&	\ref{fig:app11}	       \\ 
{\tt 4-10b}	&	0.01	&	Some	&	None	&	None & Some   &All& Some	&	\ref{fig:app12}	      \\
{\tt 5-1}	&	0.1	    &	All	    &	None	&	None & None  &All& Some	&	\ref{fig:app13}        \\
{\tt 5-2a}	&	0.01	&Some	    &	None	&	Some & Some &All   & All	&	\ref{fig:app14}	     \\
{\tt 5-2b}	&	0.1	    &Some	    &	Some	&	Some & Some & All & All	& 	\ref{fig:damping}  	\\
{\tt 5-5a}	&	0.01	&Some   	&	None	&	Some & Some & All & Some	&	\ref{fig:app15}    	\\
{\tt 5-5b}	&	0.01	&All    	&	None	&	Some & Some& All & Some	&	\ref{fig:app16}	    	\\
{\tt 5-5c}	&	0.01	&All    	&	None	&	Some & All& All & Some	&	\ref{fig:app17}	    	\\
{\tt 6-0.5}	&	1	    &Some	    &	Some	&	Some & Some& All & All	&	\ref{fig:app18}	    	\\
{\tt 6-1}	&	1	    &All    	&	Some	&	Some & Some  & None$^{\rm c}$ & All	&	\ref{fig:app19}   	\\ 
\hline
\end{tabular}
\label{tab:res}
\tablecomments{Resonant behavior of configurations. Column 1: Configuration name. Column 2:  $\Sigma_{30}$ used in the illustrative simulation. Column 3--7:  libration of arguments: ``all'' indicates that each adjacent pair librates, ``some'' that one or more adjacent pairs librate, and ``none'' that no pairs librate. The arguments are: 2:1 $\varpi_i$ ( libration of the 2:1 resonant argument involving $\varpi_i$), other MMR (a different mean motion resonant argument, such as the 3:2, e.g., Fig. \ref{fig:app18}, row 4). three-body (three-body resonance, such as Laplace), $\Delta \varpi$ (separation of longitudes of periapse), and post-gas (whether the arguments librating when gas damping is imposed continue to librate after gas damping is shut off). Commensurate indicates that the period ratio is within 10\% of a first order resonance, 5\% of a second order resonance, 2.5\% of a third order resonance, 1.25\% of a fourth order resonance, or 0.625\% of a fifth order resonance. The column Fig. refers to the figure showing the angles tabulated here. \\
a Strong resonant repulsion in damped simulation. \\
b Resonant arguments had switched to circulating just before the end of the damped simulation.\\
c Quickly went unstable without gas damping.\\}
\end{table}

\clearpage

\clearpage


\clearpage

\end{document}